%% file: text.tex
\documentclass[11pt,a4paper]{article}
\pdfoutput=1
\usepackage{jcappub}

\usepackage[english]{babel}
\usepackage{xspace} 
\usepackage[tight]{subfigure} 
\usepackage{url}

\usepackage{cosmodefs.JCAP}

\usepackage{amsmath}
\usepackage{amssymb}
\usepackage{ifthen,xcolor}
\begin{document}

\date{\today}
\title{Enhanced Preheating after Multi-Field Inflation: On the Importance of being Special}
\author{Thorsten Battefeld$^{1)}$}
\emailAdd{tbattefe@astro.physik.uni-goettingen.de}
\author{Alexander Eggemeier$^{1)}$}
\emailAdd{a.eggemeier@stud.uni-goettingen.de}
\author{John T.~Giblin,~Jr.$^{2),3)}$}
\emailAdd{giblinj@kenyon.edu}
\affiliation{1) Institute for Astrophysics,
University of Goettingen,
Friedrich Hund Platz 1,
D-37077 Goettingen, Germany}
\affiliation{2) Department of Physics, Kenyon College, Gambier, OH 43022, U.S.A.}
\affiliation{3) Department of Physics, Case Western Reserve University, Cleveland, OH 44106, U.S.A.}

\abstract{We discuss preheating after multi-field inflation in the presence of several preheat matter fields that become light in the vicinity of (but not at) the inflatons' VEV, at distinct extra-species-points (ESP); this setup is motivated by inflationary models that include particle production during inflation, e.g.~trapped inflation, grazing ESP encounters or modulated trapping, among others. While de-phasing of inflatons tends to suppress parametric resonance, we find two new effects leading to efficient preheating: particle production during the first in-fall (efficient if many preheat matter fields are present) and a subsequent (narrow) resonance phase (efficient if an ESP happens to be at one of several  distinct distances from the inflatons' VEV). Particles produced during the first in-fall are comprised of many species with low occupation number, while the latter are made up of a few species with high occupation number.

We provide analytic descriptions of both phases in the absence of back-reaction, which we test numerically. We further perform lattice simulations to investigate the effects of back-reaction. We find resonances to be robust and the most likely cause of inflaton decay in multi-field trapped inflation if ESP distributions are dense.} 

\maketitle

\section{Introduction}

After inflation the Universe is empty and cold, in dire need to be reheated so that primordial nucleosynthesis can take place. The decay of the inflaton(s) can be perturbative (reheating) or explosive (preheating), depending on the inflationary model (see \cite{Bassett:2005xm,Kofman:2008zz,Allahverdi:2010xz} for reviews).

Preheating after inflation is often treated phenomenologically, since the degrees of freedom to which the inflaton(s) couple are unknown. Exceptions are certain low scale inflationary models, such as reheating after inflation within the MSSM \cite{Allahverdi:2011aj} or Higgs inflation \cite{Bezrukov:2008ut,GarciaBellido:2008ab,Figueroa:2009jw}. It was soon discovered that non-perturbative effects, such as parametric resonance \cite{Dolgov:1989us,Traschen:1990sw,Kofman:1997yn} or tachyonic instabilities \cite{Felder:2000hj,Dufaux:2006ee}, can lead to a fast transfer of energy from the inflaton sector to a preheat matter field. These effects, if present, can dominate over the decays channels of the old theory of reheating \cite{Dolgov:1982th,Abbott:1982hn}\footnote{This perturbative decay is not complete \cite{Braden:2010wd} and can lead to troublesome long-lived scalar particles, as in the cosmological moduli problem \cite{Banks:1993en,de Carlos:1993jw}} and require subsequent lattice simulations \cite{Prokopec:1996rr,Khlebnikov:1996zt,Khlebnikov:1996mc} \footnote{Common public codes are LATTIEEASY \cite{Felder:2000hq}, Defrost \cite{Frolov:2008hy}, PSpectRe \cite{Easther:2010qz}, HLATTICE \cite{Huang:2011gf} and PyCOOL \cite{Sainio:2012mw}.} to study thermalization \cite{Felder:2000hr,Podolsky:2005bw} and ultimately reheating. The subsequent decay of preheat matter fields is seldom considered, since it requires a concrete particle physics implementation of inflation as in \cite{Bezrukov:2008ut,GarciaBellido:2008ab,Allahverdi:2011aj,Figueroa:2009jw}.
An observable by-product of preheating can be gravitational waves \cite{Khlebnikov:1997di,GarciaBellido:1998gm,Easther:2006vd,Easther:2007vj,Dufaux:2007pt,Easther:2006gt,Dufaux:2008dn,Price:2008hq,Dufaux:2010cf,Giblin:2010sp}. The majority of these phenomenological studies
focus on one (see the reviews \cite{Bassett:2005xm,Allahverdi:2010xz}) or several \cite{Bassett:1997gb,Bassett:1998yd,Battefeld:2008bu,Battefeld:2008rd,Battefeld:2009xw,Braden:2010wd} inflaton fields, coupled to a single degree of freedom that becomes light at the vacuum expectation value (VEV) of the inflatons. For example, in standard, single-field chaotic inflation with potential $V(\varphi)=m^2\varphi^2/2$, one can introduce another light scalar field that couples to the inflaton via the interaction Lagrangian $\mathcal{L}=-g^2\chi^2\varphi^2/2$, such that the effective mass of $\chi$ is minimal at the VEV $\varphi=0$. As the inflaton oscillates around the minimum, $\chi$-particles are produced in bursts whenever this point is traversed, leading to preheating via parametric resonance (stochastic resonance in an expanding universe).

But why is this Extra Species Point (ESP) \cite{Kofman:2004yc} $\varphi_{ESP}=0$ aligned with the minimum of $V(\varphi)$? There is no {\sl a priori} reason why they should coincide (even though they may in certain models, as in \cite{Allahverdi:2011aj}); thus,  consequences of misalignment need to be investigated. Furthermore, there are several models of inflation, notably trapped inflation in one \cite{Kofman:2004yc,Green:2009ds,Silverstein:2008sg} or higher dimensional field spaces \cite{Battefeld:2010sw}, among others \cite{Chung:1999ve,Elgaroy:2003hp,Romano:2008rr} (see Sec.~\ref{sec:ESPinMFI} for a brief overview), that employ a dense ESP distribution during inflation. These ESPs do not miraculously vanish once inflation is over, in fact, the particles associated with ESPs that are within reach during the oscillations of the inflatons can play the role of preheat matter fields. 

Our goal is to investigate the feasibility of preheating in the presence of one or several ESPs in the vicinity of the inflatons' potentials' minimum, without placing them directly at the VEV (in the latter case, resonances are generically suppressed, due to de-phasing of the inflatons \cite{Battefeld:2008bu,Battefeld:2008rd,Battefeld:2009xw,Braden:2010wd}).

Focussing on more than one inflaton ($\mathcal{N}\geq 2$) and dense ESP distributions, we find several distinct phases: during the \emph{first in-fall}, the inflationary trajectory traverses a large number of ESPs at which single bursts of particle production take place. If ESPs are dense enough, this first infall can already be sufficient to transfer a large fraction of the energy into preheat matter fields before any oscillations take place. We provide analytic expressions for the comoving particle number, Sec.~\ref{sec:analyticinfall}, which we test numerically, Sec.~\ref{sec:compnumericsfirstinfall}.

The second phase covers oscillations with average amplitude that are bigger than the average inter ESP distance. The length and efficiency of this transitory phase is strongly model dependent and  dominated by chance close-encounters of the trajectory with an ESP. 

In the final stage, the average amplitude of oscillations is smaller than the inter ESP distance, Sec.~\ref{sec:resonancephase}. Preheating in this phase can be very efficient if an ESP has the ``right'' distance from the origin: opposite to the earlier stages, narrow resonance is responsible for particle production. Based on a series of approximations, we compute analytically the  location of resonance bands, which in turn,  set the  locations of ESPs where particle production can be operational for prolonged periods, Sec.~\ref{sec:instabilitybands}; again, we compare our predictions with numerics, Sec.~\ref{sec:numericalresults}, finding good agreement. We also test the dependence of resonance effects onto the mass distribution of the inflatons: while our analytic results are derived for simple mass ratios, they remain qualitatively valid for incommensurate masses. 

We summarize and discuss our results in Sec.~\ref{sec:summaryanddiscussion}, paying particular attention to trapped inflation, since ESPs are necessarily densely distributed in this model. 

To complement the analytic and numeric results based on treating inflatons as background fields, we incorporate back-reaction in Sec.~{\ref{PRwithbackreaction}} by performing lattice-simulations (based on a modified version of LATTICEEASY \cite{Felder:2000hq}) in selected cases. We find resonance to be robust and in line with the analytic predictions; if compared with resonances of a massive matter field at the VEV of the inflaton, we find resonances to be more efficient than the approximate analytic results indicate, Sec.~\ref{sec:latticeatsingleESP}, high-lightening the need for lattice simulations.

To set the stage for the above studies, we start with a brief introduction to the role of ESPs in multi-field inflation in Sec.~\ref{sec:ESPinMFI}. Our general setup is described in Sec.~\ref{sec:PRsetup}. We follow with a brief review of preheating with a single ESP at the VEV of the inflatons to introduce our notation and provide a reference point for our more complicated scenario.

Readers familiar with ESPs may skip Sec.~\ref{sec:ESPinMFI}, while readers familiar with preheating may want to skip Sec.~\ref{sec:review}.

\section{Setup\label{sec:setup}} 
In this section, we motivate the presence of ESP during and after inflation and provide the field content used in this paper.

\subsection{The role of ESPs in Multi-Field Inflation \label{sec:ESPinMFI}}
An Extra Species Point (ESP) is a point, or more generally an extended location (ESL), in field space at which additional degrees of freedom become light; consequently, they have to be incorporated into the low energy effective field theory, if a trajectory in field space comes close to the ESP, to allow for particle production and back-reaction of the produced particles. 

Such ESPs are generic in moduli spaces of string theory, see \cite{Seiberg:1994rs,Seiberg:1994aj,Witten:1995im,Intriligator:1995au,Strominger:1995cz,Witten:1995ex,Katz:1996ht,Bershadsky:1996nh,Witten:1995gx} for some examples, with many applications, such as moduli trapping \cite{Kofman:2004yc,Watson:2004aq,Patil:2004zp,Patil:2005fi,Cremonini:2006sx,Greene:2007sa}, effects on inflation \cite{Chung:1999ve,Elgaroy:2003hp,Romano:2008rr}, trapped inflation \cite{Kofman:2004yc,Green:2009ds,Silverstein:2008sg,Battefeld:2010sw} (see also \cite{Bueno Sanchez:2006eq,Bueno Sanchez:2006ah,Brax:2009hd,Matsuda:2010sm,Brax:2011si,Lee:2011fj}), modulated trapping \cite{Langlois:2009jp,Battefeld:2011yj} and preheating, among others.

These additional degrees of freedom are often associated with enhanced symmetries; for example, if two D-branes come close to each other (the distance between them is the modulus field), strings stretching between them become light (the new degree of freedom) leading to an enhanced gauge symmetry \cite{Witten:1995im}. In essence, we are dealing with the string-Higgs effect \cite{Bagger:1997dv,Watson:2004aq}.  

The positions of ESPs/ESLs are characteristic for the low energy effective field theory at hand, and little can be said in general about their distribution, which can range from dense, as in trapped inflation \cite{Kofman:2004yc,Green:2009ds,Silverstein:2008sg,Battefeld:2010sw}, to sparse, as in \cite{Battefeld:2011yj}. As a consequence, the effect of ESPs are often studied from a phenomenological point of view \cite{Kofman:2004yc,Battefeld:2011yj}, supplemented with concrete case studies (for example, monodromy inflation \cite{Silverstein:2008sg} is a realization of trapped inflation \cite{Kofman:2004yc,Green:2009ds}). 

In this paper, we model the presence of ESPs during preheating by the inclusion of additional scalar fields $\chi_{\vec{\alpha}}$  with a negligible bare mass  that we couple to the inflatons via quadratic interactions; the vector $\vec{\alpha}$ parametrizes the position of the ESP in an $\mathcal{N}$-dimensional field space, see Sec.~\ref{sec:PRsetup}. More realistic setups could involve the inclusion of gauge fields, as in \cite{Barnaby:2011qe}, or more general couplings, which we postpone to future studies.

Before turning our attention to preheating, we provide a brief summary of three exemplary inflationary setups that rely on the presence of ESPs: trapped inflation \cite{Kofman:2004yc,Green:2009ds,Silverstein:2008sg,Battefeld:2010sw}, grazing ESP encounters \cite{Battefeld:2011yj} and modulated trapping \cite{Langlois:2009jp,Battefeld:2011yj} (this selection is not complete  and chosen for illustrative purposes).   In these examples, the inclusion of several ESPs (not aligned with the VEV of the inflatons) during preheating is not an option, but mandatory, motivating the distributions we consider in Sec.~\ref{sec:PRsetup}. However, the reader should be aware that ESPs are a common occurrence in almost all effective field theories and their presence is at the core of any preheating study - it is just assumed that their location coincides with the inflaton's VEV, an assumption that commonly lacks justification.

\subsubsection{Trapped Inflation \label{sec:trapped}}
In trapped inflation \cite{Kofman:2004yc,Green:2009ds,Silverstein:2008sg,Battefeld:2010sw}, the backreaction of the produced particles onto the inflaton(s) is taken into account, leading to a modification of the low energy effective potential: as the ESP is traversed, particles are produced which lead to an attractive force towards the ESP. An encounter can lead to brief temporary trapping if enough particles are produced (brief, since the number of particles and thus backreaction redshifts with $a^{-3}$), but even if particle production is minor, i.e.~when the ESP is grazed with a reasonably large impact parameter, it can still lead to observational signals, see Sec.~\ref{sec:grazing}.

The term trapped inflation is commonly used if several ESPs are either encountered head on \cite{Kofman:2004yc,Green:2009ds,Silverstein:2008sg}, as in monodromy inflation \cite{Silverstein:2008sg,Brandenberger:2008kn}, or generically grazed \cite{Kofman:2004yc,Battefeld:2010sw} if the field space is higher dimensional. The repeated encounter with ESPs and the resulting extra contributions to the potential can prevent the inflatons from rolling down fast, prolonging inflation and consequently rendering steeper potentials admissible. If the number of inflatons is large $\mathcal{N}\gg 1$, and ESPs are ubiquitous, a terminal velocity in field space results that is independent of the potential \cite{Battefeld:2010sw}. 

As an example, consider modelling the additional degrees of freedom by massless fields quadratically coupled to the inflatons, $g^2(\vec{\varphi}-\vec{\varphi}^{\vec{\alpha}}_{ESP})^2\chi_{\vec{\alpha}}^2/2$, at ESPs with average inter ESP distance $x$; further, assume a quadratic potential for all inflatons with the mass set by the COBE normalization and start at a distance of $~0.2\,M_p$ from the origin (regular slow roll inflation is over at this point). If $g\approx 0.1$, $x\approx 5\times 10^{-5}\,M_p$ and $\mathcal{N}\geq 37$ (see \cite{Battefeld:2010sw} for details), the terminal velocity $v_t\approx gx^2$ is small enough to drive  an additional sixty e-folds of inflation and satisfy all observational constraints. The same mechanism can work with almost all steep potentials, as well as higher dimensional extra species locations \cite{Battefeld:2010sw}, opening up regions of field space for inflationary model building, previously deemed unsuitable. 

Observational consequences are varied: the leading order contribution to correlation functions are altered \cite{Battefeld:2010sw} while retaining the successes of slow roll inflation:  trapping effects slow the fields down, requiring the replacement of potential slow roll parameters with the Hubble slow evolution parameters, such as $\varepsilon\sim\dot{H}/H^2$, which in turn are set by the back-reaction of the light particles. If the trajectory turns, additional contributions arise due to the interaction with isocurvature modes. Further, once the trajectory moves away from the ESP, the formerly light particles become heavy again. Their backscattering onto the inflaton condensate leads to an infra-red cascade, causing additional bump like features in the power-spectrum and non-Gaussianities of peculiar shape \cite{Barnaby:2009mc,Barnaby:2009dd,Barnaby:2010ke,Barnaby:2010sq}.

After trapped inflation terminates, either because fields start to evolve fast or because the energy density in the additional degrees of freedom becomes comparable to the one in the inflatons, preheating commences in the presence of many ESPs. 

As we shall see in this paper, if ESP distributions are dense, particle production during the first in-fall towards the VEV can be efficient before even a single oscillation of the inflatons is complete; further, any remaining energy can be drained in a subsequent resonance phase. This qualitatively different type of preheating is the expected scenario for trapped inflation in higher dimensions \cite{Battefeld:2010sw}. 

\subsubsection{Grazing ESP Encounters \label{sec:grazing}}

If ESPs are less dense, the inflationary trajectory may still encounter one or several of them during the last sixty e-folds of inflation \cite{Battefeld:2011yj}. Back-reaction generically causes the trajectory  to bend and slow down, leading either to temporary trapping or a slightly prolonged phase of inflation. 

Which one occurs depends on the coupling strength to the light fields and the impact parameter $\mu$, that is the distance of closest approach to the ESP of the unperturbed trajectory. If the value of the coupling constant depends on other fields (generic in string theory), modulated trapping results \cite{Langlois:2009jp}, see Sec.~\ref{sec:modulated}; but even if the couplings are taken to be constant, additional observational signatures are possible: if the coupling is strong, $g\sim 1$, and the impact parameter is not too big, temporary trapping occurs: the resulting trajectory is chaotic \cite{Battefeld:2011yj}, and as a consequence, the duration of trapping depends sensitively on the impact parameter, which in turn fluctuates due to isocurvature perturbations. Since the time being trapped is not a smooth function of $\mu$, the resulting fluctuations in the power-spectrum are usually in excess of the COBE normalization; thus, most temporary trapping events are ruled out during the last sixty e-folds of inflation. However, if the trajectory bends slightly and slows down without turning around, the dependence of the number of e-folds on $\mu$, and thus isocurvature perturbations, is smooth, leading to additional contributions to the power-spectrum and non-Gaussianities, consistent with current bounds.
In addition to these signatures, localised and oscillatory features are also present via IR-cascading \cite{Barnaby:2009mc,Barnaby:2009dd,Barnaby:2010ke,Barnaby:2010sq}. 

Again, ESPs do not vanish once inflation terminates, but need to be incorporated into the study of preheating. As we shall see in this paper, even a single ESP at the right location can lead to prolonged resonances, which are cut short if the location of the ESP coincides with the VEV of the inflatons.

\subsubsection{Modulated Trapping\label{sec:modulated}}

To conclude our brief overview, we would like to comment on modulated trapping \cite{Langlois:2009jp,Battefeld:2011yj}, which is the application of modulated preheating \cite{Kofman:2003nx,Dvali:2003em}  to an ESP-encounter during inflation. If the coupling constant(s) depend on other fields, their fluctuations can be converted to curvature fluctuations during the encounter: instead of a dependence on isocurvature fields via the impact parameter, the effectiveness of trapping, or more precisely the decrease in speed of the inflatons, is modulated by fluctuations in additional field(s) via the coupling constant. 

The process of converting fluctuations is similar but not identical to  a grazing ESP encounter, ultimately leading to similar observational signatures \cite{Battefeld:2011yj}. Naturally, both effects are likely to coexist in realistic models of inflation.

To summarize, ESPs are ubiquitous in moduli spaces of string theory and depending on their concrete distribution, a wide array of observational signatures, such as non-Gaussianities, are possible. In the absence of a thorough understanding of their distribution in the region of field space that inflaton(s) traversed during the last sixty e-folds in our universe, it is prudent to investigate their effect onto the end of inflation, particularly preheating.

\subsection{A Generalized Setup for Preheating \label{sec:PRsetup}} 
Having motivated the presence of ESPs during/after inflation,  we specify the concrete phenomenological implementation used throughout.

\subsubsection{Inflaton fields} 
Consider $\mathcal{N}$ inflaton fields $\varphi_i$, $i=1\dots \mathcal{N}$, with canonical kinetic terms and quadratic potentials 
\begin{eqnarray}
V(\varphi_1,\dots,\varphi_\mathcal{N})=\sum_{i=1}^{\mathcal{N}}\frac{1}{2}m_i^2\varphi_i^2 \,. \label{potential}
\end{eqnarray}
This potential need not be valid during inflation - we just assume that we can expand the inflationary potential and diagonolize the mass matrix to describe the oscillations of the inflatons around the minimum. In general, the masses conform to a distribution that differs from $m_i=m_j$ for all $i,j$; for example, in N-flation \cite{Dimopoulos:2005ac} they conform to the Marcenkov-Pastur distribution \cite{Easther:2005zr}, spread around $m_{\mbox{\tiny COBE}} \approx 6.2 \times 10^{−6}/\sqrt{8\pi G}\approx 1.2\times 10^{−6} M_P$ which is set by the COBE normalization \cite{Komatsu:2010fb}. Occasionally, we set
\begin{eqnarray}
M_p^2=\frac{1}{G}\equiv 1\,.
\end{eqnarray}
Since we do not focus on a particular inflationary model, we consider a simpler mass distribution: after arranging the fields according to their masses such that $m_i<m_j$ if $i<j$, we spread the $m_i$ around 
\begin{eqnarray}
m\equiv 10^{-6}M_P\,.
\end{eqnarray}
If not stated otherwise, we distribute the masses over the interval $1\leq m_i/m\leq 1.5$, that is
\begin{equation}
m_i = m\left(1+\frac{(i-1)}{2(\mathcal{N}-1)}\right)\equiv \sqrt{\beta_i}m\,,\label{masses1}
\end{equation}
so that mass ratios are simple fractions (enabling a simple analytic description of some resonance effects). 

To check for a  possible dependence of results on the chosen  distribution, we occasionally use
\begin{equation}
\tilde{m}_i^2 = \frac{\mathcal{N}+2i-3}{2(\mathcal{N}-1)} m^2 \equiv \tilde{\beta}_i m^2\,,\label{masses2}
\end{equation}
as in \cite{Battefeld:2009xw}, so that the average square mass is
\begin{equation}
\left<\tilde{m}^2\right> =\frac{1}{\mathcal{N}}\sum_{I=1}^\mathcal{N}\tilde{\beta}_i m^2 
= \frac{1}{\mathcal{N}}\sum_{i=1}^\mathcal{N} \frac{(\mathcal{N}+2i-3)}{2(\mathcal{N}-1)} m^2 = \frac{\mathcal{N}^2+\mathcal{N}(\mathcal{N}+1)-3\mathcal{N}}{2\mathcal{N}(\mathcal{N}-1)}m^2= m^2\,,\label{masses}
\end{equation}
the lowest one  is $\tilde{m}_1^2=m^2/2$, and the greatest one is $m_{N}^2=3m^2/2$ (masses are incommensurate). Note that (\ref{masses1}) and (\ref{masses2}) are mutually exclusive definitions of the coefficients $\beta_i$ and $\tilde{\beta}_i$. 
One can motivate such narrow mass-distributions if (\ref{potential}) is also valid during inflation and all fields have comparable initial energy. In this case the lightest fields dominate the energy density at the end of inflation, and are thus responsible for preheating, because heavy fields have already relaxed to the minimum of their potential during inflation \cite{Battefeld:2008bu}. 

We also need to specify the field values at the end of inflation, that is the initial values for preheating. In the absence of a concrete model, we provide each of the inflatons in (\ref{potential}) with the same energy at the end of inflation summing up to $m^2 \varphi_{pr}^2/2$,
where we set 
\begin{eqnarray}
\varphi_{pr}\equiv \left|\vec{\varphi}(t_{pr})\right| \equiv 0.193 \, M_p\,,
\end{eqnarray}
 in line with single-field models \cite{Kofman:1997yn}. These choices yield
\begin{equation}
\label{initvalues}
\varphi_i(t_{pr}) =\sqrt{ \frac{1}{\mathcal{N}} \frac{1}{\beta_i} }\varphi_{pr}\,,
\end{equation} 
(analogously for $\tilde{\beta}_i$) as initial values for the inflatons.
For simplicity, we further set all initial velocities to zero
\begin{eqnarray}
\dot{\varphi_i}(t_{pr}) = 0\,,
\end{eqnarray}
since the speed towards the end of inflation is still slow roll suppressed.  While we use these initial conditions most of the time,
 we occasionally deviate from them for illustrative purposes, for example, setting 
\begin{eqnarray}
\varphi_i(t_{pr}) \equiv \varphi_{pr}\equiv \varphi_0\,.
\end{eqnarray}
 We would like to point out that our conclusions turn out to be largely insensitive to the choice of masses as well as the initial conditions.

\subsubsection{Matter Fields\label{sec:matterfields}}
We wish to include several preheat ``matter'' fields that become light at different ESPs in the vicinity of $|\vec{\varphi}|=0$; we deliberately avoid putting an ESP at   $|\vec{\varphi}|=0$, since preheating of such a field is well understood (see  \cite{Traschen:1990sw,Shtanov:1994ce,Kofman:1997yn,Bassett:1997gb,Bassett:1998yd,Battefeld:2008bu,Battefeld:2009xw,Braden:2010wd} for a small, non-representative collection, with particular focus on multiple inflatons) and generally suppressed if $\mathcal{N}>\mbox{few}$ \cite{Battefeld:2008bu,Battefeld:2008rd,Battefeld:2009xw, Braden:2010wd}.

We consider three types of ESP-distributions:
\begin{enumerate}
\item Along a grid
\begin{eqnarray}
\varphi^{\vec{\alpha}}_{i,\mbox{\tiny ESP}}=\left(\frac{1}{2}+\alpha_i\right)x
\end{eqnarray}
with $\alpha_i \in \mathbb{Z}$. This distribution is almost always used for numerics in the absence of back-reaction.
\item A random distribution, placing ESPs randomly over the accessible field space up until a desired average inter ESP distance is reached; used to check if the grid-like distribution leads to artefacts (it does not).
\item A Gaussian distribution centred at $\vec{\varphi}=0$ with variance $\sigma$ and desired $x$ near the origin; used for lattice simulations, focussing on the late stages of preheating when the oscillation amplitude is smaller than $\sigma$.
\end{enumerate}

For simplicity, we ignore interactions between matter fields and their bare masses (a large bare mass suppresses resonances); we further assume that they couple with the same strength 
\begin{eqnarray}
g\equiv 5\times 10^{-4}
\end{eqnarray}
to the inflatons. This coupling is chosen to enable easy comparison with prior preheating studies such as \cite{Kofman:1997yn,Battefeld:2009xw,Braden:2010wd}. Note that this coupling is often assumed to be much bigger for models that employ ESPs during inflation, in order to yield measurable signatures \cite{Battefeld:2010sw,Battefeld:2011yj,Kofman:2004yc,Green:2009ds,Barnaby:2009mc,Barnaby:2009dd,Barnaby:2010ke,Barnaby:2010sq}. The qualitative features of preheating discussed in this article are not contingent on a finely tuned $g$ and remain valid in more general cases. 

All in all, the complete Lagrangian reads
\begin{eqnarray}
  {\cal L} &=& -\frac{1}{2}\sum_{i=1}^{{\cal N}}g^{\mu\nu}\partial_{\mu}\varphi_i\,\partial_{\nu}\varphi_i - \frac{1}{2}\sum_{i=1}^{{\cal
      N}}m_i^2\varphi_i^2 \nonumber \\
  &&- \frac{1}{2}g^2\sum_{\vec{\alpha}}\left(\vec{\varphi}-\vec{\varphi}_{\mbox{\tiny ESP}}^{\vec{\alpha}}\right)^2\chi_{\vec{\alpha}}^2
  -\frac{1}{2}\sum_{\vec{\alpha}}g^{\mu\nu}\partial_{\mu}\chi_{\vec{\alpha}}\,\partial_{\nu}\chi_{\vec{\alpha}} \,.
\end{eqnarray}
To discuss preheating, it is  useful to define a \emph{non-adiabaticity} parameter for a given ESP
\begin{eqnarray}
\eta &\equiv& \frac{|\dot{\omega}_k|}{\omega_k^2} \label{nonadiabaticity}\\
&=&\left|\frac{g^2\dot{\vec{\varphi}}(\vec{\varphi}-\vec{\varphi}_{\mbox{\tiny ESP}})
-Hk^2/a^2}{\left(g^2(\vec{\varphi}-\vec{\varphi}_{\mbox{\tiny ESP}})^2+k^2/a^2\right)^{3/2}}\right|\,.
\end{eqnarray}
In most figures, time is  given in terms of the number of oscillations in a hypothetical field with mass $m\equiv 10^{-6}M_P$, that is
\begin{eqnarray}
N\equiv \frac{mt}{2\pi}\,,
\end{eqnarray}
which should not be confused with the number of fields $\mathcal{N}$ or the number of e-folds (not used in this paper).

\section{Preheating in the Absence of Back-reaction \label{sec:Preheatingwithoutbackreaction}}
Treating the evolution of the inflaton fields as a background for the matter fields, that is ignoring back-reaction, we would like to investigate the feasibility of preheating in the setup of Sec.~\ref{sec:setup}, with the field content and interactions of Sec.~\ref{sec:PRsetup}. Concretely, we would like to examine the effectiveness of particle production at ESPs during three distinct phases:
\begin{enumerate}
\item The \emph{First In-fall}: directly after inflation, the inflaton fields evolve rapidly towards $\vec{\varphi}=0$, traversing many ESPs along the way.
\item \emph{Large Amplitude Oscillations} around $\vec{\varphi}=0$ follow the first in-fall, with $A\gg x$. 
\item A \emph{Resonance Phase} with small amplitude oscillations   around $\vec{\varphi}=0$, with $A\ll x$.
\end{enumerate} 
For the important first and third phase, an analytic description of preheating is possible; the brief intermediate phase is strongly model dependent with at most intermittent  particle production. Therefore, we focus on the first and third phase.

Analytic results are compared with numerics, entailing the integration of the equations of motion  
\begin{eqnarray}
\ddot{\varphi}_i+3H\dot{\varphi}_i+\frac{\partial V}{\partial \varphi_i}=0\,,\\
\ddot{\chi}_k^{\vec{\alpha}}+3H\dot{\chi}_k^{\vec{\alpha}}+\omega_k^2\chi_k^{\vec{\alpha}}=0\,,\label{eomchi}
\end{eqnarray}
with the time dependent oscillation frequency
\begin{eqnarray}
\omega_k^2=\frac{k^2}{a^2}+g^2(\vec{\varphi}-\vec{\varphi}_{ESP}^{\vec{\alpha}})^2\,,
\end{eqnarray}
in conjunction with the Friedmann equation
\begin{eqnarray}
3H^2=\frac{4\pi}{M_p^2}\sum_i \left(m_i^2\varphi_i^2+\dot{\varphi}_i^2\right)\,,
\end{eqnarray}
using common C++-routines. These results are supplemented by some computationally more demanding lattice simulations in Sec.~\ref{PRwithbackreaction}, which incorporate back-reaction of the preheat matter fields onto the inflatons.

To set our notation and provide a self contained treatment, we review some well known properties of preheating at a single ESP in Sec.~\ref{sec:review}, following primarily \cite{Kofman:1997yn,Mukhanov:2005sc,Battefeld:2008bu}.

\subsection{Review of Preheating at a Single ESP \label{sec:review}}
\subsubsection{Inflaton Dynamics during Preheating}
Introducing the angle variables $\theta_i$, defined via 
\begin{eqnarray}
\dot{\varphi}_i\equiv \sqrt{\frac{3}{4\pi \mathcal{N}}}H \cos{\theta_i}\,\,\,,\,\,\,\varphi_im_i\equiv \sqrt{\frac{3}{4\pi \mathcal{N}}}H \sin{\theta_i}\,,
\end{eqnarray}
the background equations of motion become
\begin{eqnarray}
\dot{H}&=&-3H^2\frac{1}{\mathcal{N}}\sum_i\cos^{2}\theta_i\,,\\
\dot{\theta}_i&=&m_i+\frac{3}{2}H\sin 2\theta_i\,.
\end{eqnarray}
During most of the preheating phase $m_i\gg H$ holds, so that $\theta_i\approx m_it$ up to some irrelevant phase. Thus $H\approx 2/(3t)$ to leading order, so that $a\propto t^{2/3}$ and 
\begin{eqnarray}
\varphi_i(t)&\approx&\frac{1}{\sqrt{3 \pi \mathcal{N}}m_i t}\sin m_i t
\equiv \phi_i \sin m_i t\,,
\end{eqnarray}
that is, the fields oscillate with ever decreasing amplitude $\phi_i(t)\propto 1/t$.
Further, even if fields start out in phase, they quickly de-phase unless the masses are identical. 

In the following, we need the average distance of the trajectory from the origin $\bar{\varphi}$, as well as the average velocity $\bar {v}$ (averaged over one oscillation). With the above approximation we get
\begin{eqnarray}
\bar{\varphi}&\equiv & |\vec{\varphi}|\approx \sqrt{\left<|\vec{\varphi}|^2\right>}
= \frac{1}{\sqrt{12 \pi \mathcal{N}}t}
\sqrt{\sum_i m_i^{-2}}\,,\label{amplitudeovertime}\\
\bar{v}&\equiv& \left<|\dot{\vec{\varphi}}|\right>\approx \sqrt{\left<\sum_i m_i^2\varphi_i^2\right>}\approx \frac{1}{\sqrt{12\pi}t}\,,
\end{eqnarray}
where we used the virial theorem to estimate the velocity.

\begin{figure}[t]
  \centering
  \scalebox{0.9}{\input{mathieu}}
  \caption{The first two instability bands (dark blue) of the Mathieu-equation for small values of $q$ (computed analytically for $q\ll 1$). $A_k<0$ is not possible during preheating.}
  \label{fig:mathieu}
\end{figure}
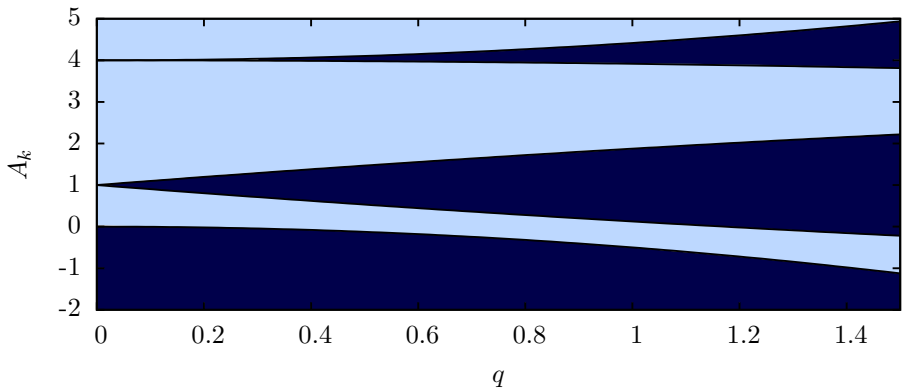

\subsubsection{Parametric Resonance at a Single ESP Driven by a Single Inflaton \label{sec:narrowresonance}}
Consider a single ESP at $\varphi=\epsilon$ so that $\mathcal L_{int}=-g^2(\varphi-\epsilon)^2\chi^2/2$. Since the oscillation frequency is time dependent, resonances can take place, increasing the particle number
\begin{eqnarray}
n_k\equiv \frac{\rho_k}{\omega_k}-\frac{1}{2}=\frac{\omega_k}{2}\left(\frac{|\dot{\chi}_k|^2}{\omega_k^2}+|\chi_k|^2\right)-\frac{1}{2}
\end{eqnarray}
explosively.
To encounter resonance, consider $\varphi\ll \epsilon$, ignore the expansion of the universe and define $A_k\equiv 4(k^2+g^2\epsilon^2)/m^2$, $q\equiv 4g^2\epsilon\phi/m^2$ and $2z\equiv mt+\pi/2$ so that (\ref{eomchi}) becomes the Mathieu-equation
\begin{eqnarray}
\chi_k^{\prime\prime}+(A_k-2q\cos{2z})\chi_k=0\,,\label{eomfloquet}
\end{eqnarray}
where a prime denotes a derivative with respect to $z$. According to the Floquet-theorem, solutions to (\ref{eomfloquet}) are of the form
\begin{eqnarray}
\chi_k=\eta(z)\exp(\mu_k z)\,,
\end{eqnarray}
 where $\eta$ is periodic with period $\pi$ and the Floquet index $\mu_k$ (in more general cases, we simply refer to $\mu_k$ as a characteristic exponent) is positive in certain bands, see Fig.~\ref{fig:mathieu}; for example, in the first band for $q\ll 1$ one finds \cite{Lachlan}
\begin{eqnarray}
\mu_k=\sqrt{\frac{q^2}{4}-\left(\frac{2k}{m}-1\right)^2}\,.
\end{eqnarray}
As a consequence, resonances occur for $m(1-q/2)/2<k<m(1+q/2)/2$. For $\varphi>g^2\epsilon/(32 \pi)$ this \emph{narrow resonance} dominates over classical reheating effects \cite{Kofman:1997yn}.

\begin{figure}[t]
  \centering
  \scalebox{1.03}{\input{chi}}
  \caption{Numerical solution of the particle number for narrow (left, $(A_k; q) = (1; 0.2)$) and broad (right, $(A_k; q) = (400; 200)$) resonance, ignoring the expansion of the universe. Time is plotted in units of $N\equiv mt/ (2\pi)$ with $m=10^{-6}M_p$.
\label{fig:narrowandbroadresonance}}
\end{figure}
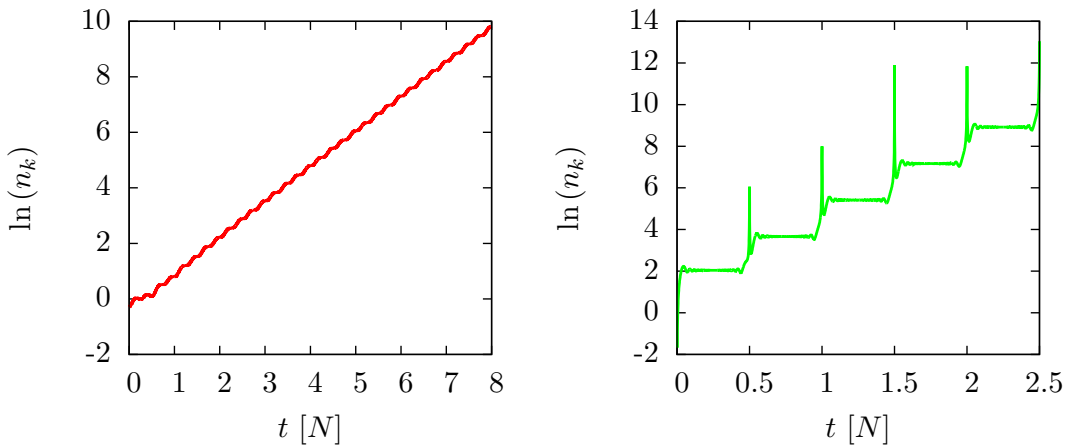

In an expanding universe, modes do not stay within a given instability band, but traverse through them to approach the stable region near $A_k=0$, $q\ll 1$. As a consequence, the effectiveness of preheating is reduced: for example, after $\Delta t\sim qH^{-1}$ a mode in the first instability band will have left said band, leading to a net amplification of the particle number by $\sim \exp(q^2m/(2H))$. Thus, a condition for effective preheating is
\begin{eqnarray}
q^2m>2H\,,
\end{eqnarray}
that is $\varphi> m\sqrt{mH/2}/(2g^2\epsilon)$. 

The above description is valid towards the end of preheating, when the oscillation amplitude is small. At earlier times $\phi\lesssim M_P$ and we need to consider $\phi \gg \epsilon$, so that
\begin{eqnarray}
\ddot{\chi}_k+(k^2+g^2\varphi^2)\chi_k=0\,,
\end{eqnarray}
if expansion effects are ignored. Defining $A_k\equiv 2q+k^2/m^2$, $q\equiv g^2\varphi^2/(4m^2)$ and $z=mt$, we arrive again at the Mathieu equation, leading to \emph{broad resonance}, see Fig.~\ref{fig:narrowandbroadresonance}.

Particle production in the broad resonance regime takes place in bursts, whenever $\varphi\sim 0$ so that the effective mass of the matter field is light. To be concrete, if the non-adiabaticity parameter in (\ref{nonadiabaticity}) becomes large
\begin{eqnarray}
\eta >1\,,\label{etabigger1}
\end{eqnarray}
particle production can take place.

\begin{figure}[t]
  \centering
  \scalebox{0.95}{\input{chi_exp}}
  \caption{Numerical solution of the particle number for the $k = 4m$ mode during stochastic resonance, that is broad resonance in an expanding universe, with $g = 5 \cdot 10^{-4}$ and $m = 10^{-6}\,M_P$.}
  \label{fig:chi_exp}
\end{figure}
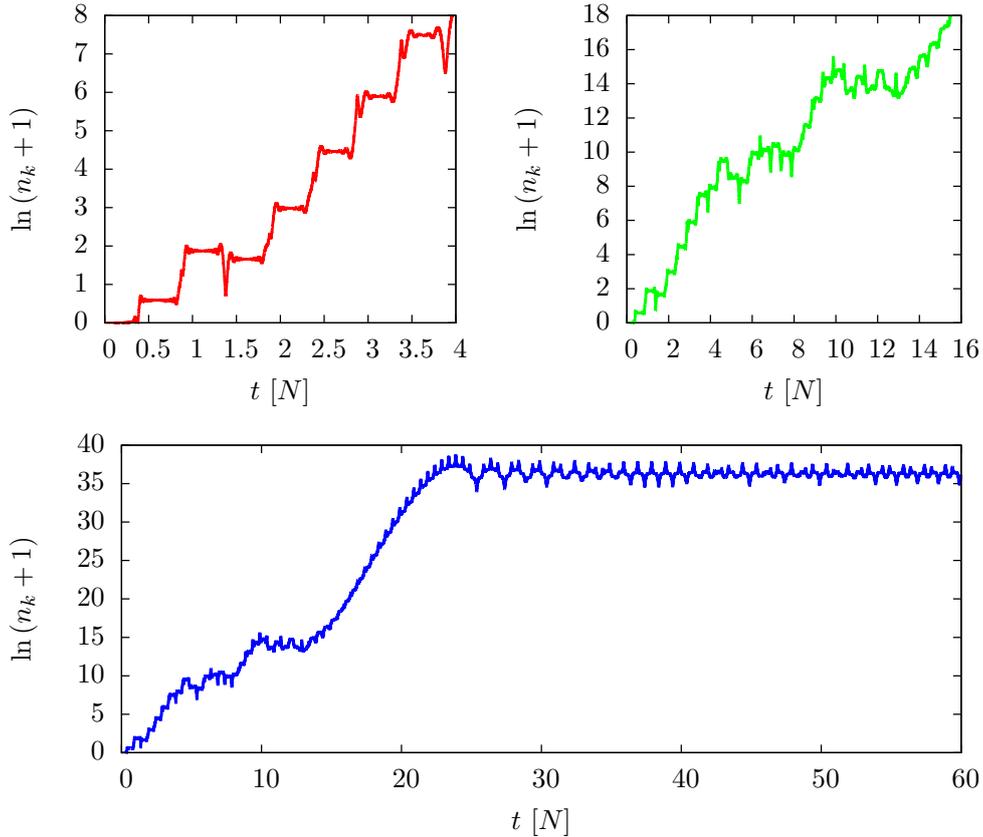

To include the expansion of the universe, it is convenient to define $X\equiv a^{3/2}\chi_k$ so that the equation of motion becomes
\begin{eqnarray}
\ddot{X}_k+\omega_k^2X_k=0\,, \label{eomX}
\end{eqnarray}
with $\omega_k^2\approx k^2/a^2+g^2\phi^2\sin^2(mt)$ and $a\propto t^{2/3}$. The resulting solutions still show ample particle production that appears to be stochastic, see Fig.~\ref{fig:chi_exp}, and carries therefore the name \emph{stochastic resonance}. It should be noted that this process is fully deterministic, and can be understood analytically \cite{Kofman:1997yn}: in the WKB approximation, the adiabatic solutions ($\eta\ll 1$) to (\ref{eomX}) are
\begin{eqnarray}
X_k(t)=\frac{\alpha_k}{\sqrt{2\omega_k}}e^{i \int \omega_k \mbox{\tiny d}t}+\frac{\beta_k}{\sqrt{2\omega_k}}e^{-i \int \omega_k \mbox{\tiny d}t}\,,\label{adiabaticsolution}
\end{eqnarray} 
where the time dependent Bogolioubov coefficients have to satisfy $|\alpha_k|^2-|\beta_k|^2=1$. The comoving particle number is $n_k=|\beta_k|^2$. Expanding $\varphi(t)$ around the $j$'s particle production event and defining  $\tau\equiv k_\star(t-t_j)$ as well as $\kappa\equiv k/(ak_\star)$ yields
\begin{eqnarray}
\frac{\partial^2 X_k}{\partial \tau^2}+(\kappa^2+\tau^2) X_k=0\,, \label{eomX2}
\end{eqnarray}
 which describes scattering at parabolic potentials. A computation of the transfer-coefficients relates the Bogolioubov coefficients before and after the particle production event (while $\eta\gg 1$), leading to \cite{Kofman:1997yn}
\begin{eqnarray}
n_k^{j+1}=e^{-\pi\mathcal{\kappa}^2}+(1+2e^{-\pi\kappa^2})n_k^j-2e^{-\pi\kappa^2/2}\sqrt{1+e^{-\pi\kappa^2}}\sqrt{n_k^j(1+n_k^j)}\sin\theta^j_{tot}\,,\label{analyticnj}
\end{eqnarray}
where $\theta^j_{tot}$ is a time and $k$-dependent total phase. Thus, the step like change of the particle number as well as the occasional drop can be understood analytically (a drop takes place if $\sin{\theta^j_{tot}}>e^{-\pi\xi^2/2}$).

As evident, resonances can efficiently increase the particle number in an expanding universe.  Naturally, once the particle number is large and the energy density in the matter field becomes comparable to the energy of the inflaton, one needs to incorporate back-reaction effects. However, to decide whether or not parametric resonance is important, the simplified treatment of this section is fully satisfactory.

\subsubsection{Parametric Resonance at a Single ESP Driven by Several Inflatons \label{sec:dephasing}}
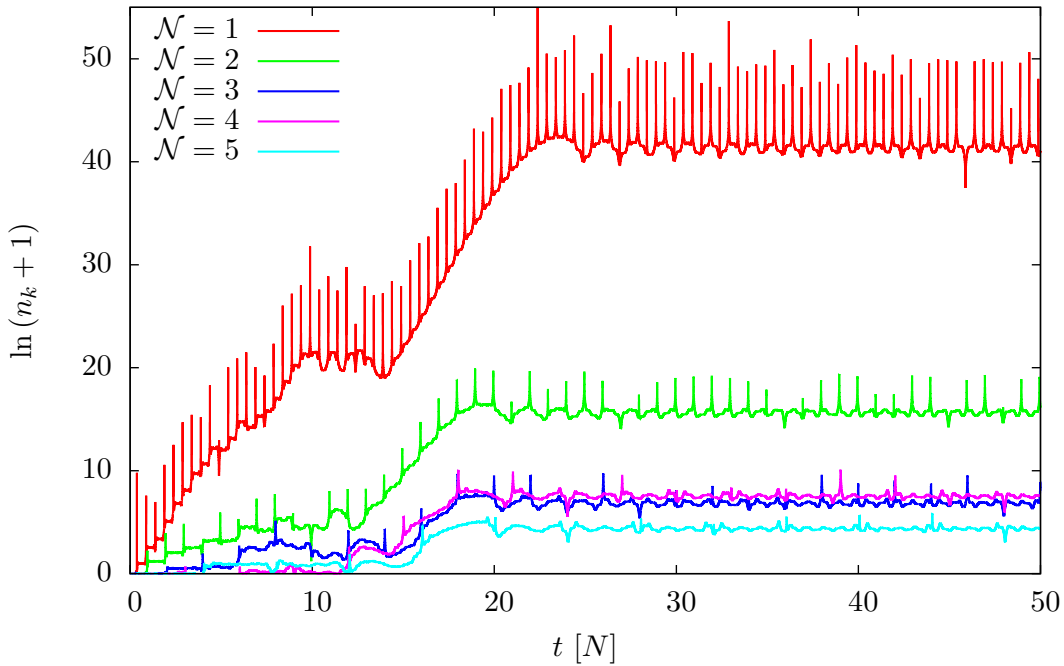
\begin{figure}[t]
  \centering
  \scalebox{1.03}{\input{psupp}}
  \caption{Evolution of the $k=0$ mode for different numbers of inflatons $\mathcal{N}$ with $\vec{\varepsilon} = 0$ and $g=5 \cdot 10^{-4}$. Resonaces are suppressed as $\mathcal{N}$ is increased due to the dephasing of fields (inflaton masses are chosen according to (\ref{masses})). \label{fig:psupp}}
\end{figure}
How does preheating change if several inflatons are present? A straightforward generalization of the interaction Lagrangian to
\begin{eqnarray} 
\mathcal{L}_{int}=-\frac{1}{2}g^2(\vec{\varphi}-\vec{\epsilon})^2\chi^2\,,
\end{eqnarray}
yields an, at first sight, surprising result for $|\vec{\epsilon}|=0$: resonances are strongly suppressed if more than a few inflatons with different masses are present \cite{Battefeld:2008bu,Battefeld:2008rd,Battefeld:2009xw, Braden:2010wd}, see Fig.~\ref{fig:psupp}. However, this can be understood easily: inflatons generically run out of phase due to differences in their masses. As a consequence, the trajectory never crosses the position where the ESP is located (the origin), the non-adiabaticity parameter remains small and particle production is suppressed.  

In case of two inflatons, an additional effect can keep the efficiency up to some degree: if the mass ratio of the inflatons is not a rational number, the driving force is only quasi periodic. As a consequence, stability bands dissolve into a no-where dense set, leading to a positive generalized Floquet index for almost all modes \cite{Bassett:1997gb,Bassett:1998yd} (Cantor preheating); this result is based on spectral theory \cite{JMoser, lachlan, gihman, Pastur, Avron, JAvron, JAvronBSimon}, and numerics. Unfortunately, the amplitude of the characteristic exponents can not be computed analytically and it appears that most of them are generically small; as a consequence, the suppression due to de-phasing wins if more than two inflatons are included \cite{Battefeld:2008bu,Battefeld:2009xw}.

\subsection{Overview: The Three Phases of Preheating \label{sec:preheatingphases}}

\begin{figure}[t]
  \centering
  \includegraphics[scale=1.2]{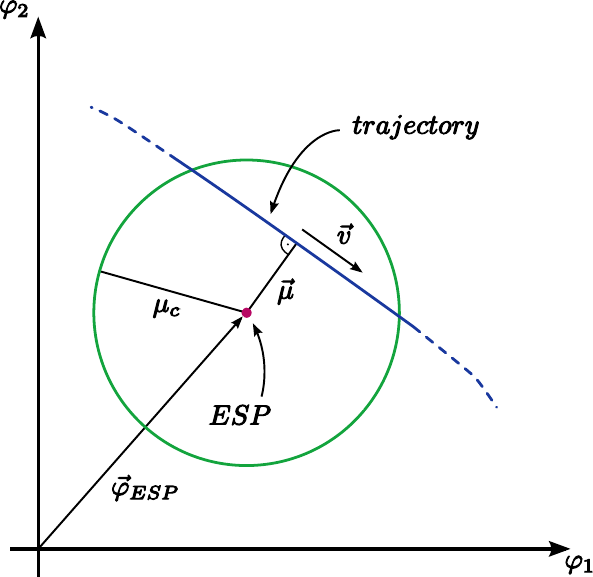}
  \caption{Schematic of an ESP-encounter in a two dimensional field space. If the trajectory comes close to the ESP, $\mu<\mu_c=(|\dot{\vec{\varphi}}|/g)^{1/2}$, the adiabaticity condition $\eta<1$ can be violated, and particle production can take place. }
  \label{fig:esp}
\end{figure}

In the last section, we discussed preheating in the presence of a single ESP at the origin: to allow for explosive particle production, we saw in (\ref{etabigger1}) that $\eta>1$ is a sufficient condition ($\eta$ is the non-adiabaticity parameter in (\ref{nonadiabaticity}) and we considered small $k$-modes). During encounters with a non-zero impact parameter $\mu$ (see Fig.\ref{fig:esp}) this condition becomes
\begin{eqnarray}
\mu<\mu_c\,,
\end{eqnarray}
where we defined the critical impact parameter
\begin{eqnarray}
\mu_c\equiv \sqrt{\frac{|\dot{\vec{\varphi}}|}{g}}\,.\label{def:muc}
\end{eqnarray}
Even though $v\equiv |\dot{\vec{\varphi}}|$ is slow roll suppressed during inflation, particle production may still be strong enough to lead to trapped inflation \cite{Kofman:2004yc,Green:2009ds,Silverstein:2008sg,Battefeld:2010sw} or additional contributions to observables \cite{Battefeld:2011yj}, if ESPs are dense or an encounter is close (see Sec.~\ref{sec:trapped} and \ref{sec:grazing}). During preheating, the critical impact parameter increases sharply, making ESPs in wider regions of field space relevant for particle production.

If we expand the trajectory near the ESP encounter $\vec{\varphi}\approx \vec{\varphi}_{\mbox{\tiny ESP}}+\vec{\mu}-\vec{v}t$, we can write the equations of motion for $X_k$ as
\begin{eqnarray}
\frac{\partial X_k}{\partial \tau^2}+(\kappa^2+\tau^2) X_k=0\,, \label{eomX3}
\end{eqnarray}
with $\tau\equiv \sqrt{gv}t$ and 
\begin{eqnarray}
\kappa^2\equiv \frac{1}{gv}\left(\frac{k^2}{a^2}+g^2\mu^2\right)\,;\label{defkappa}
\end{eqnarray}
 (\ref{eomX3}) is formally identical to (\ref{eomX2}), with the above redefinitions of $\tau$ and $\kappa$. Therefore all prior results, such as the particle number $n_k^j$ in (\ref{analyticnj}), carry over (the index $j$ now counts close encounters with the same ESP, as opposed to oscillations of the inflaton).

\begin{figure}[t]
  \centering
  \scalebox{1.03}{\input{trajectory}}
  \caption{Post inflationary trajectory in a two-dimensional field space, $\varphi_2/M_p$ over $\varphi_1/M_p$ with initial conditions $\varphi_1(t_0)=\varphi_2(t_0)=0.2\, M_p$, $\dot{\varphi}_1(t_0)=\dot{\varphi}_2(t_0)=0$ and masses $m_1 = 10^{-6}\,M_P$ and $m_2 = 1.5 \cdot 10^{-6}\,M_P$. The dotted blue circles denote regions that the trajectory does not leave on average after $t/N=0.5,8,16$, corresponding to the three phases: \emph{first in-fall} ($t/N\lesssim 0.5$), \emph{large amplitude oscillations} ($0.5\lesssim t/N\lesssim 16$) and \emph{resonance} ($16\lesssim t/N\lesssim 50$). The corresponding total particle number in the $k=0$-mode of a grid-ESP distribution with inter ESP distance $x=0.0033\,M_p$ is plotted in Fig.~\ref{fig:ph_ablauf}.}
  \label{fig:trajectory}
\end{figure}
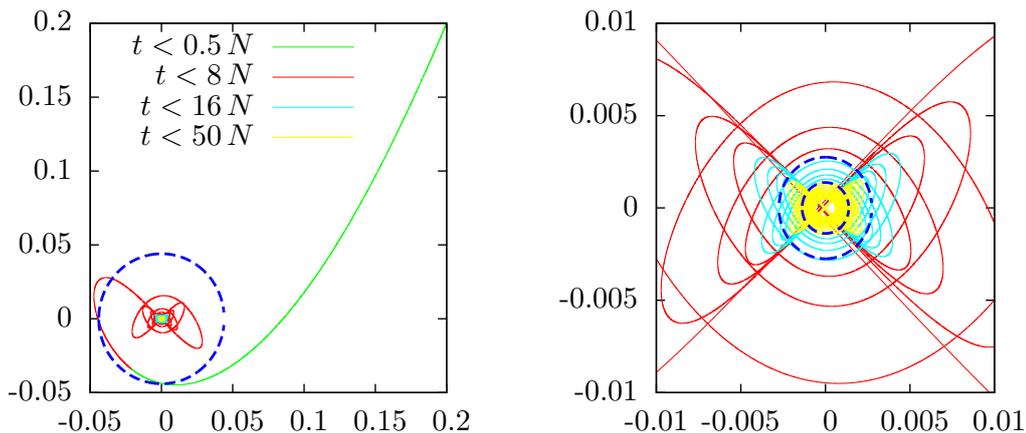

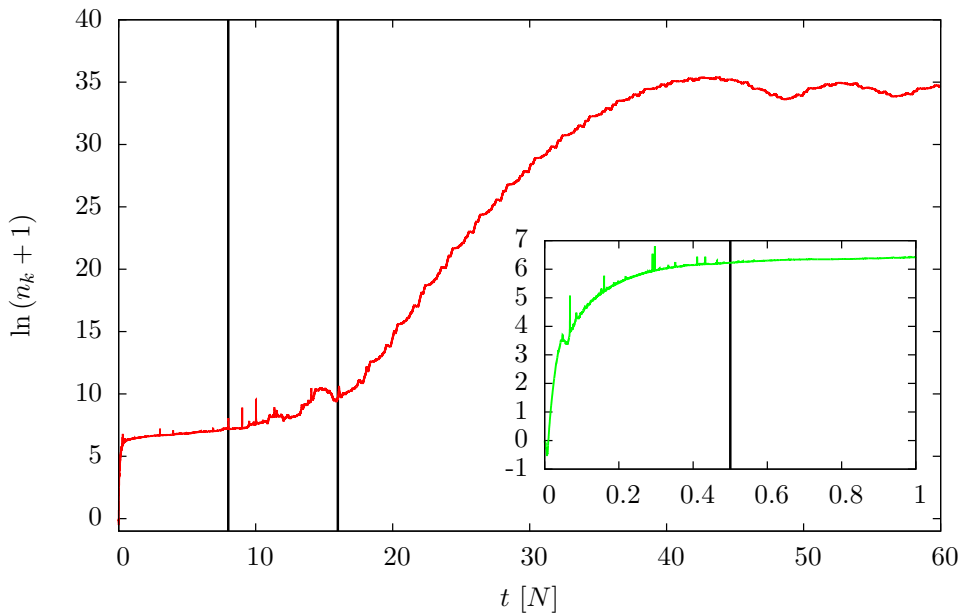
\begin{figure}[t]
  \centering
  \scalebox{0.93}{\input{ph_ablauf}}
  \caption{Occupation number of the $k=0$ mode in an ESP distribution along a grid with inter-ESP distance $x=0.0033\,M_p$ with two inflatons evolving as in Fig.~\ref{fig:trajectory}. Solid vertical lines separate the phases of preheating, as in Fig.~\ref{fig:trajectory}.}
  \label{fig:ph_ablauf}
\end{figure}

Equipped with this knowledge, we are ready to discuss particle production for actual trajectories: in Fig.~\ref{fig:trajectory} we plot an exemplary trajectory in a two dimensional field space, with the corresponding particle production at ESPs distributed along a grid with inter ESP distance $x=0.003\,M_p$ in Fig.~\ref{fig:ph_ablauf}.

 We initialize $\varphi_1(t_0)=\varphi_2(t_0)=0.2\, M_p$, $\dot{\varphi}_1(t_0)=\dot{\varphi}_2(t_0)=0$, and follow the fields evolution in an expanding universe. We chose zero initial speed to mimic the low speed at the end of inflation. In the absence of large Hubble friction, the speed grows quickly, increasing the critical impact parameter. Since $\mu_c\gg x$, many ESPs are within reach during the \emph{first in-fall} and a seizable quantity of particles can be produced (we lump all particle species together in the particle number depicted in Fig.~\ref{fig:ph_ablauf}), accounting for the increase to $\ln(n_k+1)\approx 7$ in the insert \footnote{If ESPs are dense and $
\mathcal{N}$ is large (larger than in this section), back-reaction can lead to trapped inflation at a terminal velocity even for the sub-Planckian field values considered here (see the worked example in Sec. III.C of \cite{Battefeld:2010sw}). Once inflation ends, a seizable amount of $\chi$-particles are already present in this scenario; the subsequent evolution is identical to the one discussed above.}. Even though the particle number at each individual ESP is low because each one is only passed once, the resulting overall increase in $n_k$, and thus the corresponding energy transfer, can be considerable.
The end of this phase, which is absent in the single ESP case, is marked by the blue dotted circle on the left of Fig.~\ref{fig:trajectory} and the vertical line in the insert of Fig.~\ref{fig:ph_ablauf}.

During the subsequent \emph{large amplitude oscillations} the fields slow down $v\propto 1/t$, $\mu_c$ decreases and particle production takes place only sporadically in bursts, reminiscent of stochastic resonance: in our example, $n_k$ stays roughly constant until $t/N=8$;  bursts become more frequent thereafter, leading up to the resonance phase.  The duration and effectiveness of this intermediate phase is model dependent and somewhat unpredictable; it ends once the amplitude of oscillations falls below the inter-ESP distance $\phi<x$, indicated by the small blue circle on the right side of Fig.~\ref{fig:trajectory} and the vertical line at $t/N=16$ in Fig.~\ref{fig:ph_ablauf}.

During the third \emph{resonance phase}, the oscillation amplitude of the inflatons is small. This phase is reminiscent of the narrow resonance regime, inducing a steady increase of the particle number; as we show in Sec.~\ref{sec:resonancephase}, particles are produced at a few ESPs that have the ``right'' location to enable prolonged resonances. Particle production ends, once the amplitude/speed become too small, around $t/N=50$ in our example (see Sec.~\ref{sec:discres}), and the particle number remains constant thereafter.

The three phases, especially the first in-fall and the resonance phase, are generic to multi-field preheating in the presence of dense ESP distributions, $x\ll M_p$. We provide the analytic explanation of these phases in Sec.~\ref{sec:firstinfall} and Sec.~\ref{sec:resonancephase}, which contain the main new insights gained in this paper (see Sec.~\ref{sec:discfirstinfall} and \ref{sec:discres} for a summary of the results).

Efficient preheating requires that the energy density in the different matter particle species becomes comparable to the one in the inflatons before the resonance phase ends. In this case one has to include back-reaction, see Sec.~\ref{PRwithbackreaction}, as well as the decay of preheat matter fields \cite{Figueroa:2009jw,Allahverdi:2011aj} (not covered in this paper). 

An important result of such multi-field preheating is the presence of two distinct particle classes: those produced at many ESPs with low occupation number (first in-fall and intermediate phase) and particles produced in great number at a few ESPs (resonance phase); it is tempting to identify the subsequent decay products of particles in these two classes with dark matter and standard model particles respectively. However, one would need to consider a specific scenario to concretize this speculation, which goes beyond the scope of this study.

\subsection{The First In-fall \label{sec:firstinfall}}
\subsubsection{Analytic Description \label{sec:analyticinfall}}
\begin{figure}[t]
  \centering
  \includegraphics[scale=1.0]{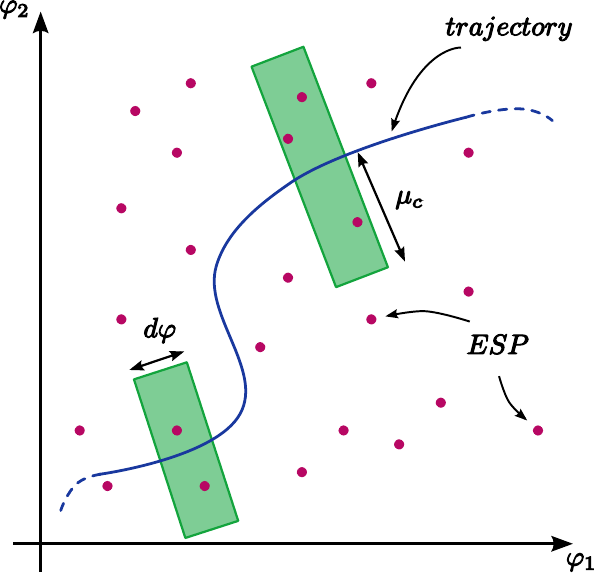}
  \caption{Schematic of a trajectory during the first in-fall. ESPs in the green shaded regions are close enough to allow for the violation of the adiabaticity condition $\eta<1$, and thus particle production. $\mu_c=(|\dot{\vec{\varphi}}|/g)^{1/2}$ is the impact parameter up to which we consider ESPs ($\mu\approx \mu_c$ for most ESPs in the green region; this approximation becomes better, the larger $\mathcal{N}$ is).}
  \label{fig:first_infall1}
\end{figure}
During the first in-fall, many ESPs are close enough to the trajectory to allow for adiabaticity to be violated and particles to be produced, see Fig.~\ref{fig:first_infall1} for a schematic. We would like to compute the occupation number in the $k=0$ mode for comparison with numerics, as well as the total particle number
\begin{eqnarray}
n_\chi=\int n_k \mbox{d}^3k\,,
\end{eqnarray}
where we combine all the different matter species. We closely follow  \cite{Battefeld:2010sw} in this section. Since $\chi$-fields start out in the vacuum, only the first term in (\ref{analyticnj}) contributes to the particle numbers during the first in-fall,
\begin{eqnarray}
n_k^{\vec{\alpha}}\approx \exp\left(-\frac{\pi}{gv}\left(k^2+g^2\mu^2_{\vec{\alpha}}\right)\right)\,,\label{defnk}
\end{eqnarray}
where $\mu_{\vec{\alpha}}$ is the impact parameter of the ESP at $\vec{\varphi}_{\mbox{\tiny ESP}}^{\vec{\alpha}}$, we used (\ref{defkappa}) and ignored expansion effects ($a\equiv 1$), since inflation already terminated. To compute the total particle number, we replace the discrete ESP distribution by a homogeneous ESP density  $\rho_{\mbox{\tiny ESP}}=x^{-\mathcal{N}}$ so that  
\begin{eqnarray}
\mbox{d}n_k&=&\left<n_k\right>\rho_{\mbox{\tiny ESP}}\mbox{d}V\\
&=&\left<n_k\right>\rho_{\mbox{\tiny ESP}}V_{\mathcal{N}-1}\mbox{d}\varphi\,,
\end{eqnarray}
where $V_{\mathcal{N}-1}$ is the volume of an $\mathcal{N}-1$ dimensional hyper-sphere, and $\left<n_k\right>$ denotes the average particle number at ESPs in the infinitesmal volume element $\mbox{d}V$ shown in Fig.~\ref{fig:first_infall1}. As a rough estimate for the radius of the hyper-sphere, we take the critical impact parameter $\mu_c=\sqrt{v/g}$, since particle production at ESPs further away is strongly suppressed\footnote{See \cite{Battefeld:2010sw} for an improved estimate in the large $\mathcal{N}$ limit; the simpler estimate above is satisfactory for our purposes (we slightly underestimate particle production).}, leading to
\begin{eqnarray}
\mbox{d}n\approx \frac{\pi^{(\mathcal{N}-1)/2}\mu_c^{\mathcal{N}-1}}{\Gamma(\frac{\mathcal{N}+1}{2})x^{\mathcal{N}}}\left<n_k\right>\mbox{d}\varphi\,,
\end{eqnarray}
where $\Gamma$ is the Gamma function.

\begin{figure}[t]
  \centering
  \scalebox{1.03}{\input{vel_int}}
  \caption{Left: comparison of $\left<n_0\right>$ with the large $\mathcal{N}$-limit $\exp(-\pi)$ (taking $\mu=\mu_c$ for all ESPs in $dV$).
Right: velocity-integral  ${\cal F}({\cal
      N},\alpha)$ for different values of $\alpha$ with the corresponding fits.} 
  \label{fig:vel_int}
\end{figure}
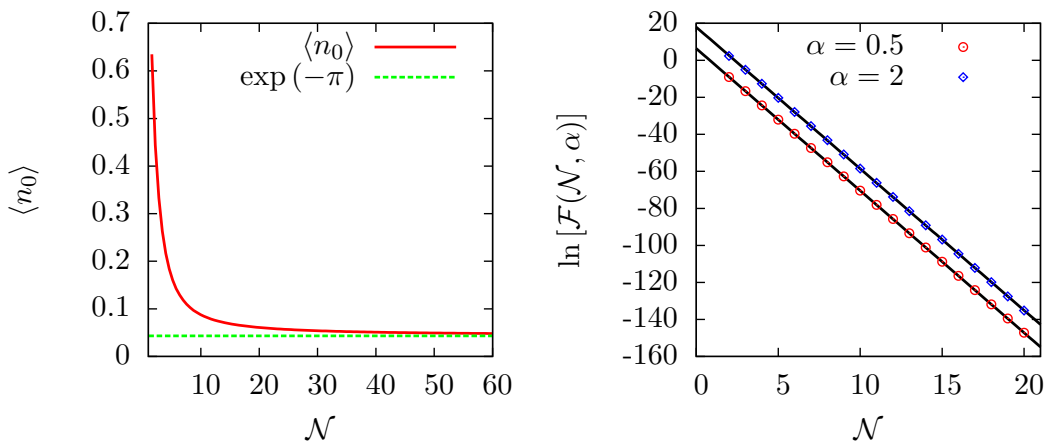

Let us focus on the $k=0$ mode to see how rapidly $\left<n_0\right>$ approaches the limit value $\exp(-\pi)$ 
as $\mathcal{N}$ is increased. Based on the definition of the average, we get 
\begin{eqnarray}
\left<n_0\right>&=&\frac{\int\exp\left(-\frac{\pi g}{v}\mu^2\right)\mbox{d}V_{\mathcal{N}-1}}{\int\mbox{d}V_{\mathcal{N}-1}}
\\
&=&\frac{\mathcal{N}-1}{2}\pi^{-\frac{\mathcal{N}-1}{2}}\Gamma\left(\frac{\mathcal{N}-1}{2},\pi\right)\,,\label{avgn0}
\end{eqnarray}
which we plot in the left panel of Fig.~\ref{fig:vel_int}. Evidently, the limit is approached rapidly, providing an excellent approximation for $\mathcal{N}$ as low as $20$, while still giving the right order of magnitude for lower $\mathcal{N}$. This is not surprising, since almost all of the volume in a sphere is near the surface if $\mathcal{N}\gg 1$, so that the impact parameter to almost all relevant ESPs is close to $\mu_c$.

Using this approximation, we can compute the total occupation number in the zero-mode by integration over the first in-fall to
\begin{eqnarray}
n_0^{tot}=\int\mbox{d}n_0 &=& \rho_{ESP}\,\left<n_0\right> \cdot \int_0^{t_*}V_{{\mathcal N}-1}\,v\,\mbox{d}t 
\\ 
 &=& g^{-\frac{{\mathcal N}-1}{2}}\,\frac{\Gamma\left(\frac{{\mathcal N}-1}{2},\pi\right)}{\Gamma\left(\frac{{\cal N}-1}{2}\right)}\,x^{-{\mathcal N}} \mathcal{F}(\mathcal{N},1/2) \,.\label{analyticn0}
\end{eqnarray}
where we defined the velocity integral 
\begin{eqnarray}
\mathcal{F}(\mathcal{N},\alpha)\equiv \int_0^{\frac{\pi}{m_1}} v^{\frac{{\cal N}}{2}+\alpha}(t)\,\mbox{d}t\,.\label{velocityintegral}
\end{eqnarray}
The latter can usually be  approximated by
\begin{eqnarray}
\mathcal{F}(\mathcal{N},\alpha)\approx e^{c_1(\alpha)+c_2(\alpha)\mathcal{N}}\,,\label{approxF}
\end{eqnarray}
with constant coefficients $c_1,c_2$. For example, using the mass distribution of Eq.~\ref{masses1} and the initial conditions specified in
Sec.~\ref{sec:preheatingphases} for all of the $\mathcal{N}$ inflatons leads to the coefficients in Tab.~\ref{tab:reg}. The corresponding
regression is shown in the right panel of Fig.~\ref{fig:vel_int}.
\begin{table}[t]
  \centering
  \begin{tabular}{|c||c|c|}
    \hline
    $\alpha$ &  $c_1$ & $c_2$ \\ \hline \hline
    $0.5$ & $-7.6777(4)$ & $6.360(5)$ \\ \hline
    $2$ & $-7.653(4)$ & $17.95(4)$ \\ \hline
  \end{tabular}
  \caption{Coefficients of the approximation in (\ref{approxF}) for the velocity-integral in (\ref{velocityintegral}) via linear regression.}
  \label{tab:reg}
\end{table}

The analytic expression of  $n_0^{tot}$ is easily compared to numerical results, providing a check for the validity of the approximations, see Sec.~\ref{sec:compnumericsfirstinfall}.

\begin{figure}[t]
  \centering
  \scalebox{1.03}{\input{fi_eff}}
  \caption{ The total comoving occupation number after the first in-flall over the number of fields $\mathcal{N}$ for different inter ESP distances $x$. The solid black line corresponds to $n_0$ at the end of preheating in presence of a single field and an ESP at the origin, see Fig.~\ref{fig:psupp}.}
  \label{fig:fi_eff}
\end{figure}
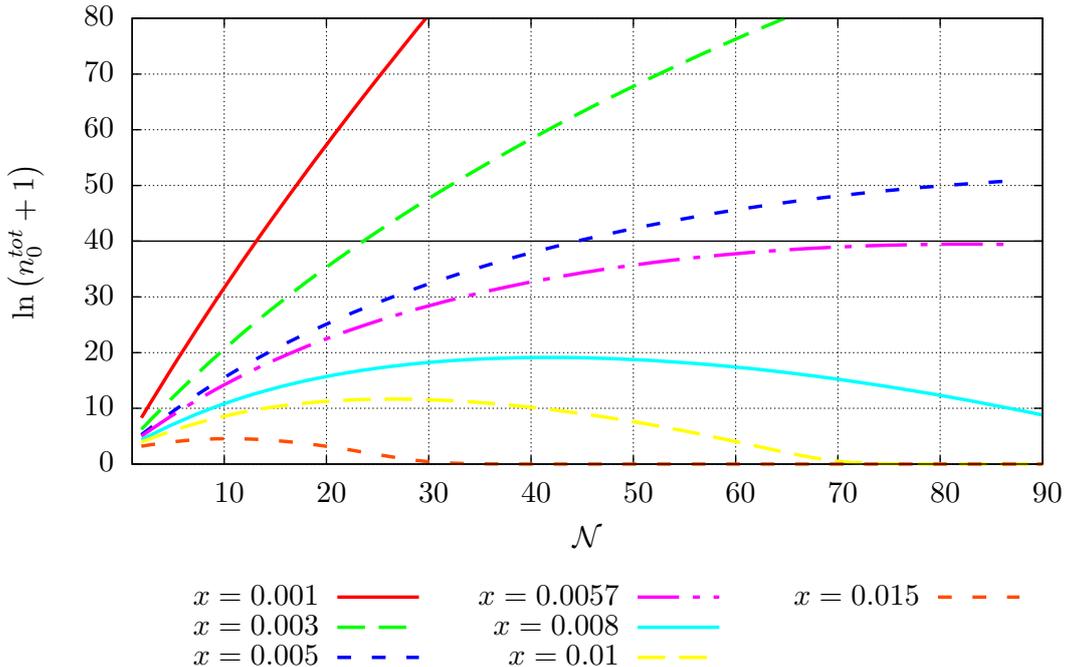

It is instructive to plot the total co-moving particle number in the zero mode at all ESPs after the first in-fall over the number of fields $\mathcal{N}$ for varying  inter-ESP distances $x$, see Fig.~\ref{fig:fi_eff}: the solid black line corresponds to the particle number $n_0$ generated at the end of the entire preheating phase with $\mathcal{N}=1$ and a single ESP at the origin (the canonical example of preheating); thus, this line serves as an indicator whether or not preheating is efficient. As evident, for dense ESP distributions and many inflatons, $x\lesssim 0.0057 M_p$ and $\mathcal{N}\geq 40$, more particles are produced during the first in-fall than during the entire preheating phase in the corresponding single field model. If ESPs are denser, less inflatons suffice. Thus, the first in-fall alone has the potential to compensate for the (potentially) decreased efficiency of preheating due to de-phasing effects \cite{Battefeld:2008bu,Battefeld:2009xw,Braden:2010wd}. Furthermore, as we shall see in Sec.~\ref{sec:resonancephase}, the late resonance phase can be prolonged if an ESP happens to be at the ``right'' distance from the origin. 

Since the total comoving particle number $n_\chi$ is related to the energy in the $\chi$-sector, and thus better suited to gauge the effectiveness of preheating, we would also like to provide an analytic approximation for $n_{\chi}$.
Using the definition of $n_k$ in (\ref{defnk}) we get
\begin{eqnarray}
\left<n_k\right>=\left<n_0\right>e^{-\frac{\pi}{gv}k^2}\,,
\end{eqnarray}
so that 
\begin{eqnarray}
\nonumber  n_{\chi} &=& \int n_k^{tot}\,\frac{\text{d}\vec{k}}{(2\pi)^3} = \frac{4\pi}{(2\pi)^3}\,\int_0^{\infty}k^2\,n_k^{tot}\,\text{d}k \\
 &=& \frac{4\pi}{(2\pi)^3}\,g^{-\frac{{\cal N}-1}{2}}\,\frac{\Gamma\left(\frac{{\cal N}-1}{2},\pi\right)}{\Gamma\left(\frac{{\cal
          N}-1}{2}\right)}\,x^{-{\cal N}}\,\int_0^{\frac{\pi}{m_1}}v^{\frac{{\cal N}+1}{2}}\,\underbrace{\left(\int_0^{\infty}k^2\,\text{e}^{-\frac{\pi}{g v} 
        k^2}\,\text{d}k\right)} \,\text{d}t\,. \\
\nonumber &&  \hspace{8.4cm} \frac{1}{4\pi}\,(g v)^{3/2}
\end{eqnarray}
With the definition of the velocity-integral, we arrive at 
\begin{eqnarray}
  \label{eq:fi_nchi}
  n_{\chi} = g^{-\frac{{\cal N}-4}{2}}\,\frac{\Gamma\left(\frac{{\cal N}-1}{2},\pi\right)}{\Gamma\left(\frac{{\cal N}-1}{2}\right)}\,x^{-{\cal N}}
  \cdot \frac{{\cal F}({\cal N},2)}{(2\pi)^3}\,.
\end{eqnarray}
(\ref{eq:fi_nchi}) provides an analytic understanding of particle production during the first in-fall and is our first main result. In its derivation, we made a series of approximations: we ignored expansion effects (justified since the first in-fall typically takes much less than a Hubble time), did not incorporate back-reaction (valid as long as occupation numbers do not grow too much) and considered particle production at ESPs up to the critical impact parameter  $\mu_c=\sqrt{g/v}$ only. 

Naturally, to gauge the ultimate effectiveness of preheating, back-reaction effects need to be taken into account once particle numbers in the $\chi$-fields are high (lattice simulations are needed) or if the corrections to the effective potential brought forth by the produced particles rivals that of the bare one\footnote{For example, if inflation is driven at the terminal velocity in a higher dimensional field space, as in \cite{Battefeld:2010sw}, back-reaction is important throughout; further, inflation ends once $\rho_\chi$ becomes comparable to the energy in the inflatons, leading to a smooth transition to preheating.} (the WKB approximation can be used).

\subsubsection{Comparison with Numerics\label{sec:compnumericsfirstinfall}}
We would like to test the validity of the analytic expressions in (\ref{analyticn0}) and (\ref{eq:fi_nchi}), which required a series of approximations. 

\begin{figure}[t]
  \centering
  \scalebox{1.03}{\input{first_infall2}}
  \caption{Comoving particle number in the $k=0$ mode, $n_0^{tot}$, after the first in-flall for different inter-ESP distances $x$ and number of inflatons $\mathcal{N}$. Solid lines: analytic approximation in (\ref{analyticn0}); filled circles: numerical result with ESPs distributed along a rectangular lattice, see Sec.~\ref{sec:matterfields}; open circles: random ESP distribution with the same average inter ESP distance as the grid.} 
  \label{fig:first_infall2}
\end{figure}
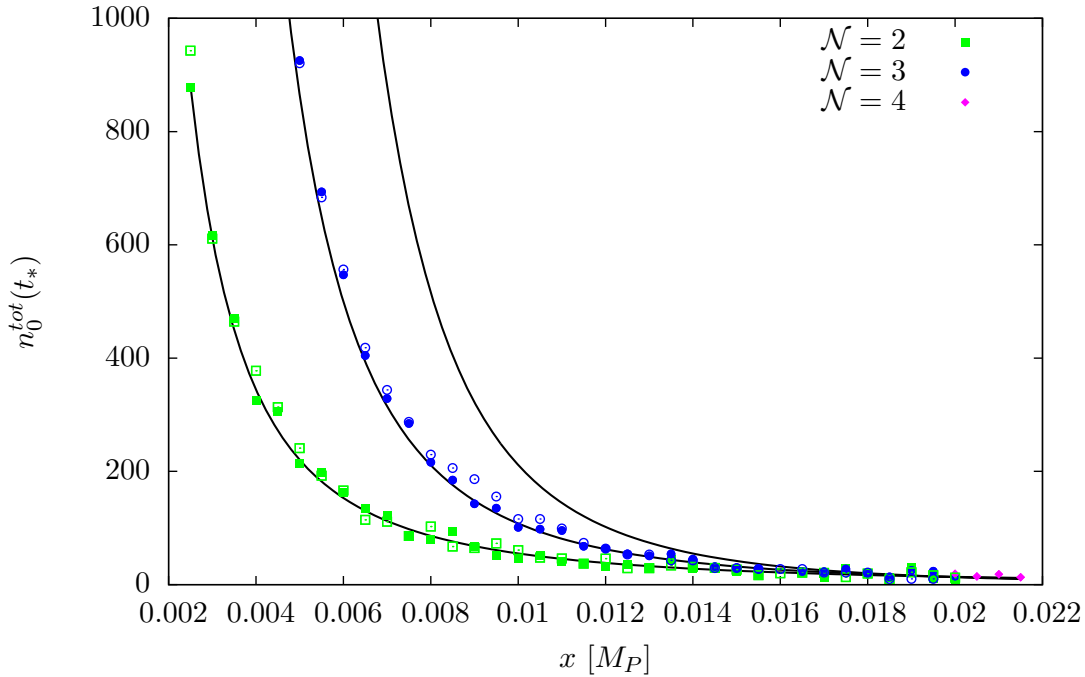

To this end, we compute numerically the comoving occupation number in the zero mode $n_0^{tot}$ at the end of the first in-fall, $t=t_*$, for different inter-ESP distances $x$ and distributions (grid like and random, see Sec.~\ref{sec:matterfields}) for up to four inflatons $\mathcal{N}=2,3,4$ (for $\mathcal{N}=4$, we only consider inter ESP distances down to $x=0.02\,M_p$ due to increasing computational demand), see Fig.~\ref{fig:first_infall2}.

We observe that the analytic approximation does a good job in explaining particle production during the first in-fall. The scatter around the analytic result is caused primarily by evaluating $n_0^{tot}$ at a fixed time, which is sensitive to the small spikes in Fig.~\ref{fig:ph_ablauf} (one could average over the first plateau to reduce the scatter). These spikes are caused by chance,  close ESP encounters. A concrete example is present for a regular grid with $x=0.016$ in Fig.~\ref{fig:first_infall2}: our initial conditions put the trajectory right on top of an ESP, leading to increased particle production (as a consequence, the corresponding data-point falls out of the range depicted in Fig.~\ref{fig:first_infall2}). Of course, such chance encounters can not be recovered by our analytic result.

To check whether or not ESP positioning along a grid influences the outcome qualitatively, we also performed numerical runs with randomly placed ESPs that have the same average inter-ESP distance in the region of field space of interest. Discrepancies are small and of the same order as the scatter around the analytic expression, which is expected.

\begin{figure}[t]
  \centering
  \scalebox{1.03}{\input{fi_iv}}
  \caption{Dependence of particle production during the first in-fall on initial conditions. Left: comoving occupation number is plotted over time for ${\cal N}=2$,
    $x=0,005\,M_P$ and several initial conditions $(\varphi_0;\,\dot{\varphi}_0)$ in units of $(M_P; 10^{-6}\,M_P^2)$ for all inflatons, except in the third and forth case from above, for which we also set $\varphi_1(0) = 0,2\,M_P$ and $\dot{\varphi}_1(0) = 0$. Right: The corresponding value of the velocity integral $\mathcal{F}(\mathcal{N},0.5)$, compared to the one resulting from our ``standard'' initial conditions. Changes in particle production are present (as expected) but minor.} 
  \label{fig:fi_iv}
\end{figure}
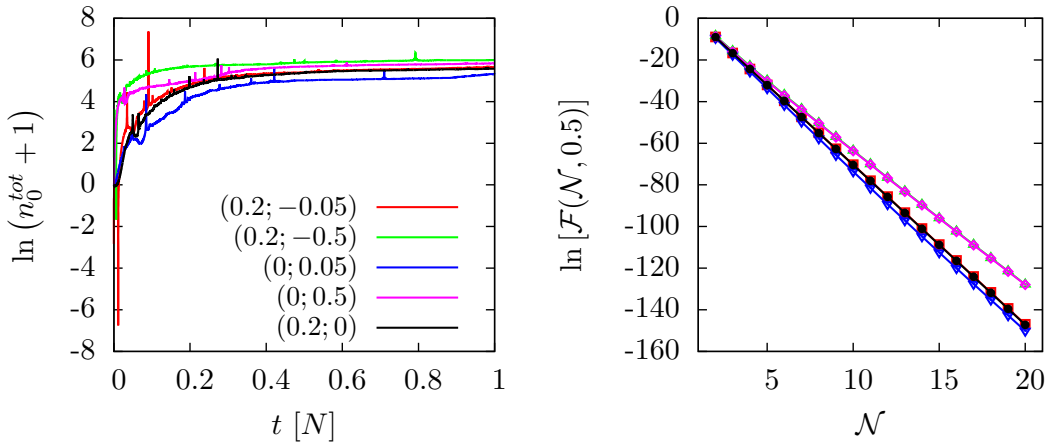

Changes in the initial conditions lead to altered analytical as well as numerical results, for two reasons: firstly, choosing a different starting point can prolong/shorten the first in-fall. Secondly, changing the initial speed alters the value of the velocity integrals $\mathcal{F}$: an increased speed leads to a larger $\mu_c$ and thus boosts particle production. Both effects are evident in Fig.~\ref{fig:fi_iv}, but they are ultimately minor, leaving the qualitative features unchanged. 

Thus, the analytic expressions in Sec.~\ref{sec:firstinfall} provide an excellent description of particle production during the first in-fall as long as back-reaction effects are negligible.

\subsection{The Resonance Phase \label{sec:resonancephase}}
After a brief intermediate phase, the oscillation amplitude is smaller than the inter-ESP distance $x$, and resonances can occur. Opposite to the first in-fall, these resonances take place at a few ESPs only. Thus, let us focus at an exemplary ESP located at 
\begin{eqnarray}
\vec{\varphi}_{\tiny ESP}=\frac{x}{2}\vec{1}\,,\label{locesp}
\end{eqnarray}
and investigate how much particle production in the $k=0$ mode occurs until preheating terminates, i.e.~when the final plateau in Fig.~\ref{fig:trajectory} is reached. Note that the distance to the origin is
\begin{eqnarray}
d\equiv \sqrt{\mathcal{N}}\frac{x}{2}\,.
\end{eqnarray}

\begin{figure}[!t]
  \centering
  \scalebox{1.03}{\input{occ_number2}}
  \caption{Comoving occupation number of the $k=0$ mode over the distance of a single ESP from the origin, (\ref{locesp}), for two and three inflatons. Top: masses are distributed according to (\ref{masses1}), such that ratios are simple fractions, i.e.~for $\mathcal{N}=2$ we have $m_2/m_1=3/2$ and for $\mathcal{N}=3$ we get $m_2/m_1=5/4$, $m_3/m_1=3/2$.  Bottom: masses are distributed according to (\ref{masses2}), such that ratios are irrational, i.e.~for $\mathcal{N}=2$ we have $m_2/m_1=\sqrt{3}$ and for $\mathcal{N}=3$ we get $m_2/m_1=\sqrt{2}$, $m_3/m_1=\sqrt{3}$. The first peak at $x=0.00141\,M_p$ in the top-left panel is truncated. }
  \label{fig:occ_number}
\end{figure}
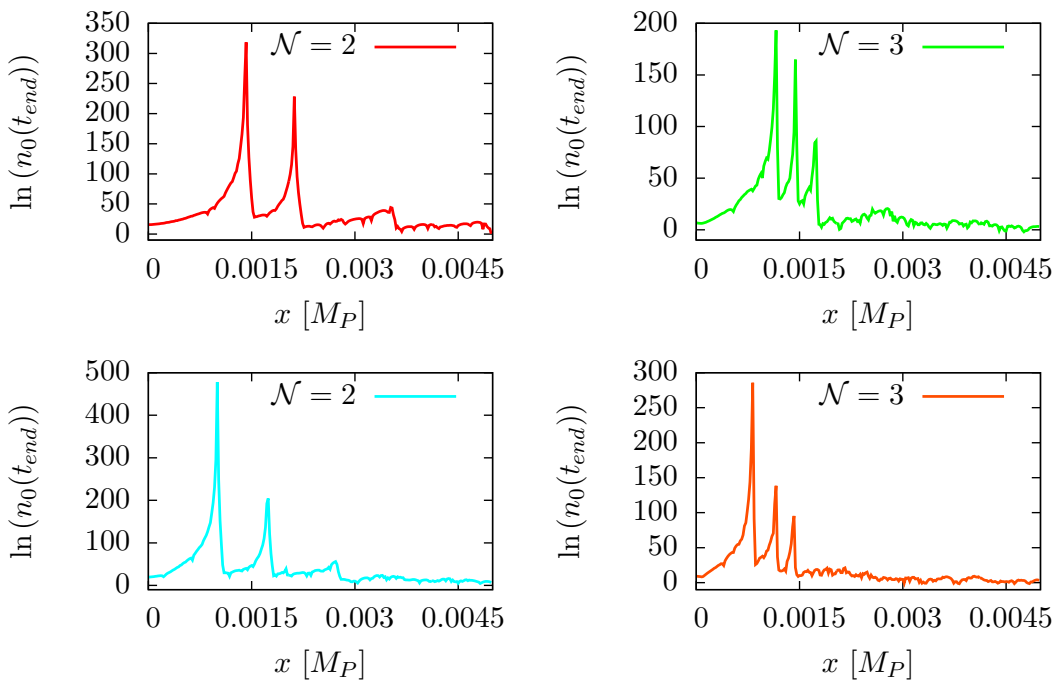

To get an overview of possible outcomes, we first compute $n_0$ numerically, as in the previous section, for different values of $x$, see Fig.~\ref{fig:occ_number}. The plotted value is the result of performing a linear fit over the final plateau (the interval is chosen to be long enough so that the average slope is close to zero). The top plots are for the mass distribution in (\ref{masses1}) (simple mass ratios), while the bottom one follows (\ref{masses2}) (incommensurate masses). 

Our intuition based on Sec.~\ref{sec:firstinfall} indicates that an increasing $x$ should result in decreased particle production: the impact parameter to the ESP is on average larger, resulting in a suppressed particle number according to (\ref{defnk}). This intuition is wrong, as evident in Fig.~\ref{fig:occ_number}: increasing $x$ from zero leads to a steady increase of particle production with spikes for certain values of $x$. These resonances appear to be more frequent and closer to $x=0$ as the dimensionality of field space is increased, but also weaker, in line with the results of Sec.~\ref{sec:dephasing}. Only for large values of $x$ is a steady suppression evident, in accord with ones intuition.

Our intuition is flawed, because it is based on adiabaticity and violation thereof, i.e.~we argued that $\eta \ll 1$ needs to be violated for particle production to be efficient. This point of view is valid if the effective mass of the matter field $m_\chi^{eff}=g|\vec{\varphi}-\vec{\varphi}_{\mbox{\tiny ESP}}|$,  and thus the oscillation frequency $\omega_k$, is considerably larger than the inflaton masses; this is only the case in the early stages of preheating: the preheat matter field oscillates many times during a single oscillation of even the heaviest inflaton field; since $|\chi_k|\propto \omega_k^{1/2}$, see (\ref{adiabaticsolution}), the particle number can only change during the brief moments when $|\vec{\varphi}-\vec{\varphi}_{\mbox{\tiny ESP}}|<\mu_c$. However, once the amplitude of the effective mass becomes comparable to the inflaton masses, it is impossible to separate the evolution of $n_k$ into adiabatic/non-adiabatic regimes (exponential increases in particle number are possible during the latter, as seen in Sec.~\ref{sec:narrowresonance}). Since stochastic resonance arises from the interplay of adiabatic and non-adiabatic regimes, it can't be operational here and we conclude that narrow resonance must be responsible for the spikes visible in Fig.~\ref{fig:occ_number}.

We confirm this expectation in the next subsections, and, in the case of simple mass ratios, compute the location of these spikes as well as the generalized Floquet index responsible for the particle number's exponential increase. 

This description remains qualitatively valid for incommensurable masses, but we will not be able to provide exact analytic expressions.

\subsubsection{Instability Bands \label{sec:instabilitybands}}
To perform a stability analysis for $X_k$ in case of simple mass ratios, (\ref{masses1}), we ignore the expansion of the universe so that
\begin{eqnarray}
X_k'' + \frac{1}{m_1^2}\left[ k^2 + g^2\,\sum_{i=1}^{{\cal N}}\left( \phi_{0}^{(i)}\,\cos{(m_it)} - \frac{x_i}{2}\right)^2
  \right]\,X_k = 0\,,
\end{eqnarray}
where we introduced $z\equiv m_1t$, a prime denotes a derivative with respect to $z$, we used $\varphi_i(t)=\phi_{0}^{(i)}\cos(m_it)$ (static universe) and put the ESP at $\vec{\varphi}_{ESP}=\vec{x}/2$. Defining
\begin{eqnarray}
  \label{eq:parameter}
  A_k \equiv \frac{4k^2 + g^2\sum_{i=1}^{{\cal N}}(2\phi_{0}^{(i)2}+x_i^2)}{4m_1^2}\,, \hspace{0.5cm} q_1^{(i)} \equiv -\frac{g^2\phi_0^{(i)2}}{4m_1^2}\,, \hspace{0.5cm} q_2^{(i)} \equiv
  \frac{g^2\phi_0^{(i)} x_{i}}{2m_1^2}\,,
\end{eqnarray}
we arrive at an equation similar to the Mathhieu-equation
\begin{eqnarray}
  \label{eq:qpm}
  X_k'' + \omega_k^2 X_k = 0 
\end{eqnarray}
with 
\begin{eqnarray}
\omega_k^2\equiv A_k - 2\sum_{i=1}^{{\cal N}}q_1^{(i)}\cos{\left(2\frac{m_i}{m_1}z\right)} - 2\sum_{i=1}^{{\cal N}}q_2^{(i)}\cos{\left(\frac{m_i}{m_1}z\right)}\,;
\end{eqnarray}

The differences to the Mathieu-equation are the presence of several oscillators and the appearance of new parameters, $q_1^{(i)}$ and $q_2^{(i)}$. As a consequence, we can not directly apply Floquet-theory even in the narrow resonance regime. Nevertheless, it is illuminating to investigate the instability regions dependence on $\phi_0^{(i)}$ to decide whether or not resonance are likely. Since $\phi_0^{(i)}$ are decreasing in an expanding universe, $\phi_0^{(i)}\propto t^{-1}$, a certain mode will not remain in an instability region, but scan several bands -- knowledge of these bands  enables us to predict the occurrence and length of preheating \cite{Zlatev:1997vd}. 

The main difference to prior studies such as \cite{Kofman:1997yn} is the position of the ESP (it is not at the origin any more), leading to qualitative differences. Its main effect is the addition of a new term $\propto x_i^2$ in $A_k$ (indistinguishable from a bare $\chi$-mass), as well as the appearance of $q_2\propto x_i$ (unlike a bare mass). Thus, conclusions in the small $q_i$ limit carry over directly to multi-field preheating with a massive $\chi$-field that becomes light, but not massless, at the origin (see Sec.~\ref{sec:latticeatsingleESP}). Surprisingly, this case has not received much attention in the literature either (see however \cite{Zlatev:1997vd}), primarily based on the intuition that a bare mass suppresses broad resonances (it does) and that narrow resonances are always irrelevant in an expanding universe (they are not, as we shall see below).

Let us examine the case $q_1^{(i)},q_2^{(i)}\ll A_k$, that is small oscillation amplitudes and an ESP close to (but not at) the origin. Note that even though the presence of the ESP breaks spherical symmetry, it is regained in this regime since 
\begin{eqnarray}
 A_k \equiv \frac{4k^2 + g^2{\cal N}(2\phi_0^2+x^2)}{4m_1^2}
 \end{eqnarray}
 depends only on the distance of the ESP to the origin,
\begin{eqnarray}
d\equiv \sqrt{\cal N}\frac{x}{2}\equiv |\vec{\varphi}_{ESP}|= \sqrt{\sum_{i=1}^{\cal N} \left(\frac{x_i}{2}\right)^2}\,,
\end{eqnarray}
and the total oscillation amplitude
\begin{eqnarray}
\sqrt{\cal N}\phi_0\equiv\sqrt{\sum_{i=1}^{\cal N}\phi_0^{(i)2}}\,.
\end{eqnarray} 
Since we work in the limit of small amplitudes and we are particularly interested in the case $\phi_0<x/\sqrt{2}$ and low $k$, we approximate 
\begin{eqnarray}
A_k\approx \frac{4k^2+g^2{\cal N}x^2}{4m_1^2}
\end{eqnarray}
to remain consistent. We comment on the breaking of spherical symmetry by the presence of $q_2^{(i)}$ in Sec.~\ref{sec:latticesveralESP}.

Defining $s_1\equiv X_k$ and $s_2\equiv X_k^\prime$, we get a simple system of first order differential equations
\begin{eqnarray}
  \label{eq:qpmatrix}
  \frac{\text{d}}{\text{d}z}\vec{s} = {\cal M}(z)\,\vec{s} \hspace{0.5cm} \text{with} \hspace{0.5cm} {\cal M}(z) = \begin{pmatrix} 0 & 1 \\ -w_k^2(z) &
    0\end{pmatrix}\,.
\end{eqnarray}
Given initial conditions $\vec{s}_0$, its solution 
$\vec{s}(z) = {\cal U}(z,z_0)\vec{s}(z_0)$ is given by a linear  \emph{evolution operator} ${\cal U}(z,z_0)$ that solves
\begin{eqnarray}
  \frac{\text{d}}{\text{d}z}{\cal U}(z,z_0) = {\cal M}(z)\,{\cal U}(z,z_0)\,,
\end{eqnarray}
and satisfies ${\cal U}(z_2,z_0) = {\cal U}(z_2,z_1)\,{\cal U}(z_1,z_0)$.

Let us denote the smallest common multiple of $T_i\equiv 2\pi m_1/m_i$, that is the period of $\omega_k^2$, by $T$. For  the mass distribution in (\ref{masses1}) we get\footnote{Only the frequencies of $q_2$ need to be considered, since they are (up to a factor of $2$) identical to the ones of $q_1$.}
\begin{eqnarray}
T=4\pi(\mathcal{N}-1)\,.
\end{eqnarray}
If mass ratios are incommensurable, $\omega_k^2$ is only quasi-periodic and this period does not exist. However, the presence of spikes in Fig.~\ref{fig:occ_number} is not contingent on $T$ being the exact period, but merely the time-scale after which oscillators tend to be (approximately) in phase again; practically, it appears that truncating the mass ratio to one digit and computing the resulting period provides a good estimate of this time-scale for our purposes \footnote{A precise definition of $T$ for incommensurable masses is of course desirable, but we were not able to find one. In all cases that we investigated numerically, such a relevant time-scale $T$ clearly exists.}.

Given this ``period'' and keeping in mind that the following is only exact for commensurable masses as in (\ref{masses1}), we get for $j \in \mathbb{N}$
\begin{eqnarray}
\vec{s}(jT+z_0)=(\mathcal{U}_T(z_0))^j\vec{s}(z_0)\,,
\end{eqnarray}
where we defined $\mathcal{U}_T(z_0)\equiv \mathcal{U}(z_0+T,z_0)$; based on this property, we can perform a simple stability analysis, focussing on the eigenvalues of $\mathcal{U}$: if they are imaginary, solutions are stable, while real eigenvalues indicate an instability. The eigenvalues are
\begin{eqnarray}
\lambda_{\pm}=\frac{\mbox{Tr }\mathcal{U}_T\pm\sqrt{(\mbox{Tr }\mathcal{U}_T)^2-4 \,\mbox{det }\mathcal{U}_T}}{2}\,.
\end{eqnarray}
According to Jacobi's identity for the derivative of a determinant, we conclude that $\mbox{det }\mathcal{U}_T=\mbox{const.}$, since the trace of $\mathcal{M}(z)$ vanishes. Further, since $\mathcal{U}_T(z_0,z_0)$ is the identity matrix, we find $\mbox{det }\mathcal{U}_T=1$, from which we can deduce the stability condition
\begin{eqnarray}
|\mbox{Tr }\mathcal{U}_T|\leq 2\,.\label{stabilitycond}
\end{eqnarray}
In general, it is impossible to determine the evolution operator analytically; however, for negligible small $q_i$, the equations of motion are the ones of a harmonic oscillator with constant frequency $\sqrt{A_k}$ and known solutions.  Since the trace is invariant under a change of basis, we can choose the linearly independent initial value problems $\vec{s}(0)=(1,0)$ and $\vec{s}(0)=(0,1)$, leading to
\begin{eqnarray}
  {\cal U}(z,0) = \begin{pmatrix} \cos{\left(\sqrt{A_k}\,z\right)} & \frac{1}{\sqrt{A_k}}\,\sin{\left(\sqrt{A_k}\,z\right)} \\
    -\sqrt{A_k}\,\sin{\left(\sqrt{A_k}\,z\right)} & \cos{\left(\sqrt{A_k}\,z\right)} \end{pmatrix}\,.
\end{eqnarray}
According to (\ref{stabilitycond}), the transition from stability to instability regions occurs at
\begin{eqnarray}
  \label{eq:akstab}
  \hspace{0.5cm} \left| \cos{\left(\sqrt{A_k}\,T\right)} \right| = 1 \hspace{0.5cm} \Leftrightarrow \hspace{0.5cm} A_k =
  \left(\frac{n\,\pi}{T}\right)^2\,, \hspace{0.25cm} n \in \mathbb{N}\,.
\end{eqnarray}
The limit of negligible $q_1^{(i)},q_2^{(i)}$ becomes a good approximation in the late stages of preheating, since they are redshifted by the expansion of the universe. Thus, the asymptotic value of $A_k$, which is set primarily by the ESPs position parametrized by $x$ (momenta also redshift), determines the effectiveness of preheating and we expect the strongest particle production at\footnote{The index $n$ labels the $n$'th distance of the ESP from the origin at which particle production is efficient, $d_n=\sqrt{\cal N}x_n/2$; the $x_n$ should not be confused with the coordinates of the ESPs position in field space, $x_i$ where $i=1\dots {\cal N}$. }
\begin{eqnarray}
  \label{eq:xkrit}
  \frac{g^2{\cal N}}{4\,m_1^2}\,x_n^2 = \left(\frac{n\,\pi}{T}\right)^2 \hspace{0.5cm} \Rightarrow \hspace{0.5cm} x_n = \frac{2\,m_1}{g \sqrt{{\cal N}}}
  \cdot \frac{n\,\pi}{T}\,.
\end{eqnarray}
As we argued above, the above is valid for arbitrary positions of the ESP at distance $d_n = \sqrt{{\cal N}}\,x_n/2$ from the origin, since only this distance entered in our derivation. 

For the concrete mass distribution in (\ref{masses1}), that is for masses distributed in the intervall $[1,1.5] \cdot 10^{-6}\,M_P$, we get for ${\cal N} = 2,3$ 
\begin{eqnarray}
  x_n({\cal N}=2) &= 7,071 \cdot 10^{-4}\,M_P \cdot n\,, \label{xnNis2}\\
  x_n({\cal N}=3) &= 2,887 \cdot 10^{-4}\,M_P \cdot n\,.
\end{eqnarray}
In the interval $0.001 M_p$ to $0.002 M_p$ we find indeed peaks at these positions, see top panels in Fig.~\ref{fig:occ_number}, but not for smaller or larger values. This is not particularly surprising, since we made some harsh approximations along the way. For incommensurate masses, we find qualitatively the same behaviour, given that $T$ is not identified with an exact period, but a characteristic time-scale common to the oscillators, as explained above. Unfortunately, the amplitude of the peaks is not as easily computed analytically, requiring a numerical integration.

\subsubsection{Numerical Results \label{sec:numericalresults}}
\begin{figure}[t]
  \centering
  \scalebox{1.03}{\input{iband}}
  \caption{The characteristic exponent $\mu$ of the $k=0$-mode for different ESP distances from the origin, plotted over the initial values of the inflatons $\phi_0$ (top) and time (bottom).}
  \label{fig:iband}
\end{figure}
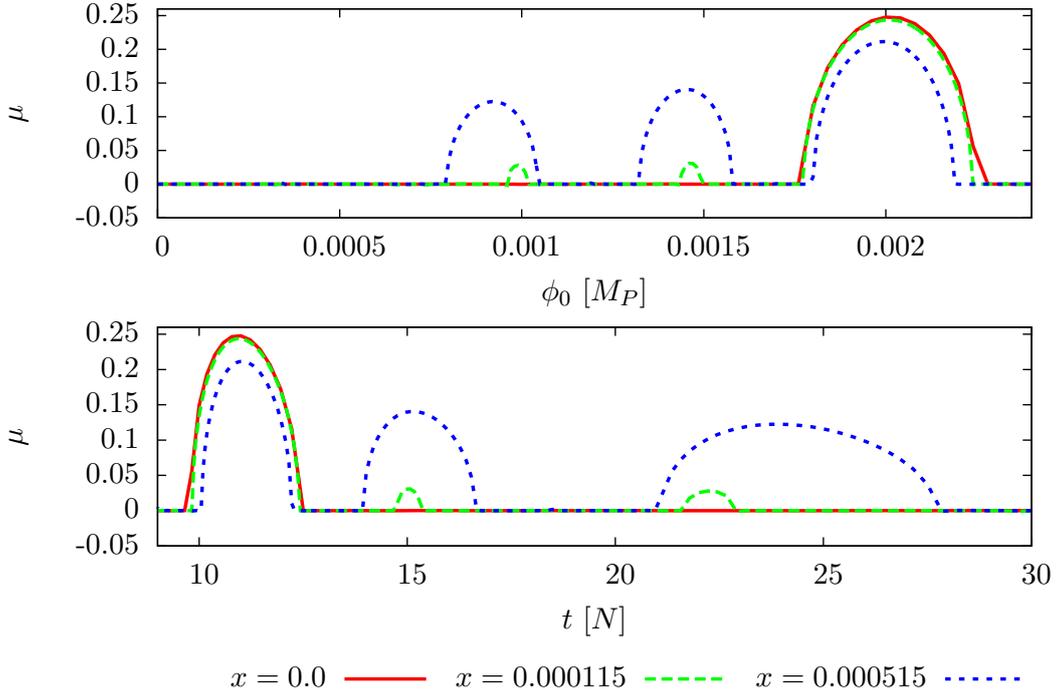

To go beyond analytic estimates, we compute numerically the characteristic exponent $\mu$ of the zero mode, $k=0$,  for varying distances of the ESP from the origin, $x$, oscillation amplitude of the inflatons, $\phi_0<x$, and mass ratios in a static universe. We chose $\phi_i^{(0)} \equiv \phi_0$ and $x_i\equiv x$ for all $i=1\dots {\cal{N}}$, keeping in mind that results are valid more generally since spherical symmetry is recovered in the small amplitude limit. 

\begin{figure}[t]
  \centering
  \scalebox{1}{\input{first_band}}
  \caption{The first instability-band of the $k=0$-mode in a two dimensional field space, $\mathcal{N}=2$, for different ESP distances from the origing, plotted over the initial values of the inflatons $\phi_0$.}
  \label{fig:first_band}
\end{figure}
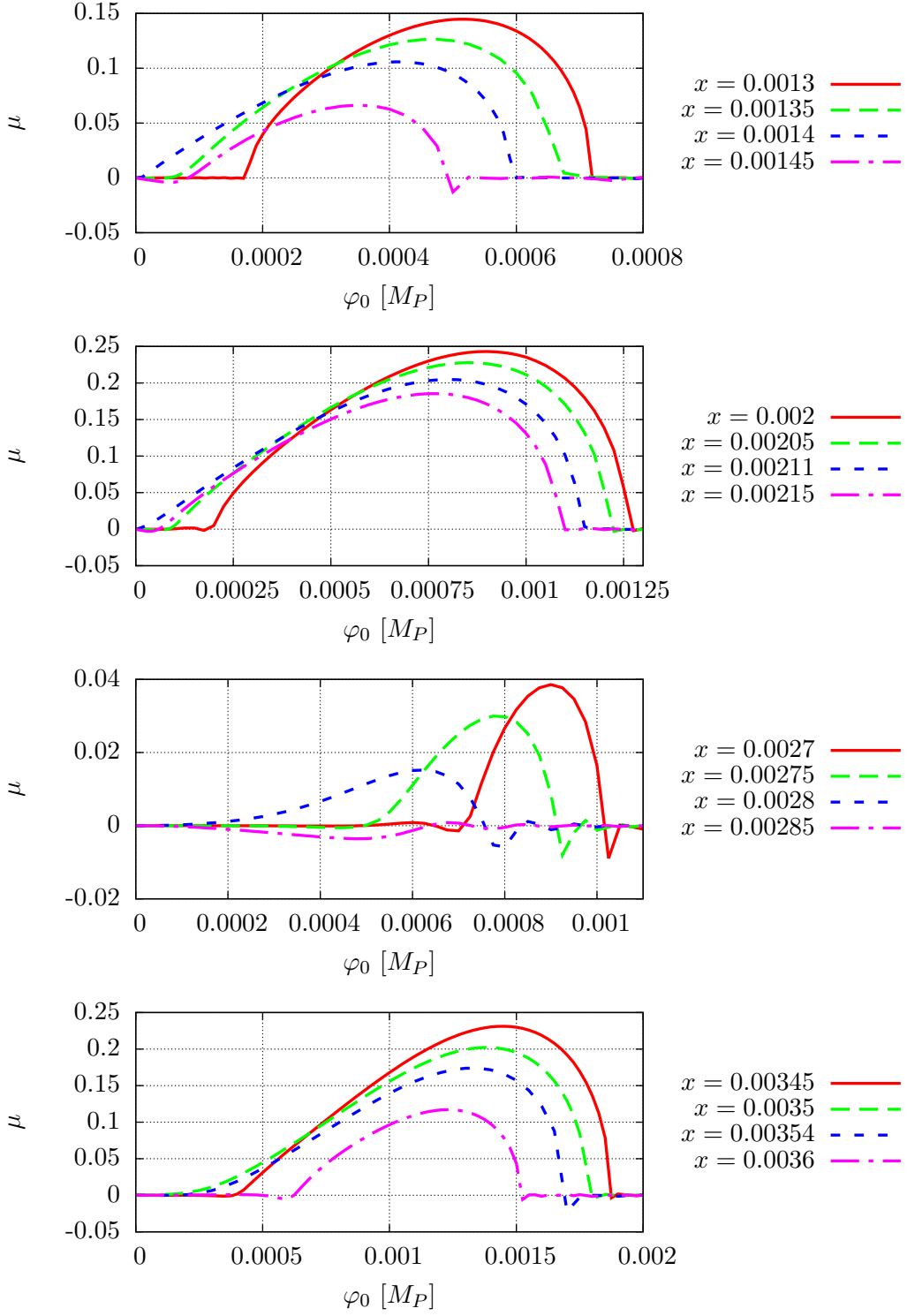

To compute $\mu$ we perform a linear fit to $\ln(n_k+1)$ whose slope, divided by $2\pi$, provides the desired exponent. The appearance of resonance bands (positive $\mu$) in plots of $\mu$ over $\phi_0$, see Fig.~\ref{fig:iband} to Fig.~\ref{fig:mass_dep},  indicates that the zero mode gets amplified; in an expanding universe, the amplitude $\phi_0$ decreases over time, such that a mode is amplified for a limited time only while the oscillation amplitude lies within an instability band.

In Fig.~\ref{fig:iband} we increase the distance of the ESP from the origin by changing $x$ from $x=0$ over $x=0.00015\,M_p$ to $0.000515\,M_p$. We observe the appearance of more instability bands as the ESP is shifted away from the origin, as well as an increase in $\mu$ for the first bands as $x$ increases. This explains the increase of particle production as $x$ shifts away from zero in Fig.~\ref{fig:occ_number}. It is possible to compute the position of these bands based on (\ref{eq:akstab}) with (\ref{eq:parameter}), by determining the instability boundaries of $\phi_0$ from the corresponding ones of $A_k$, but the approximation of small $q_i$ becomes unreliable as $x$ and $\phi_0$ grow. Thus, our analytic estimates provide a qualitative (and for small $q_i$ quantitative) explanation of the instability charts.

The appearance of additional instability bands can be understood as well: by increasing $x$, we can no longer neglect the presence of $q_2$, leading to a doubling of the period $T$ and correspondingly the number of instability bands.

The observed growth of $\mu$ with increasing $x$ in the first bands leads to the dominance of the first one, and to a lesser degree the second one, during the late stages of preheating. To see this more clearly, it is instructive to associate a time $t$ with each $\phi_0$ according to (\ref{amplitudeovertime}), that is to identify the average oscillation amplitude $\left<|\vec{\varphi}|\right>$ with $\phi_0$ so that 
\begin{eqnarray}
t \equiv \frac{1}{\phi_0}\frac{\sqrt{\sum m_i^2}}{ \sqrt{12\pi \mathcal{N}}}\,.\label{endtime}
\end{eqnarray}
As evident from the lower panel in Fig.~\ref{fig:iband}, the zero mode spends the longest time in the first band. Thus, the first band can dominate preheating even though higher bands might be broader and have a larger characteristic exponent. From this plot we can also deduce the time at which preheating ends, since it corresponds to leaving the first resonance band once $\phi_0$ becomes too small.

We also observe a shift of instability bands to lower values of $\phi_0$ as $x$ is increased, resulting in an increased efficiency of preheating. We expect strong preheating should an instability band shift all the way down to $\phi_0=0$. To see at which ESP positions such a shift takes place, we compute the first instability band in a two dimensional field space while increasing $x$ from $x=0.0013\,M_p$ to $x=0.0036\,M_p$ in $16$ increments, see Fig.~\ref{fig:first_band}. Evidently, the distance of the first band from $\phi=0$ is minimal for certain values of $x_*$, leading to a prolonged (and thus strong) phase of particle production. As $x$ is increased further, the band moves away from $\phi_0=0$ again, with ever decreasing $\mu$ up until it vanishes; finally, the next band repeats this sequence of events.

Focussing on the first instability band, we see that it is closest to $\phi_0=0$ for $x_*\approx 0.00141\,M_p$, in accord with our analytic prediction in (\ref{xnNis2}) for $n=2$. 

However, we observed in Fig.~{\ref{fig:occ_number} the absence of spikes for $n\geq 4$, opposite to the prediction of (\ref{xnNis2}). To investigate the origin for their absence, we plot the instability bands corresponding to $n=4$ and $n=5$ in the lower two panels of Fig.~\ref{fig:first_band}: even though a given instability band approaches the origin as before, we observe an increased suppression of $\mu$ (and thus preheating) for small values of $\phi_0$ that are particularly relevant during the late stages of the resonance phase. This explains the absence of spikes in Fig.~{\ref{fig:occ_number}. Similarly, no instability bands are present for $n=1$, that is $x_1=7.071 \times 10^{-4}\, M_p$, and in case of three inflatons for $n\geq 4$. 

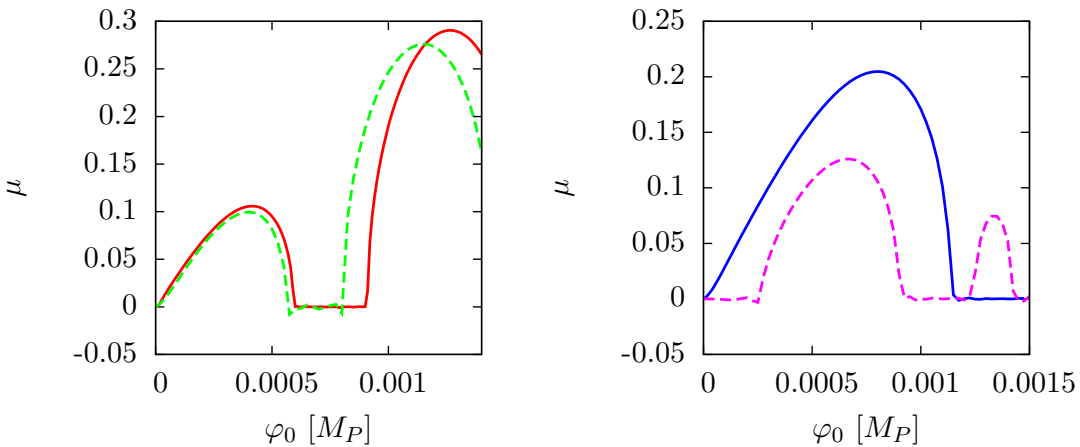
\begin{figure}[t]
  \centering
  \scalebox{1.03}{\input{mass_dep}}
  \caption{Instability-bands of the $k=0$ mode for 
${\cal N}=2$ and $x=0,0014\,M_P$ (left) as well as
    $x=0,00211\,M_P$ (right). Solid line: simple mass ratio $m_2/m_1 = 1.5$; dotted line: incommensurable mass ratio $m_2/m_1 = \sqrt{2}$.}
  \label{fig:mass_dep}
\end{figure}

All of our analytic estimates were based on simple mass ratios, so that the period $T$ is well defined. However, realistic models are expected to have more complicated mass spectra, such as the Marcenko-Pastur distribution in $N$-flation \cite{Easther:2005zr}. 

How do our estimates fare with more complicated mass ratios? In Fig.~\ref{fig:occ_number}, we observed the same qualitative behaviour, i.e.~the presence of spikes, in line with our analytic expectation if $T$ is not identified with the period, but a characteristic time-scale of the oscillators. Thus, we still expect to see instability bands. To test this hypothesis, we computed the characteristic exponent in a two field model for two different mass ratios, $m_2/m_1=1.5$ (simple) and $m_2/m_1=\sqrt{2}$ (incommensurate) while keeping the overall scale unchanged, see Fig.~\ref{fig:mass_dep}. 

For small $x$, we observe a close correspondence between bands, especially for small $\phi_0$. Hence, the late stages of preheating are nearly indistinguishable in both cases. However, this correspondence becomes worse as $x$ is increased and we loose the ability to predict the location of instability bands. Nevertheless, the qualitative features (clear instability bands are present that shift to lower $\phi_0$ as $x$ is increased) remain and we expect little change in the overall effectiveness of the resonance phase.

\section{Summary and Discussion: Is Preheating Efficient? \label{sec:summaryanddiscussion}}
What is meant by efficiency? A pragmatic definition considers whether or not a  seizable fraction of the inflatons' energy  can be converted rapidly (compared to the Hubble time) into other degrees of freedom. In the presence of several inflatons, parametric resonance of a preheat matter-field $\chi$ that becomes light at the VEV of the inflatons is generally strongly suppressed due to de-phasing of inflatons \cite{Battefeld:2008bu,Battefeld:2008rd,Battefeld:2009xw,Braden:2010wd}; such de-phasing leads to a large additional contribution to the effective mass of $\chi$, preventing resonances. 

The main goal of this work was to investigate if this suppression prevails if one or more preheat matter fields become light at locations (ESPs) different from the VEV of the inflatons; the presence of several ESPs is generic for moduli spaces in string theory and often unavoidable in concrete models of inflation, e.g.~in trapped inflation \cite{Kofman:2004yc,Green:2009ds,Silverstein:2008sg,Battefeld:2010sw}.

In Sec.~\ref{sec:Preheatingwithoutbackreaction} we observed and explained two main effects that can compensate for the suppression due to de-phasing,
\begin{enumerate}
\item particle production during the the first in-fall (ESPs have to be densely distributed),
\item particle production at certain ESPs (special locations) during a prolonged (narrow) resonance regime.
\end{enumerate}
We shall now summarize and discuss our main findings, before turning to lattice simulations to investigate the effects of back-reaction and double check our analytic results.

\subsection{The First In-fall \label{sec:discfirstinfall}} 

During the first in-fall, particles are produced due to the violation of the adiabaticity  condition $\eta<1$  with $\eta$ in (\ref{nonadiabaticity}) at all ESPs that are within a certain distance of the trajectory in field space, roughly $|\vec{\varphi}-\vec{\varphi}_{ESP}|\lesssim \mu_c$, where $\mu_c$ is the critical impact parameter in (\ref{def:muc}). We computed analytically the co-moving occupation number in the zero mode in (\ref{analyticn0}) as well as the total number density in (\ref{eq:fi_nchi}). We compared these results with numerics (no back-reaction), see Fig.~\ref{fig:first_infall2}, and we found good agreement. Based on the analytic results, we deduced that the suppression due to de-phasing can be more than compensated if ESPs are distributed densely and several inflatons are present, see Fig.~\ref{fig:fi_eff} where we plot the total number density in the zero mode which we compare to the canonical single field preheating result ($g=5\times 10^{-4}$ and inflaton masses close to the COBE mass; for simplicity, we took the same coupling $g$ between all pairs of $\chi_i$ and $\varphi_i$). As a rule of thumb, the inter ESP distance should be below $x\lesssim 0.005 M_p$ for  $\mathcal{N}\sim 40$: including more inflatons first increases and then decreases the total particle number (too many or too few inflatons suppress $n_0^{total}$); further, denser ESP distributions lead to more particle production. We checked that these results are insensitive to the concrete distribution of ESPs (only the average inter ESP distance matters) or initial conditions of the inflatons (deviations are primarily caused by chance close encounters with individual ESPs, that boost particle production briefly).

A word of caution: once the particle number becomes large, back-reaction needs to be incorporated: it is possible for the inflatons to slow down considerably, extending the inflationary phase; this effect is explained in detail in \cite{Battefeld:2010sw} and therefore not repeated here. 
It should be noted that lattice simulations are not feasible during the first in-fall, due to the plethora of preheat matter fields that need to be considered. Thus, back-reaction can at most be incorporated via the WKB approximation, as in \cite{Battefeld:2010sw}.

 Nevertheless, we can make some important, general observations: firstly, if this phase is efficient, a seizable fraction of the energy in the inflatons is distributed over many distinct preheat matter fields, but the occupation number in each $\chi_i$ is low. 
 Secondly, preheating can not be completed in this regime: once the speed along the trajectory falls below $v\sim gx^2$, less than one ESP is (on average) close enough to allow for violation of the adiabaticity condition. Thus, at the very least energy of order $v^2\sim g^2x^4$ remains in the inflaton sector after the first in-flall. This energy needs to be transferred to other fields subsequently. A slow decay boosts the relative importance of these decay products, while a rapid decay, e.g.~by resonance, retains the initial energy density ratio.

\subsection{The Resonance Phase \label{sec:discres}}
After the first in-fall, a transitory phase with large (compared to the inter ESP distance $x$) amplitude oscillations follows, during which random bursts of particle production can occur whenever the trajectory comes close to an ESP, see Fig.~\ref{fig:ph_ablauf}. These intermittent bursts are usually not sufficient to increase the particle number in preheat matter fields considerably. Once the oscillation amplitude falls below the inter ESP distance, a distinction into adiabatic/non-adiabatic regimes ceases to be meaningful, indicating that broad/stochastic resonances are impossible. Surprisingly, we found that a type of narrow resonance can be efficient in creating many particles in the zero mode if ESPs are at certain distances $d_n=x_n\sqrt{\mathcal{N}}/2$ (with $x_n$ in (\ref{eq:xkrit})) from the VEV of the inflatons, see Sec.~\ref{sec:instabilitybands}. These distances depend on the number of inflatons, their masses and the coupling(s) to the preheat matter fields. Further, a period $T$ entered, which is only well defined for simple mass ratios (for incommensurate masses, it needs to be replaced by a characteristic time-scale of the oscillations). 

The presence of resonances can be understood intuitively by computing the characteristic exponent $\mu$ ($\mu>0$ indicates an exponential increase of $n_0$) for particle production at a single ESP in a static universe, see Fig.~\ref{fig:iband}-\ref{fig:mass_dep}, where we plot $\mu$ over the initial amplitude of inflatons $\phi_0$ ($\phi_0$ is defined such that the total  amplitude is $A=\varphi_0\sqrt{\mathcal{N}}$). As the location of the ESP moves away from the origin, instability bands move towards lower values of $\varphi_0$ (whenever a band touches $\varphi_0=0$, the ESP is at one of the special distances $d_n$). In an expanding universe, $A$ decreases in time, scanning different bands in the process; naturally, if a band reaches all the way down to $\varphi_0=0$, resonant particle production can operate for a long time and preheating is efficient. 
However, not all $d_n$ lead to efficient preheating, since the instability bands can become exceedingly flat (see the discussion at the end of Sec.~\ref{sec:numericalresults}); as a rule of thumb, only low $n$ lead to strong resonances. 

The resonance phase ends once the oscillation amplitude becomes so small that the first resonance band is left\footnote{If the first band reaches all the way down to $\varphi_0$, we still have $\mu(\varphi_0)\rightarrow 0$ smoothly, and particle production still becomes inefficient at some point.}, see Fig.~\ref{fig:iband}. Thus, to compute the time at which the plateau in Fig.~\ref{fig:ph_ablauf} is reached, one needs to identify first the ESP at which particle production is most efficient, compute the instability chart as in Fig.~\ref{fig:iband}, read off the lower boundary of the first band, and finally plug this $\varphi_0$ into (\ref{endtime}). 

A few comments are in order: particle production in this regime is dominated by a few ESPs that just happen to be at the right positions. If the average inter ESP distance in some random distribution is of order $x_1$ or smaller, we expect that a few of them are at ``right'' distances. For the concrete case studies in Fig.~\ref{fig:occ_number}, we find $x_1\sim 10^{-3}$, similar to the inter ESP distances at which the first in-fall is also efficient; as a consequence, we expect to see two distinct preheat matter classes: the first class consists of many different species with low occupation numbers, while the second one encompasses few preheat matter fields with high occupation numbers. 
 
In trapped inflation, ESP distributions need to be dense in order to cause any seizable back-reaction during inflation. As a consequence, the first in-fall is smoothly connected to the inflationary regime and we expect preheating in such models to be of the type examined in this article, including an efficient resonance phase. To decide if the inflatons decay completely after trapped inflation via resonances, one needs to perform lattice simulations; based on what we learned, we conclude that one does not need to consider all ESPs, but only those near  the critical distances $d_n$, which greatly reduces computational costs. We will not follow this line of research here: using a concrete implementation of trapped inflation in string theory is needed to include the necessary decay of preheat matter fields, which goes beyond the scope of the current paper.

Nevertheless, we wish to put the above predictions with regard to the resonance phase to the test by means of lattice simulations in a few simple scenarios in Sec.~\ref{PRwithbackreaction}, primarily to make sure that resonances persevere in the presence of back-reaction, but also to check some of the approximations that went into the analytic results.

\section{Lattice Simulations \label{PRwithbackreaction}}
\begin{figure}[t] 
   \centering
   \includegraphics[width=3in]{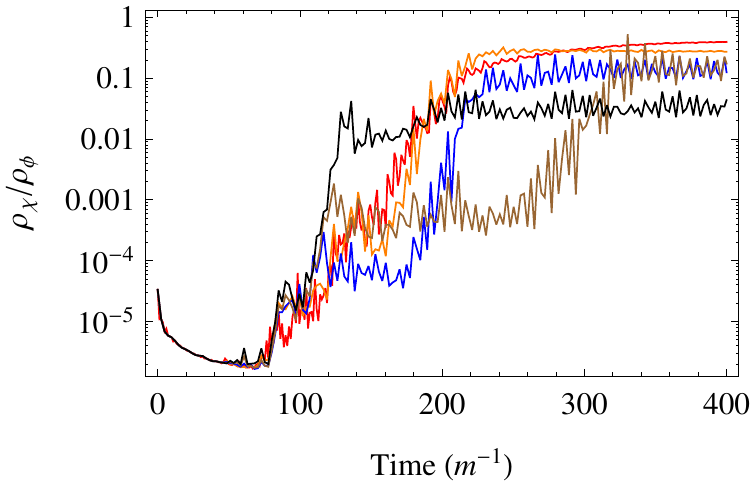} 
   \caption{The ratio of the energy in the $\chi$ field compared to the energy in $\phi$ for $\mathcal{N}=2$.  The color coding shows $x=0.00106\,M_p$ (red), $x=0.00141\,M_p$ (orange), $x=0.00177\,M_p$ (blue), $x=0.00212\,M_p$ (brown), and  $x=0.00247\,M_p$ (black).}
   \label{fig:1A}
\end{figure}
To solve the coupled system of the full field equations in an expanding background, we use the same modified LATTICEEASY code as in \cite{Battefeld:2009xw} (see Appendix A of \cite{Battefeld:2009xw} for details on the code implementation as well as initial conditions for inhomogeneous modes). As mentioned in Sec.~\ref{sec:matterfields}, we use a coupling of $g_i\equiv g=5\times 10^{-4}$ and start the matter field(s) in their vacuum state. Since we will not be studying any tachyonic preheating scenarios as were studied in  \cite{Battefeld:2009xw}, we neglect the  self interaction for the matter fields, $\lambda \chi_i^4$. With this tool at hand, we wish to discuss two setups:
\begin{enumerate}
\item  Particle production at a single ESP, with varying number of inflatons, initial conditions and masses, to test the predictions of Sec.~\ref{sec:resonancephase}, particularly the viability of a prolonged narrow resonance regime as an efficient preheating mechanism.
\item Particle production at several ESPs, Gaussianly distributed near the origin (see Sec.~\ref{sec:matterfields}), to test if preheating is indeed dominated by a few matter fields.  
\end{enumerate}

\subsection{Particle Production at a Single ESP \label{sec:latticeatsingleESP}}

The first question we would like to address is whether or not the prolonged period of narrow resonance during small amplitude inflaton oscillations is sufficient to transfer $\mathcal{O}(1)$ of the inflatons energy into a matter field. We choose the same setup as in Sec.~\ref{sec:resonancephase}, varying the ESP position by changing $x$ over an  interval that covers the spikes in Fig.~\ref{fig:occ_number}. To test the efficiency, we compute the ratio of the energy in the matter field $\rho_\chi$ over the total energy in the inflaton sector $\rho_\varphi$.

All simulations in this section use a $128^3$ lattice, each side has initial length $L_0 = 5 m^{-1}$, and we run the simulations until $t=400m^{-1}$ unless otherwise noted.

\begin{figure}[t] 
   \centering
   \includegraphics[width=3in]{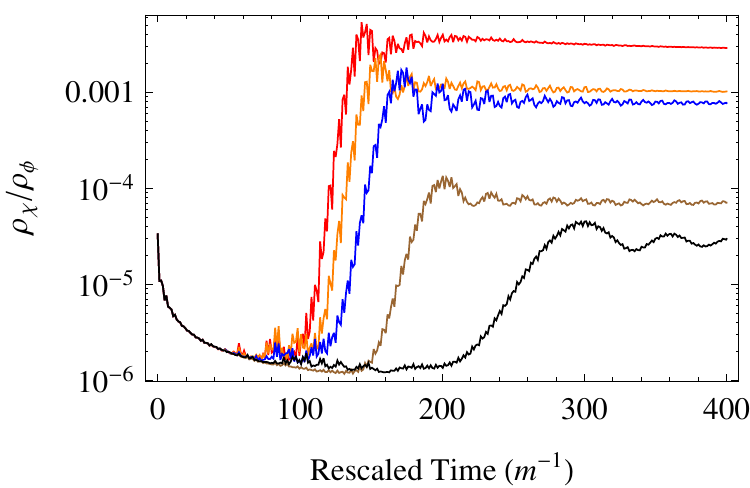} 
   \caption{The ratio of the energy in the $\chi$ field compared to the energy in $\phi$ for $\mathcal{N}=2$ and a single ESP at the origin, but a bare mass of the matter field, $m_{\chi {\rm eff}}^2 = g^2x^2/2$.  The color coding shows $x=0.00106\,M_p$ (red), $x=0.00141\,M_p$ (orange), $x=0.00177\,M_p$ (blue), $x=0.00212\,M_p$ (brown), and  $x=0.00247\,M_p$ (black).}
   \label{fig:1E}
\end{figure} 

In Fig.~\ref{fig:1A}, we plot the energy ratio for two inflatons with $m_2/m_1=3/2\,,\, m_1=m=10^{-6}\, M_P$, we use the initial field values defined by Eq.~(\ref{initvalues}), zero initial speed and a single ESP 
at $\vec{\varphi}_{\mbox{\tiny ESP}}=(x,x)/2$. 
We vary $x$ to cover the region around the first two peaks at $x\approx 0.00141\,M_P$ and $x\approx 0.00212\,M_P$ in Fig.~\ref{fig:occ_number} top left panel (see equation (\ref{eq:xkrit}) for the analytic value). We observe that as long as $x$ is in the vicinity of the peaks, resonances are indeed sufficient to raise the ratio $\rho_\chi/\rho_\varphi$ to order one, at which point the fields fragment and subsequently enter a turbulent regime. The exact value of $x$ has little effect onto preheating, as long as it is in the region of the peaks. This is expected, since altering $x$ changes primarily the maximal duration of the narrow resonance regime, without significantly changing the generalized Floquet index. Hence, as long as resonances last long enough to bring this ratio to order one, preheating operates similarly. However, once the ESP is too far away, i.e.~$x$ is to the right of the second peak in Fig.~\ref{fig:occ_number} top left panel, resonances shut of before preheating is completed (black line in Fig.~\ref{fig:1A}). It should be noted that an ESP at the origin can still lead to effective preheating for $\mathcal{N}=2$, see i.e.~Fig.~1A in \cite{Battefeld:2009xw} or Fig.~\ref{fig:psupp}, such that prolonged narrow resonances may not be crucial if only two inflatons are present.

It is instructive to compare the observed resonances to the ones with a single ESP at the origin, but allowing for a non-zero mass of the matter field $m_{\chi {\rm eff}}^2 \equiv g^2x^2/2$, see Fig.~\ref{fig:1E}; the initial stages of parametric resonance are also described by the analytic result in (\ref{eq:xkrit}), since we took the limit $\Phi_0\rightarrow 0$ and consequently ignored $q_1^{(i)}$ and $q_2^{(i)}$. As expected, we observe that particle production becomes less efficient as the mass of the $\chi$-field grows, and ceases to be important if
\begin{eqnarray}
m_{\chi {\rm eff}}^2 >\mbox{few}\times g^2 x_1^2/2= \mbox{few}\times \frac{m_1^2 \pi^2}{{\cal{N}}T^2}\,.
\end{eqnarray}
In our case resonances become more and more suppressed once the corresponding $x$ is to the right of the first peak in Fig.~\ref{fig:occ_number} (top left panel); for $x \gtrsim 0.0025\,M_p$ ($x$ to the right of the second peak) resonances are hardly present at all. Interestingly, preheating is efficient in Fig.~\ref{fig:1A} with the corresponding choice of ESP position: thus, to explain $\rho_\chi/\rho_\varphi\sim \mathcal{O}(1)$ in Fig.~\ref{fig:1A}, the inclusion of $q_1^{(i)}$, $q_2^{(i)}$ and/or backreaction are important, while our simple analytic treatment remains only qualitatively valid.  We conclude that lattice simulations are crucial: the presence of resonances in perturbative analytic calculations are necessary, but not sufficient, indicators to ascertain the efficiency of preheating.

\begin{figure}[t] 
   \centering
   \includegraphics[width=3in]{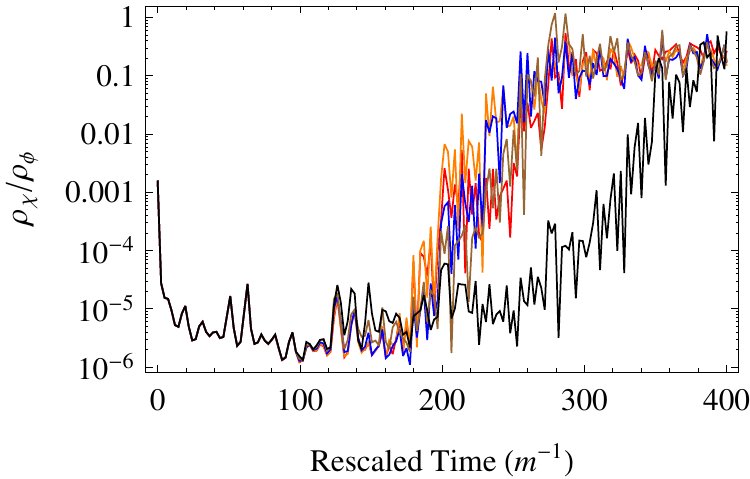} 
   \caption{The ratio of the energy in the $\chi$ field compared to the energy in $\phi$ for $\mathcal{N}=3$.  The color coding shows $x=0.00115\,M_p$ (red), $x=0.00130\,M_p$ (orange), $x=0.00144\,M_p$ (blue), $x=0.00159\,M_p$ (brown), and $x=0.00173\,M_p$ (black).}
   \label{fig:1B}
\end{figure}

Let us turn our attention to the $\mathcal{N}=3$ case, Fig.~\ref{fig:1B}, for which parametric resonance with an ESP at the origin is strongly suppressed due to dephasing, see Fig.~1A in \cite{Battefeld:2009xw} or Fig.~\ref{fig:psupp}.
Choosing $m_2/m_1=5/4, m_3/m_1=3/2, m_1=m=10^{-6}\, M_P$ and $\vec{\varphi}_{ini}=(1,1,1)x/2$, corresponding the top right panel in Fig.~\ref{fig:occ_number}, we observe again efficient, practically indistinguishable, preheating for $x$ near the peaks in Fig.~\ref{fig:occ_number}. 
Once the last peak is traversed ($x\geq 0.00173\,M_p$, the black line in Fig.~\ref{fig:1B}), narrow resonances weaken, shut off earlier and preheating becomes inefficient.

\begin{figure}[t] 
   \centering
   \includegraphics[width=3in]{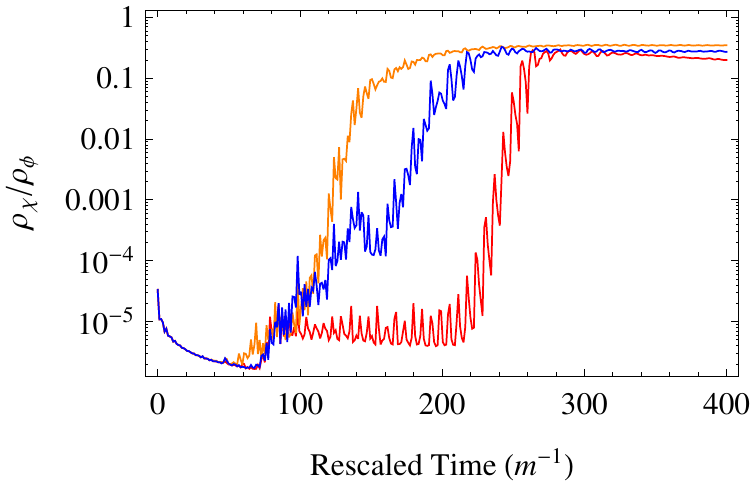} 
   \caption{The ratio of the energy in the $\chi$ field compared to the energy in $\phi$ for $\mathcal{N}=2$.  The color coding shows $\vec{x}=(0.00141,0)/\sqrt{2}\,M_p$ (red), $\vec{x}=(0,0.00141)/\sqrt{2}\,M_p$ (blue), $x=(0.00141,0.00141)/2\,M_p$ (orange).}
   \label{fig:1C}
\end{figure}

 So far we kept the ESP at $\vec{\varphi}_{\mbox{\tiny ESP}}=(x,\dots,x)/2$, just changing the value of $x$. Based on our analytic derivation leading to the  prediction of the peaks' positions in (\ref{eq:xkrit}), we do not expect a strong dependence on the angular position of the ESP if the oscillation amplitude and $x_i$ are small enough to ignore $q_2^{(i)}$ (defined in (\ref{eq:parameter})). We can confirm this expectation in Fig.~\ref{fig:1C}, where we keep the distance of the ESP to the origin fixed (at the first peak), but change its angular position. While some angular dependence is evident regarding the onset of resonance, the final amplitude of $\rho_\chi/\rho_\varphi$ remains the same. If the ESP is pushed out further, we expect  to see a stronger breaking of spherical symmetry, see Sec.~\ref{sec:latticesveralESP}. 
 
\begin{figure}[t] 
   \centering
   \includegraphics[width=3in]{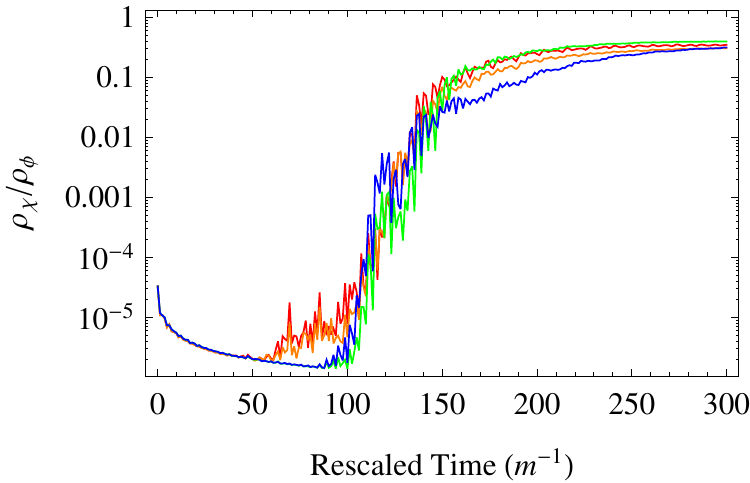} 
   \caption{The ratio of the energy in the $\chi$ field compared to the energy in $\phi$ for $\mathcal{N}=2$.  The color coding shows case (a) where $m_2/m_1=3/2$ for $x=0.00141\,M_p$ (red) and $x=0.001\,M_p$ (orange) and case (b) where $m_2/m_1 = \sqrt{3}$ for $x=0.00141\,M_p$ (green) and $x=0.001\,M_p$ (blue).  }
   \label{fig:1D}
\end{figure} 
 
The analytic prediction of the peaks position in (\ref{eq:xkrit}) was based on the assumption of commensurate masses. However, as we argued in Sec.~\ref{sec:instabilitybands} and saw i.e.~in the lower panels of Fig.~\ref{fig:occ_number}, the qualitative features of the resonance phase remain valid for incommensurate masses; this feature remains true on the lattice, see Fig.~\ref{fig:1D}, where efficient preheating via narrow resonance is still present.  

\subsection{Particle Production at Gaussianly Distributed ESPs \label{sec:latticesveralESP}}

\begin{figure}[ht] 
   \centering
   \includegraphics[width=3in]{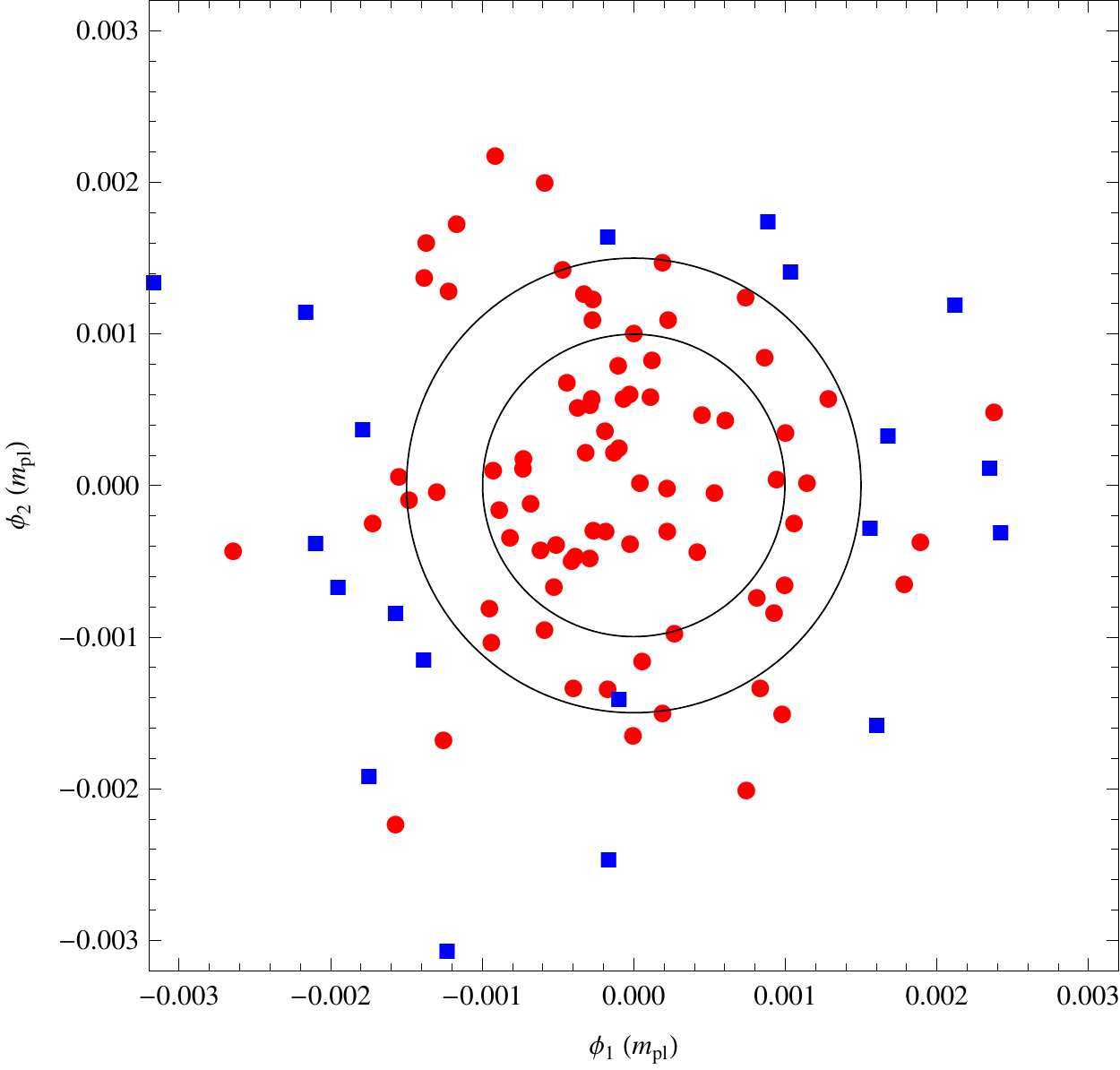} \includegraphics[width=3in]{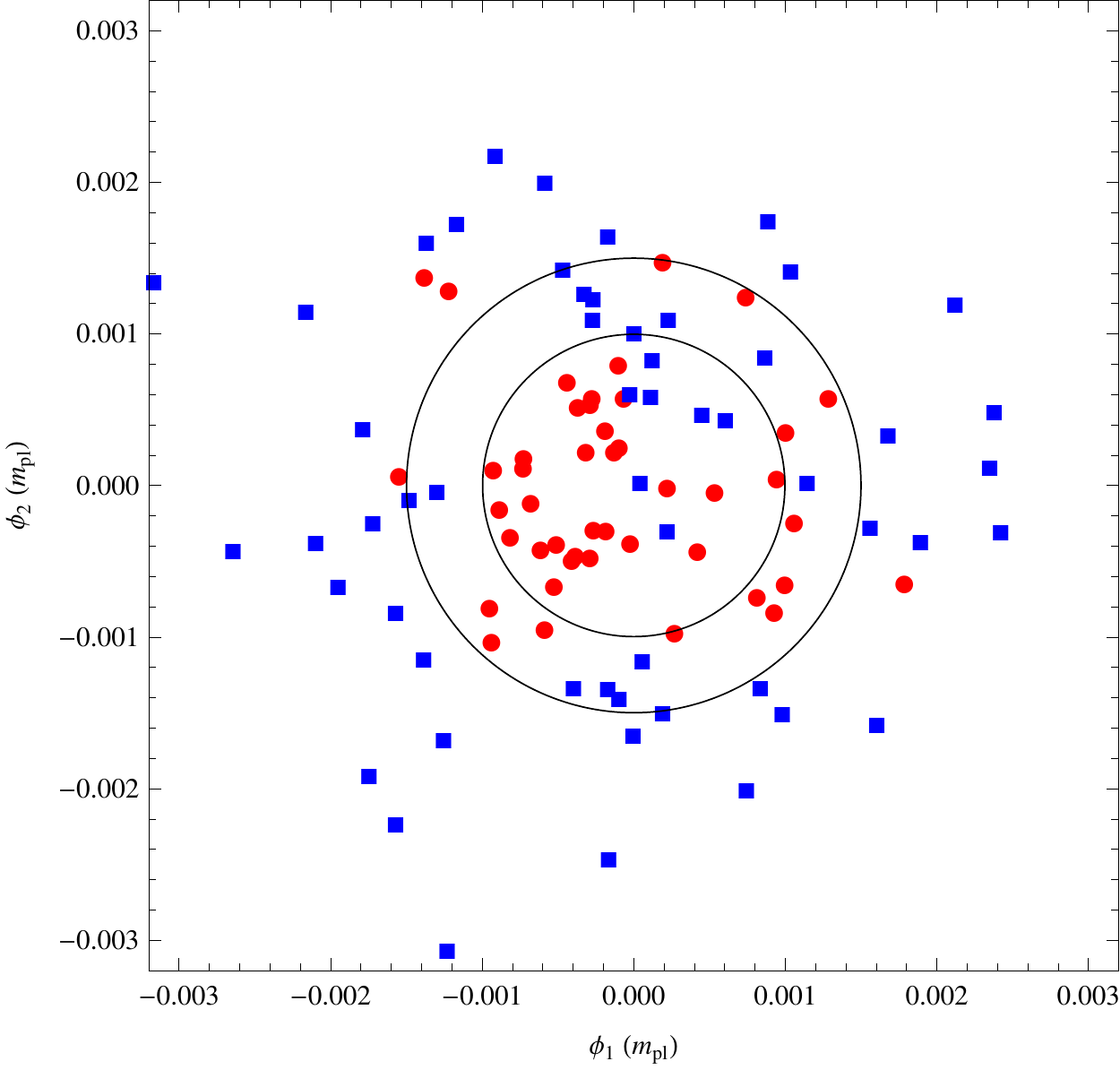} 
   \caption{The location of ESPS in ten different simulations containing ten ESPs each.  The left hand panel shows the most-significant ESPs that contribute to 60\% of the total energy of the $\chi$-fields (blue squares) in each simulation compared to the less energetic ESPs (red circles).  The right hand panel shows the most-significant ESPs that contribute to 90\% of the total energy of the $\chi$-fields in each simulation (blue squares) compared to the less energetic ESPs (red circles).  For comparison, the two circles represent the locus of points that correspond to $x=0.00141$ (inner) and $x=0.00212$ (outer).}
   \label{fig:twoDgaussian}
\end{figure} 

We now want to look at resonance when more than one ESP is present in the simulation.  First, we examine the case of $\mathcal{N}=2$ inflatons.  Due to time limitations on our simulations, we consider independent runs with 10-ESPs per simulation.  The locations of the ESP's are drawn from a Gaussian distribution with variance $\sigma^2=1.875\times 10^{-4}\,M_{\rm pl}^2$, so that there are a number of ESP's near the peaks of Fig.~\ref{fig:occ_number}. Again we chose the ratio of masses as $m_2/m_1=3/2\,,\, m_1=m=10^{-6}\, M_P$, with initial field values defined by Eq.~(\ref{initvalues}) and zero initial speed. 

 In Fig.~\ref{fig:twoDgaussian} the position of ESPs in ten of these simulations is plotted accumulative on the same panels. ESPs at which matter fields are predominantly produced are indicated by blue squares (making up $60\%$ and $90\%$ of the total energy in the left and right panels respectively). The two circles indicate the position of the first and second peak in Fig.~\ref{fig:occ_number} (top left). Firstly, we observe that ESPs further out dominate, particularly near the second circle, while ESPs near the origin are unimportant. This is in line with our analytic and numeric conclusions in previous sections. Secondly, we observe the breaking of spherical symmetry, particularly in the right panel: ESPs can be closer in the $\varphi_2$ direction (the heavier field) than in the $\varphi_1$ direction, while remaining important for preheating. This breaking of spherical symmetry is caused by $q_2^{(i)}$ in (\ref{eq:xkrit}), which we ignored in the analytic discussion. As we saw by comparing Fig.~\ref{fig:1A} to Fig.~\ref{fig:1E}, the presence of $q_1^{(i)}$ and $q_2^{(i)}$ is important to attain a large energy fraction. It is thus not surprising that ESPs dominating preheating are the ones that show a breaking of spherical symmetry.
 
\begin{figure}[ht] 
   \centering
   \includegraphics[width=3in]{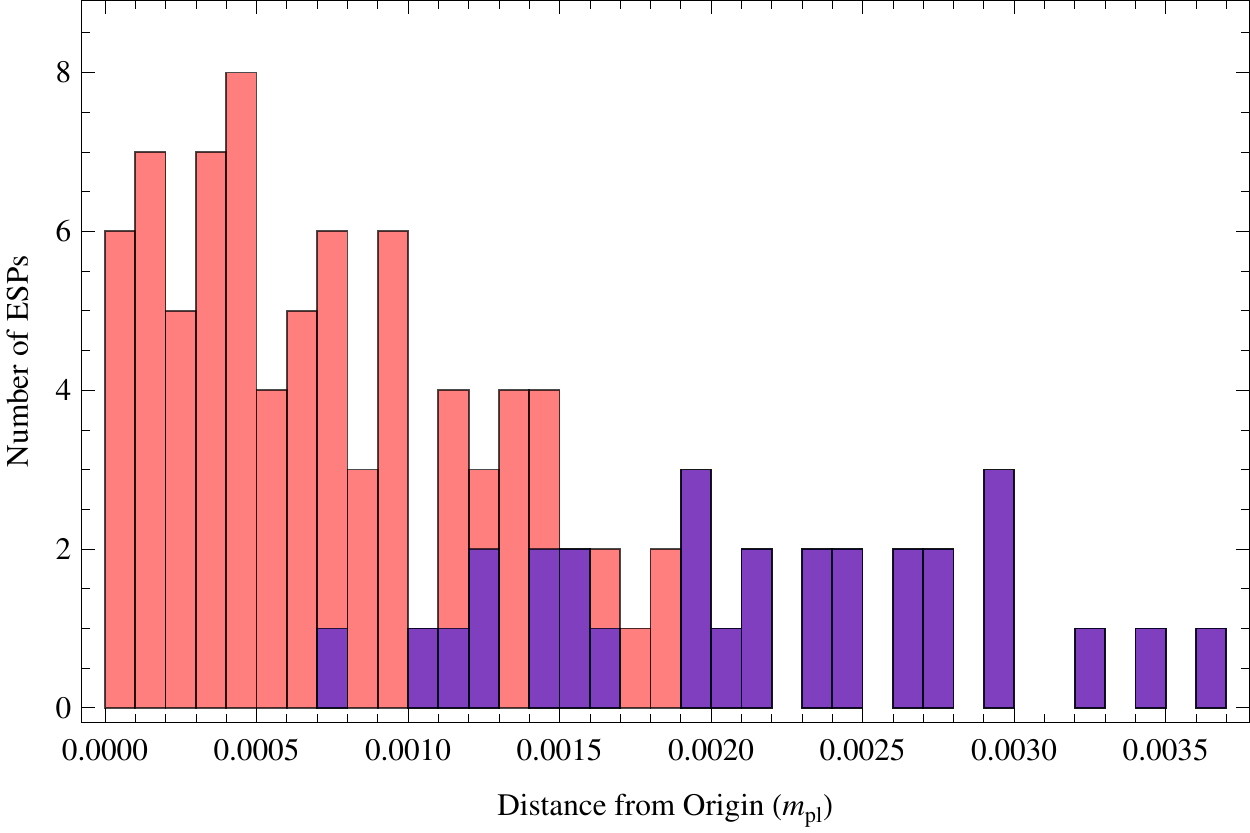} 
   \caption{A histogram showing the number of ESPS compared to the distance from the origin in field space for ten simulations containing ten ESPs each.  Each ESP is located at $\varphi_{\rm ESP} = (x,x,x)/2$, where $x$ is drawn from a Gaussian distribution. Three most-significant ESPs that contribute to the energy of any given simulation (blue) are compared to the total number of ESPS at that distance (red).}
   \label{fig:threeDgaussian}
\end{figure}  
 
 A similar breaking of spherical symmetry should be present for $\mathcal{N} \geq 3$. However, even the $\mathcal{N}=3$ case is quite time consuming to treat on a lattice, since it requires a great number of ESPs to yield an adequate filling of field space. Hence, we only investigate a Gaussian distribution of ESPs along a prescribed axis (again accumulative, with $10$ ESPs per run): in  Fig.~\ref{fig:threeDgaussian} we chose $\varphi_{\rm ESP} = (x,x,x)/2$, with $x$ Gaussianly distributed. The histogram shows the ESPs that contribute the most to preheating ($60\%$, violet), while the less important ones are coloured pink. We observe again that preheating is not dominated by ESPs at the origin, but at distances bigger than the fist peak in Fig.~\ref{fig:occ_number} (top right), in line with the discussions in previous sections. It is tempting to identify peaks in Fig.~\ref{fig:threeDgaussian} with the ones in Fig.~\ref{fig:occ_number}, but the total number of ESPs in our simulations, and thus the statistical significance, is too low for such a detailed comparison. 

\section{Conclusion}

Motivated by the presence of extra species points (ESP) in many models of inflation in string theory (i.e.~trapped inflation), we investigated preheating of several preheat matter fields that become light near, but not at, the vacuum expectation value (VEV) of the inflatons. In the presence of several inflaton fields, dephasing tends to suppress resonances, which can however be compensated by two new, hitherto overlooked, effects: first, if ESP distributions are dense, particles can be produced at many ESPs via broad resonance once slow roll ends, since the inflaton speed is not slow roll suppressed any more and the adiabaticity condition is violated for many preheat fields. We compute analytically the resulting occupation number, paying particular attention to the dependence of preheating on the number of inflatons and the density of ESPs. Preheating during this first in-fall, albeit potentially sufficient to convert the majority of the inflatons energy to preheat matter fields, is however never complete. 

After an intermediate phase of large amplitude oscillations during which additional particle production may occur, a prolonged phase of narrow resonance generically leads to the complete decay of inflatons, if the location of an ESP is at one of several ``right'' distances from the VEV. We compute these locations analytically, by means of a simple stability analysis in a static universe, providing a thorough, yet intuitive understanding of the resonance regime. We test our analytic predictions numerically by direct integration of the equations of motion (no backreaction) and complementary lattice simulations and find good agreement with the analytics. 

Our study shows that the understanding of preheating after multi-field inflation is far from complete: we found new, efficient mechanisms of preheating that can be missed easily if the model is over-simplified. For example, while the presence of ESPs in field space is often incorporated during inflation, they are usually ignored or lumped together into a single, effective preheat matter field that becomes light at the VEV of the inflatons, which completely eliminates the preheating mechanism discussed in this paper. Further, combining inflatons into an effective single field, albeit extremely useful and predictive during inflation, alters the nature of preheating qualitatively and is therefore never a prudent approach towards the end of inflation.

One of the main, remaining challenges for inflationary model building in string theory is the identification of all relevant degrees of freedom that the inflaton(s) couple to once inflation terminates. Our work is a step in this direction by incorporating the presence of ESPs, describing fields that are present in almost all moduli spaces of string theory. Consequently, the next step towards a theory of reheating after inflation in string theory is the incorporation of subsequent decay products and the computation of thermalization, which we leave for future studies.

\acknowledgments
We would like to thank D.~Battefeld and D.~Figueroa for useful discussions and comments.T.B. is thankful for hospitality at the Laboratoire Astroparticule et Cosmologie (Paris), where this work was instigated, as well the Aspen Center for Physics and NSF Grant $\# 1066293$ for hospitality during the final writing stages of this paper.  JTG is supported by the National Science Foundation, PHY-1068080, and a Cottrell College Science Award from the Research Corporation for Science Advancement.

\end{document}

%% file: mathieu.tex
\begingroup
  \makeatletter
  \providecommand\color[2][]{%
    \GenericError{(gnuplot) \space\space\space\@spaces}{%
      Package color not loaded in conjunction with
      terminal option `colourtext'%
    }{See the gnuplot documentation for explanation.%
    }{Either use 'blacktext' in gnuplot or load the package
      color.sty in LaTeX.}%
    \renewcommand\color[2][]{}%
  }%
  \providecommand\includegraphics[2][]{%
    \GenericError{(gnuplot) \space\space\space\@spaces}{%
      Package graphicx or graphics not loaded%
    }{See the gnuplot documentation for explanation.%
    }{The gnuplot epslatex terminal needs graphicx.sty or graphics.sty.}%
    \renewcommand\includegraphics[2][]{}%
  }%
  \providecommand\rotatebox[2]{#2}%
  \@ifundefined{ifGPcolor}{%
    \newif\ifGPcolor
    \GPcolortrue
  }{}%
  \@ifundefined{ifGPblacktext}{%
    \newif\ifGPblacktext
    \GPblacktexttrue
  }{}%
  \let\gplgaddtomacro\g@addto@macro
  \gdef\gplbacktext{}%
  \gdef\gplfronttext{}%
  \makeatother
  \ifGPblacktext
    \def\colorrgb#1{}%
    \def\colorgray#1{}%
  \else
    \ifGPcolor
      \def\colorrgb#1{\color[rgb]{#1}}%
      \def\colorgray#1{\color[gray]{#1}}%
      \expandafter\def\csname LTw\endcsname{\color{white}}%
      \expandafter\def\csname LTb\endcsname{\color{black}}%
      \expandafter\def\csname LTa\endcsname{\color{black}}%
      \expandafter\def\csname LT0\endcsname{\color[rgb]{1,0,0}}%
      \expandafter\def\csname LT1\endcsname{\color[rgb]{0,1,0}}%
      \expandafter\def\csname LT2\endcsname{\color[rgb]{0,0,1}}%
      \expandafter\def\csname LT3\endcsname{\color[rgb]{1,0,1}}%
      \expandafter\def\csname LT4\endcsname{\color[rgb]{0,1,1}}%
      \expandafter\def\csname LT5\endcsname{\color[rgb]{1,1,0}}%
      \expandafter\def\csname LT6\endcsname{\color[rgb]{0,0,0}}%
      \expandafter\def\csname LT7\endcsname{\color[rgb]{1,0.3,0}}%
      \expandafter\def\csname LT8\endcsname{\color[rgb]{0.5,0.5,0.5}}%
    \else
      \def\colorrgb#1{\color{black}}%
      \def\colorgray#1{\color[gray]{#1}}%
      \expandafter\def\csname LTw\endcsname{\color{white}}%
      \expandafter\def\csname LTb\endcsname{\color{black}}%
      \expandafter\def\csname LTa\endcsname{\color{black}}%
      \expandafter\def\csname LT0\endcsname{\color{black}}%
      \expandafter\def\csname LT1\endcsname{\color{black}}%
      \expandafter\def\csname LT2\endcsname{\color{black}}%
      \expandafter\def\csname LT3\endcsname{\color{black}}%
      \expandafter\def\csname LT4\endcsname{\color{black}}%
      \expandafter\def\csname LT5\endcsname{\color{black}}%
      \expandafter\def\csname LT6\endcsname{\color{black}}%
      \expandafter\def\csname LT7\endcsname{\color{black}}%
      \expandafter\def\csname LT8\endcsname{\color{black}}%
    \fi
  \fi
  \setlength{\unitlength}{0.0500bp}%
  \begin{picture}(7936.00,3400.00)%
    \gplgaddtomacro\gplbacktext{%
      \csname LTb\endcsname%
      \put(814,704){\makebox(0,0)[r]{\strut{}-2}}%
      \put(814,1051){\makebox(0,0)[r]{\strut{}-1}}%
      \put(814,1399){\makebox(0,0)[r]{\strut{} 0}}%
      \put(814,1746){\makebox(0,0)[r]{\strut{} 1}}%
      \put(814,2093){\makebox(0,0)[r]{\strut{} 2}}%
      \put(814,2440){\makebox(0,0)[r]{\strut{} 3}}%
      \put(814,2788){\makebox(0,0)[r]{\strut{} 4}}%
      \put(814,3135){\makebox(0,0)[r]{\strut{} 5}}%
      \put(946,484){\makebox(0,0){\strut{} 0}}%
      \put(1834,484){\makebox(0,0){\strut{} 0.2}}%
      \put(2722,484){\makebox(0,0){\strut{} 0.4}}%
      \put(3610,484){\makebox(0,0){\strut{} 0.6}}%
      \put(4497,484){\makebox(0,0){\strut{} 0.8}}%
      \put(5385,484){\makebox(0,0){\strut{} 1}}%
      \put(6273,484){\makebox(0,0){\strut{} 1.2}}%
      \put(7161,484){\makebox(0,0){\strut{} 1.4}}%
      \put(308,1919){\rotatebox{-270}{\makebox(0,0){\strut{}$A_k$}}}%
      \put(4275,154){\makebox(0,0){\strut{}$q$}}%
    }%
    \gplgaddtomacro\gplfronttext{%
    }%
    \gplbacktext
    \put(0,0){\includegraphics{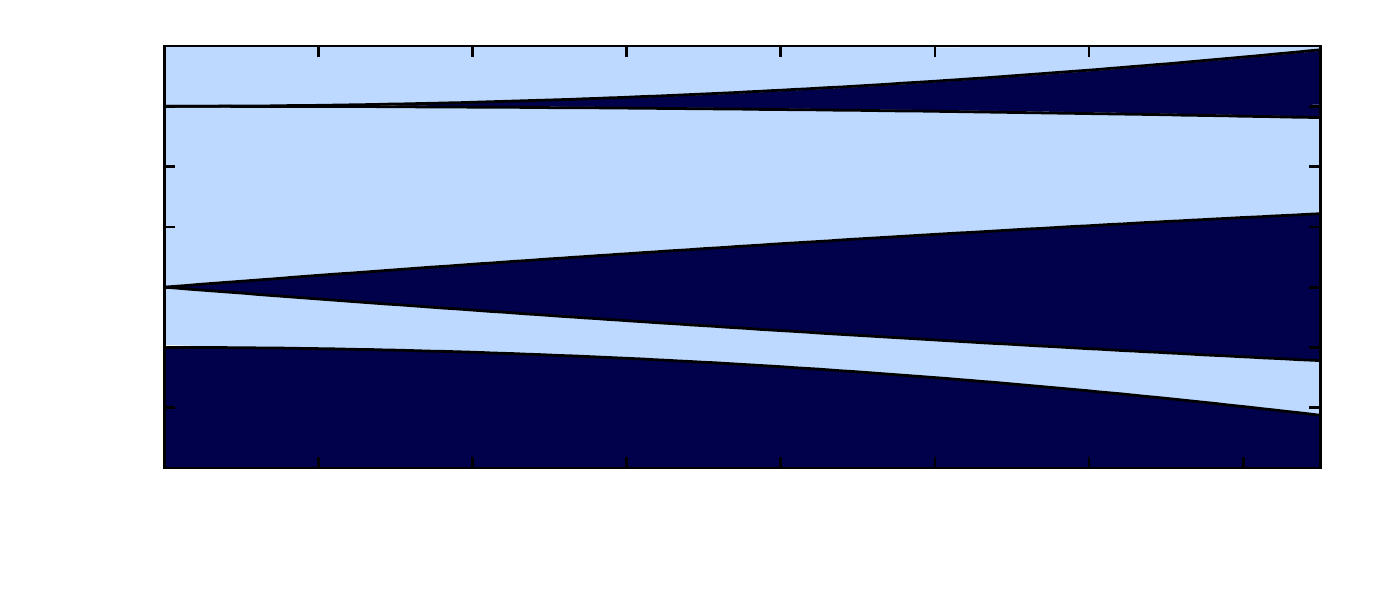}}%
    \gplfronttext
  \end{picture}%
\endgroup

%% file: chi.tex
\begingroup
  \makeatletter
  \providecommand\color[2][]{%
    \GenericError{(gnuplot) \space\space\space\@spaces}{%
      Package color not loaded in conjunction with
      terminal option `colourtext'%
    }{See the gnuplot documentation for explanation.%
    }{Either use 'blacktext' in gnuplot or load the package
      color.sty in LaTeX.}%
    \renewcommand\color[2][]{}%
  }%
  \providecommand\includegraphics[2][]{%
    \GenericError{(gnuplot) \space\space\space\@spaces}{%
      Package graphicx or graphics not loaded%
    }{See the gnuplot documentation for explanation.%
    }{The gnuplot epslatex terminal needs graphicx.sty or graphics.sty.}%
    \renewcommand\includegraphics[2][]{}%
  }%
  \providecommand\rotatebox[2]{#2}%
  \@ifundefined{ifGPcolor}{%
    \newif\ifGPcolor
    \GPcolortrue
  }{}%
  \@ifundefined{ifGPblacktext}{%
    \newif\ifGPblacktext
    \GPblacktexttrue
  }{}%
  \let\gplgaddtomacro\g@addto@macro
  \gdef\gplbacktext{}%
  \gdef\gplfronttext{}%
  \makeatother
  \ifGPblacktext
    \def\colorrgb#1{}%
    \def\colorgray#1{}%
  \else
    \ifGPcolor
      \def\colorrgb#1{\color[rgb]{#1}}%
      \def\colorgray#1{\color[gray]{#1}}%
      \expandafter\def\csname LTw\endcsname{\color{white}}%
      \expandafter\def\csname LTb\endcsname{\color{black}}%
      \expandafter\def\csname LTa\endcsname{\color{black}}%
      \expandafter\def\csname LT0\endcsname{\color[rgb]{1,0,0}}%
      \expandafter\def\csname LT1\endcsname{\color[rgb]{0,1,0}}%
      \expandafter\def\csname LT2\endcsname{\color[rgb]{0,0,1}}%
      \expandafter\def\csname LT3\endcsname{\color[rgb]{1,0,1}}%
      \expandafter\def\csname LT4\endcsname{\color[rgb]{0,1,1}}%
      \expandafter\def\csname LT5\endcsname{\color[rgb]{1,1,0}}%
      \expandafter\def\csname LT6\endcsname{\color[rgb]{0,0,0}}%
      \expandafter\def\csname LT7\endcsname{\color[rgb]{1,0.3,0}}%
      \expandafter\def\csname LT8\endcsname{\color[rgb]{0.5,0.5,0.5}}%
    \else
      \def\colorrgb#1{\color{black}}%
      \def\colorgray#1{\color[gray]{#1}}%
      \expandafter\def\csname LTw\endcsname{\color{white}}%
      \expandafter\def\csname LTb\endcsname{\color{black}}%
      \expandafter\def\csname LTa\endcsname{\color{black}}%
      \expandafter\def\csname LT0\endcsname{\color{black}}%
      \expandafter\def\csname LT1\endcsname{\color{black}}%
      \expandafter\def\csname LT2\endcsname{\color{black}}%
      \expandafter\def\csname LT3\endcsname{\color{black}}%
      \expandafter\def\csname LT4\endcsname{\color{black}}%
      \expandafter\def\csname LT5\endcsname{\color{black}}%
      \expandafter\def\csname LT6\endcsname{\color{black}}%
      \expandafter\def\csname LT7\endcsname{\color{black}}%
      \expandafter\def\csname LT8\endcsname{\color{black}}%
    \fi
  \fi
  \setlength{\unitlength}{0.0500bp}%
  \begin{picture}(7936.00,3400.00)%
    \gplgaddtomacro\gplbacktext{%
      \csname LTb\endcsname%
      \put(814,704){\makebox(0,0)[r]{\strut{}-2}}%
      \put(814,1109){\makebox(0,0)[r]{\strut{} 0}}%
      \put(814,1514){\makebox(0,0)[r]{\strut{} 2}}%
      \put(814,1920){\makebox(0,0)[r]{\strut{} 4}}%
      \put(814,2325){\makebox(0,0)[r]{\strut{} 6}}%
      \put(814,2730){\makebox(0,0)[r]{\strut{} 8}}%
      \put(814,3135){\makebox(0,0)[r]{\strut{} 10}}%
      \put(946,484){\makebox(0,0){\strut{} 0}}%
      \put(1274,484){\makebox(0,0){\strut{} 1}}%
      \put(1602,484){\makebox(0,0){\strut{} 2}}%
      \put(1930,484){\makebox(0,0){\strut{} 3}}%
      \put(2259,484){\makebox(0,0){\strut{} 4}}%
      \put(2587,484){\makebox(0,0){\strut{} 5}}%
      \put(2915,484){\makebox(0,0){\strut{} 6}}%
      \put(3243,484){\makebox(0,0){\strut{} 7}}%
      \put(3571,484){\makebox(0,0){\strut{} 8}}%
      \put(176,1919){\rotatebox{-270}{\makebox(0,0){\strut{}$\ln{(n_k)}$}}}%
      \put(2258,154){\makebox(0,0){\strut{}$t$ [$N$]}}%
    }%
    \gplgaddtomacro\gplfronttext{%
    }%
    \gplgaddtomacro\gplbacktext{%
      \csname LTb\endcsname%
      \put(4782,704){\makebox(0,0)[r]{\strut{}-2}}%
      \put(4782,1008){\makebox(0,0)[r]{\strut{} 0}}%
      \put(4782,1312){\makebox(0,0)[r]{\strut{} 2}}%
      \put(4782,1616){\makebox(0,0)[r]{\strut{} 4}}%
      \put(4782,1920){\makebox(0,0)[r]{\strut{} 6}}%
      \put(4782,2223){\makebox(0,0)[r]{\strut{} 8}}%
      \put(4782,2527){\makebox(0,0)[r]{\strut{} 10}}%
      \put(4782,2831){\makebox(0,0)[r]{\strut{} 12}}%
      \put(4782,3135){\makebox(0,0)[r]{\strut{} 14}}%
      \put(4914,484){\makebox(0,0){\strut{} 0}}%
      \put(5439,484){\makebox(0,0){\strut{} 0.5}}%
      \put(5964,484){\makebox(0,0){\strut{} 1}}%
      \put(6489,484){\makebox(0,0){\strut{} 1.5}}%
      \put(7014,484){\makebox(0,0){\strut{} 2}}%
      \put(7539,484){\makebox(0,0){\strut{} 2.5}}%
      \put(4144,1919){\rotatebox{-270}{\makebox(0,0){\strut{}$\ln{(n_k)}$}}}%
      \put(6226,154){\makebox(0,0){\strut{}$t$ [$N$]}}%
    }%
    \gplgaddtomacro\gplfronttext{%
    }%
    \gplbacktext
    \put(0,0){\includegraphics{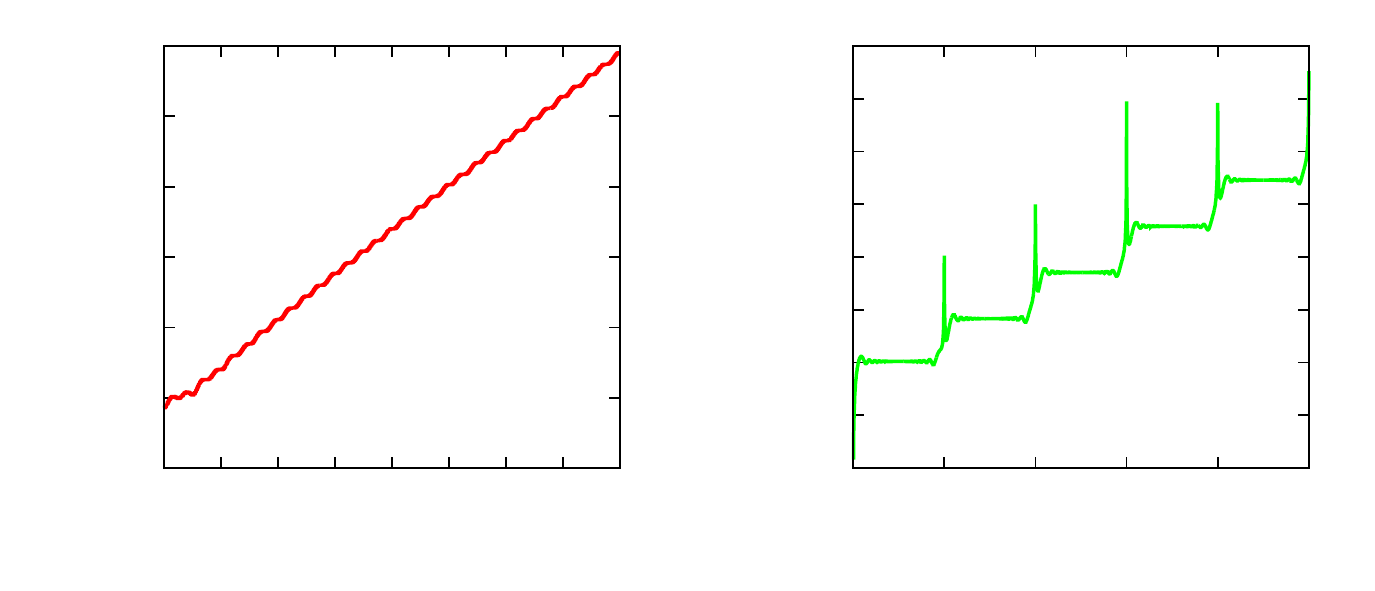}}%
    \gplfronttext
  \end{picture}%
\endgroup

%% file: chi_exp.tex
\begingroup
  \makeatletter
  \providecommand\color[2][]{%
    \GenericError{(gnuplot) \space\space\space\@spaces}{%
      Package color not loaded in conjunction with
      terminal option `colourtext'%
    }{See the gnuplot documentation for explanation.%
    }{Either use 'blacktext' in gnuplot or load the package
      color.sty in LaTeX.}%
    \renewcommand\color[2][]{}%
  }%
  \providecommand\includegraphics[2][]{%
    \GenericError{(gnuplot) \space\space\space\@spaces}{%
      Package graphicx or graphics not loaded%
    }{See the gnuplot documentation for explanation.%
    }{The gnuplot epslatex terminal needs graphicx.sty or graphics.sty.}%
    \renewcommand\includegraphics[2][]{}%
  }%
  \providecommand\rotatebox[2]{#2}%
  \@ifundefined{ifGPcolor}{%
    \newif\ifGPcolor
    \GPcolortrue
  }{}%
  \@ifundefined{ifGPblacktext}{%
    \newif\ifGPblacktext
    \GPblacktexttrue
  }{}%
  \let\gplgaddtomacro\g@addto@macro
  \gdef\gplbacktext{}%
  \gdef\gplfronttext{}%
  \makeatother
  \ifGPblacktext
    \def\colorrgb#1{}%
    \def\colorgray#1{}%
  \else
    \ifGPcolor
      \def\colorrgb#1{\color[rgb]{#1}}%
      \def\colorgray#1{\color[gray]{#1}}%
      \expandafter\def\csname LTw\endcsname{\color{white}}%
      \expandafter\def\csname LTb\endcsname{\color{black}}%
      \expandafter\def\csname LTa\endcsname{\color{black}}%
      \expandafter\def\csname LT0\endcsname{\color[rgb]{1,0,0}}%
      \expandafter\def\csname LT1\endcsname{\color[rgb]{0,1,0}}%
      \expandafter\def\csname LT2\endcsname{\color[rgb]{0,0,1}}%
      \expandafter\def\csname LT3\endcsname{\color[rgb]{1,0,1}}%
      \expandafter\def\csname LT4\endcsname{\color[rgb]{0,1,1}}%
      \expandafter\def\csname LT5\endcsname{\color[rgb]{1,1,0}}%
      \expandafter\def\csname LT6\endcsname{\color[rgb]{0,0,0}}%
      \expandafter\def\csname LT7\endcsname{\color[rgb]{1,0.3,0}}%
      \expandafter\def\csname LT8\endcsname{\color[rgb]{0.5,0.5,0.5}}%
    \else
      \def\colorrgb#1{\color{black}}%
      \def\colorgray#1{\color[gray]{#1}}%
      \expandafter\def\csname LTw\endcsname{\color{white}}%
      \expandafter\def\csname LTb\endcsname{\color{black}}%
      \expandafter\def\csname LTa\endcsname{\color{black}}%
      \expandafter\def\csname LT0\endcsname{\color{black}}%
      \expandafter\def\csname LT1\endcsname{\color{black}}%
      \expandafter\def\csname LT2\endcsname{\color{black}}%
      \expandafter\def\csname LT3\endcsname{\color{black}}%
      \expandafter\def\csname LT4\endcsname{\color{black}}%
      \expandafter\def\csname LT5\endcsname{\color{black}}%
      \expandafter\def\csname LT6\endcsname{\color{black}}%
      \expandafter\def\csname LT7\endcsname{\color{black}}%
      \expandafter\def\csname LT8\endcsname{\color{black}}%
    \fi
  \fi
  \setlength{\unitlength}{0.0500bp}%
  \begin{picture}(7936.00,6802.00)%
    \gplgaddtomacro\gplbacktext{%
      \csname LTb\endcsname%
      \put(682,4105){\makebox(0,0)[r]{\strut{} 0}}%
      \put(682,4409){\makebox(0,0)[r]{\strut{} 1}}%
      \put(682,4713){\makebox(0,0)[r]{\strut{} 2}}%
      \put(682,5017){\makebox(0,0)[r]{\strut{} 3}}%
      \put(682,5321){\makebox(0,0)[r]{\strut{} 4}}%
      \put(682,5625){\makebox(0,0)[r]{\strut{} 5}}%
      \put(682,5929){\makebox(0,0)[r]{\strut{} 6}}%
      \put(682,6233){\makebox(0,0)[r]{\strut{} 7}}%
      \put(682,6537){\makebox(0,0)[r]{\strut{} 8}}%
      \put(814,3885){\makebox(0,0){\strut{} 0}}%
      \put(1159,3885){\makebox(0,0){\strut{} 0.5}}%
      \put(1503,3885){\makebox(0,0){\strut{} 1}}%
      \put(1848,3885){\makebox(0,0){\strut{} 1.5}}%
      \put(2193,3885){\makebox(0,0){\strut{} 2}}%
      \put(2537,3885){\makebox(0,0){\strut{} 2.5}}%
      \put(2882,3885){\makebox(0,0){\strut{} 3}}%
      \put(3226,3885){\makebox(0,0){\strut{} 3.5}}%
      \put(3571,3885){\makebox(0,0){\strut{} 4}}%
      \put(176,5321){\rotatebox{-270}{\makebox(0,0){\strut{}$\ln{(n_k+1)}$}}}%
      \put(2192,3555){\makebox(0,0){\strut{}$t$ [$N$]}}%
    }%
    \gplgaddtomacro\gplfronttext{%
    }%
    \gplgaddtomacro\gplbacktext{%
      \csname LTb\endcsname%
      \put(4782,4105){\makebox(0,0)[r]{\strut{} 0}}%
      \put(4782,4375){\makebox(0,0)[r]{\strut{} 2}}%
      \put(4782,4645){\makebox(0,0)[r]{\strut{} 4}}%
      \put(4782,4916){\makebox(0,0)[r]{\strut{} 6}}%
      \put(4782,5186){\makebox(0,0)[r]{\strut{} 8}}%
      \put(4782,5456){\makebox(0,0)[r]{\strut{} 10}}%
      \put(4782,5726){\makebox(0,0)[r]{\strut{} 12}}%
      \put(4782,5997){\makebox(0,0)[r]{\strut{} 14}}%
      \put(4782,6267){\makebox(0,0)[r]{\strut{} 16}}%
      \put(4782,6537){\makebox(0,0)[r]{\strut{} 18}}%
      \put(4914,3885){\makebox(0,0){\strut{} 0}}%
      \put(5242,3885){\makebox(0,0){\strut{} 2}}%
      \put(5570,3885){\makebox(0,0){\strut{} 4}}%
      \put(5898,3885){\makebox(0,0){\strut{} 6}}%
      \put(6227,3885){\makebox(0,0){\strut{} 8}}%
      \put(6555,3885){\makebox(0,0){\strut{} 10}}%
      \put(6883,3885){\makebox(0,0){\strut{} 12}}%
      \put(7211,3885){\makebox(0,0){\strut{} 14}}%
      \put(7539,3885){\makebox(0,0){\strut{} 16}}%
      \put(4144,5321){\rotatebox{-270}{\makebox(0,0){\strut{}$\ln{(n_k+1)}$}}}%
      \put(6226,3555){\makebox(0,0){\strut{}$t$ [$N$]}}%
    }%
    \gplgaddtomacro\gplfronttext{%
    }%
    \gplgaddtomacro\gplbacktext{%
      \csname LTb\endcsname%
      \put(814,704){\makebox(0,0)[r]{\strut{} 0}}%
      \put(814,1008){\makebox(0,0)[r]{\strut{} 5}}%
      \put(814,1312){\makebox(0,0)[r]{\strut{} 10}}%
      \put(814,1616){\makebox(0,0)[r]{\strut{} 15}}%
      \put(814,1921){\makebox(0,0)[r]{\strut{} 20}}%
      \put(814,2225){\makebox(0,0)[r]{\strut{} 25}}%
      \put(814,2529){\makebox(0,0)[r]{\strut{} 30}}%
      \put(814,2833){\makebox(0,0)[r]{\strut{} 35}}%
      \put(814,3137){\makebox(0,0)[r]{\strut{} 40}}%
      \put(946,484){\makebox(0,0){\strut{} 0}}%
      \put(2045,484){\makebox(0,0){\strut{} 10}}%
      \put(3144,484){\makebox(0,0){\strut{} 20}}%
      \put(4243,484){\makebox(0,0){\strut{} 30}}%
      \put(5341,484){\makebox(0,0){\strut{} 40}}%
      \put(6440,484){\makebox(0,0){\strut{} 50}}%
      \put(7539,484){\makebox(0,0){\strut{} 60}}%
      \put(176,1920){\rotatebox{-270}{\makebox(0,0){\strut{}$\ln{(n_k+1)}$}}}%
      \put(4242,154){\makebox(0,0){\strut{}$t$ [$N$]}}%
    }%
    \gplgaddtomacro\gplfronttext{%
    }%
    \gplbacktext
    \put(0,0){\includegraphics{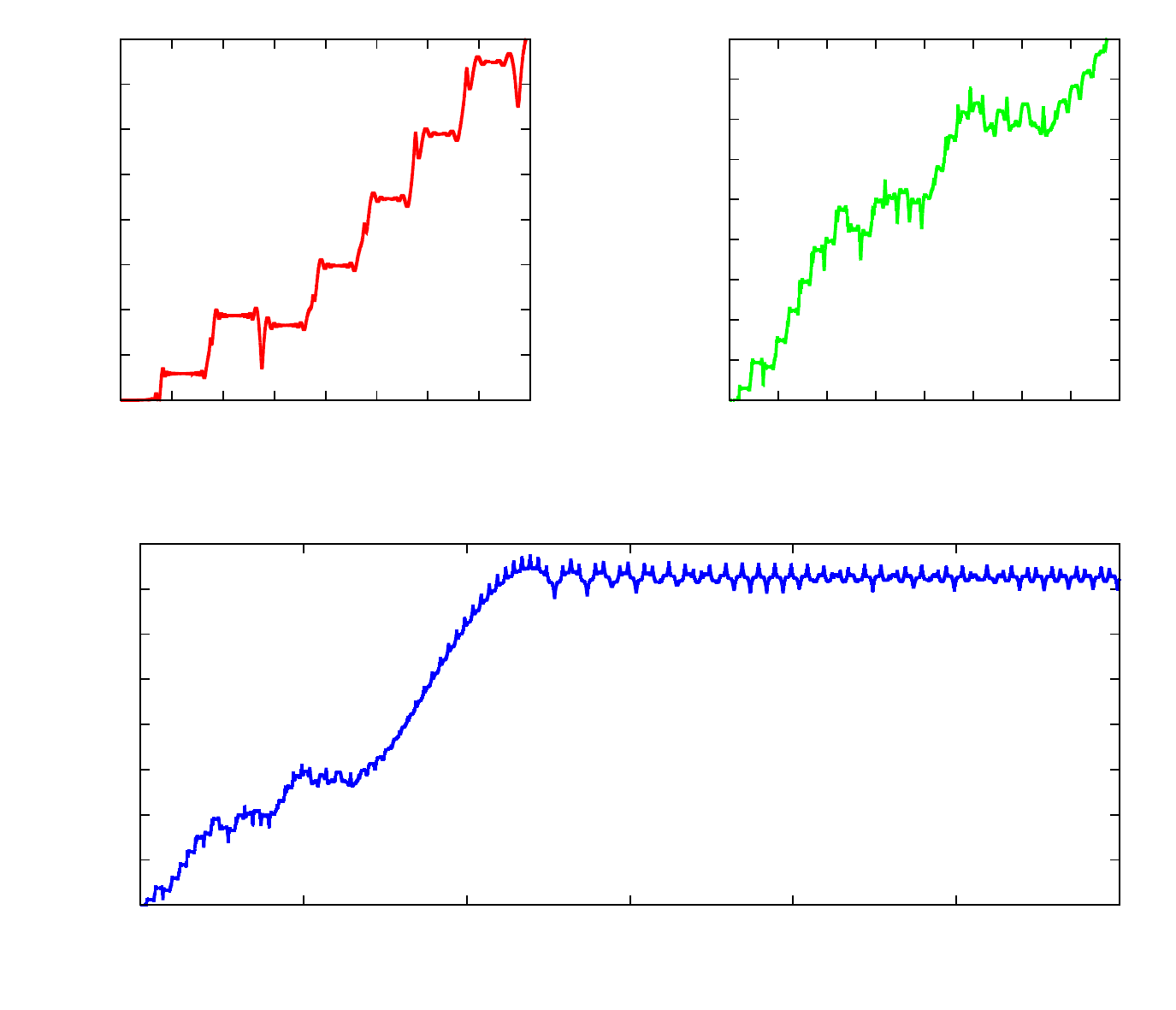}}%
    \gplfronttext
  \end{picture}%
\endgroup

%% file: psupp.tex
\begingroup
  \makeatletter
  \providecommand\color[2][]{%
    \GenericError{(gnuplot) \space\space\space\@spaces}{%
      Package color not loaded in conjunction with
      terminal option `colourtext'%
    }{See the gnuplot documentation for explanation.%
    }{Either use 'blacktext' in gnuplot or load the package
      color.sty in LaTeX.}%
    \renewcommand\color[2][]{}%
  }%
  \providecommand\includegraphics[2][]{%
    \GenericError{(gnuplot) \space\space\space\@spaces}{%
      Package graphicx or graphics not loaded%
    }{See the gnuplot documentation for explanation.%
    }{The gnuplot epslatex terminal needs graphicx.sty or graphics.sty.}%
    \renewcommand\includegraphics[2][]{}%
  }%
  \providecommand\rotatebox[2]{#2}%
  \@ifundefined{ifGPcolor}{%
    \newif\ifGPcolor
    \GPcolortrue
  }{}%
  \@ifundefined{ifGPblacktext}{%
    \newif\ifGPblacktext
    \GPblacktexttrue
  }{}%
  \let\gplgaddtomacro\g@addto@macro
  \gdef\gplbacktext{}%
  \gdef\gplfronttext{}%
  \makeatother
  \ifGPblacktext
    \def\colorrgb#1{}%
    \def\colorgray#1{}%
  \else
    \ifGPcolor
      \def\colorrgb#1{\color[rgb]{#1}}%
      \def\colorgray#1{\color[gray]{#1}}%
      \expandafter\def\csname LTw\endcsname{\color{white}}%
      \expandafter\def\csname LTb\endcsname{\color{black}}%
      \expandafter\def\csname LTa\endcsname{\color{black}}%
      \expandafter\def\csname LT0\endcsname{\color[rgb]{1,0,0}}%
      \expandafter\def\csname LT1\endcsname{\color[rgb]{0,1,0}}%
      \expandafter\def\csname LT2\endcsname{\color[rgb]{0,0,1}}%
      \expandafter\def\csname LT3\endcsname{\color[rgb]{1,0,1}}%
      \expandafter\def\csname LT4\endcsname{\color[rgb]{0,1,1}}%
      \expandafter\def\csname LT5\endcsname{\color[rgb]{1,1,0}}%
      \expandafter\def\csname LT6\endcsname{\color[rgb]{0,0,0}}%
      \expandafter\def\csname LT7\endcsname{\color[rgb]{1,0.3,0}}%
      \expandafter\def\csname LT8\endcsname{\color[rgb]{0.5,0.5,0.5}}%
    \else
      \def\colorrgb#1{\color{black}}%
      \def\colorgray#1{\color[gray]{#1}}%
      \expandafter\def\csname LTw\endcsname{\color{white}}%
      \expandafter\def\csname LTb\endcsname{\color{black}}%
      \expandafter\def\csname LTa\endcsname{\color{black}}%
      \expandafter\def\csname LT0\endcsname{\color{black}}%
      \expandafter\def\csname LT1\endcsname{\color{black}}%
      \expandafter\def\csname LT2\endcsname{\color{black}}%
      \expandafter\def\csname LT3\endcsname{\color{black}}%
      \expandafter\def\csname LT4\endcsname{\color{black}}%
      \expandafter\def\csname LT5\endcsname{\color{black}}%
      \expandafter\def\csname LT6\endcsname{\color{black}}%
      \expandafter\def\csname LT7\endcsname{\color{black}}%
      \expandafter\def\csname LT8\endcsname{\color{black}}%
    \fi
  \fi
  \setlength{\unitlength}{0.0500bp}%
  \begin{picture}(7936.00,5102.00)%
    \gplgaddtomacro\gplbacktext{%
      \csname LTb\endcsname%
      \put(814,704){\makebox(0,0)[r]{\strut{} 0}}%
      \put(814,1455){\makebox(0,0)[r]{\strut{} 10}}%
      \put(814,2207){\makebox(0,0)[r]{\strut{} 20}}%
      \put(814,2958){\makebox(0,0)[r]{\strut{} 30}}%
      \put(814,3710){\makebox(0,0)[r]{\strut{} 40}}%
      \put(814,4461){\makebox(0,0)[r]{\strut{} 50}}%
      \put(946,484){\makebox(0,0){\strut{} 0}}%
      \put(2265,484){\makebox(0,0){\strut{} 10}}%
      \put(3583,484){\makebox(0,0){\strut{} 20}}%
      \put(4902,484){\makebox(0,0){\strut{} 30}}%
      \put(6220,484){\makebox(0,0){\strut{} 40}}%
      \put(7539,484){\makebox(0,0){\strut{} 50}}%
      \put(176,2770){\rotatebox{-270}{\makebox(0,0){\strut{}$\ln{\left(n_k+1\right)}$}}}%
      \put(4242,154){\makebox(0,0){\strut{}$t$ [$N$]}}%
    }%
    \gplgaddtomacro\gplfronttext{%
      \csname LTb\endcsname%
      \put(1738,4664){\makebox(0,0)[r]{\strut{}$\mathcal{N} = 1$}}%
      \csname LTb\endcsname%
      \put(1738,4444){\makebox(0,0)[r]{\strut{}$\mathcal{N} = 2$}}%
      \csname LTb\endcsname%
      \put(1738,4224){\makebox(0,0)[r]{\strut{}$\mathcal{N} = 3$}}%
      \csname LTb\endcsname%
      \put(1738,4004){\makebox(0,0)[r]{\strut{}$\mathcal{N} = 4$}}%
      \csname LTb\endcsname%
      \put(1738,3784){\makebox(0,0)[r]{\strut{}$\mathcal{N} = 5$}}%
    }%
    \gplbacktext
    \put(0,0){\includegraphics{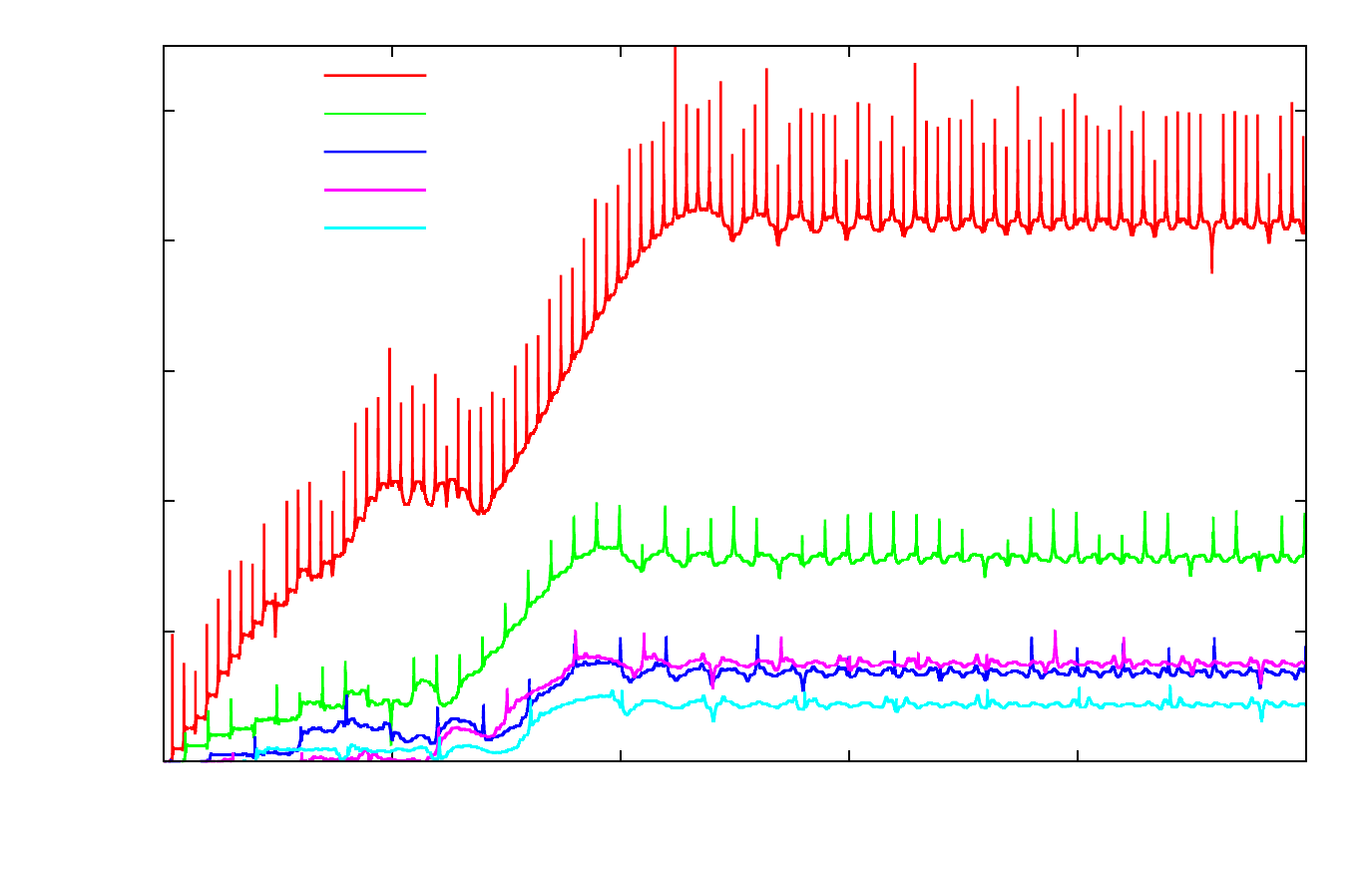}}%
    \gplfronttext
  \end{picture}%
\endgroup

%% file: trajectory.tex
\begingroup
  \makeatletter
  \providecommand\color[2][]{%
    \GenericError{(gnuplot) \space\space\space\@spaces}{%
      Package color not loaded in conjunction with
      terminal option `colourtext'%
    }{See the gnuplot documentation for explanation.%
    }{Either use 'blacktext' in gnuplot or load the package
      color.sty in LaTeX.}%
    \renewcommand\color[2][]{}%
  }%
  \providecommand\includegraphics[2][]{%
    \GenericError{(gnuplot) \space\space\space\@spaces}{%
      Package graphicx or graphics not loaded%
    }{See the gnuplot documentation for explanation.%
    }{The gnuplot epslatex terminal needs graphicx.sty or graphics.sty.}%
    \renewcommand\includegraphics[2][]{}%
  }%
  \providecommand\rotatebox[2]{#2}%
  \@ifundefined{ifGPcolor}{%
    \newif\ifGPcolor
    \GPcolortrue
  }{}%
  \@ifundefined{ifGPblacktext}{%
    \newif\ifGPblacktext
    \GPblacktexttrue
  }{}%
  \let\gplgaddtomacro\g@addto@macro
  \gdef\gplbacktext{}%
  \gdef\gplfronttext{}%
  \makeatother
  \ifGPblacktext
    \def\colorrgb#1{}%
    \def\colorgray#1{}%
  \else
    \ifGPcolor
      \def\colorrgb#1{\color[rgb]{#1}}%
      \def\colorgray#1{\color[gray]{#1}}%
      \expandafter\def\csname LTw\endcsname{\color{white}}%
      \expandafter\def\csname LTb\endcsname{\color{black}}%
      \expandafter\def\csname LTa\endcsname{\color{black}}%
      \expandafter\def\csname LT0\endcsname{\color[rgb]{1,0,0}}%
      \expandafter\def\csname LT1\endcsname{\color[rgb]{0,1,0}}%
      \expandafter\def\csname LT2\endcsname{\color[rgb]{0,0,1}}%
      \expandafter\def\csname LT3\endcsname{\color[rgb]{1,0,1}}%
      \expandafter\def\csname LT4\endcsname{\color[rgb]{0,1,1}}%
      \expandafter\def\csname LT5\endcsname{\color[rgb]{1,1,0}}%
      \expandafter\def\csname LT6\endcsname{\color[rgb]{0,0,0}}%
      \expandafter\def\csname LT7\endcsname{\color[rgb]{1,0.3,0}}%
      \expandafter\def\csname LT8\endcsname{\color[rgb]{0.5,0.5,0.5}}%
    \else
      \def\colorrgb#1{\color{black}}%
      \def\colorgray#1{\color[gray]{#1}}%
      \expandafter\def\csname LTw\endcsname{\color{white}}%
      \expandafter\def\csname LTb\endcsname{\color{black}}%
      \expandafter\def\csname LTa\endcsname{\color{black}}%
      \expandafter\def\csname LT0\endcsname{\color{black}}%
      \expandafter\def\csname LT1\endcsname{\color{black}}%
      \expandafter\def\csname LT2\endcsname{\color{black}}%
      \expandafter\def\csname LT3\endcsname{\color{black}}%
      \expandafter\def\csname LT4\endcsname{\color{black}}%
      \expandafter\def\csname LT5\endcsname{\color{black}}%
      \expandafter\def\csname LT6\endcsname{\color{black}}%
      \expandafter\def\csname LT7\endcsname{\color{black}}%
      \expandafter\def\csname LT8\endcsname{\color{black}}%
    \fi
  \fi
  \setlength{\unitlength}{0.0500bp}%
  \begin{picture}(7936.00,3400.00)%
    \gplgaddtomacro\gplbacktext{%
      \csname LTb\endcsname%
      \put(858,440){\makebox(0,0)[r]{\strut{}-0.05}}%
      \put(858,979){\makebox(0,0)[r]{\strut{} 0}}%
      \put(858,1518){\makebox(0,0)[r]{\strut{} 0.05}}%
      \put(858,2057){\makebox(0,0)[r]{\strut{} 0.1}}%
      \put(858,2596){\makebox(0,0)[r]{\strut{} 0.15}}%
      \put(858,3135){\makebox(0,0)[r]{\strut{} 0.2}}%
      \put(990,220){\makebox(0,0){\strut{}-0.05}}%
      \put(1506,220){\makebox(0,0){\strut{} 0}}%
      \put(2022,220){\makebox(0,0){\strut{} 0.05}}%
      \put(2539,220){\makebox(0,0){\strut{} 0.1}}%
      \put(3055,220){\makebox(0,0){\strut{} 0.15}}%
      \put(3571,220){\makebox(0,0){\strut{} 0.2}}%
    }%
    \gplgaddtomacro\gplfronttext{%
      \csname LTb\endcsname%
      \put(2178,2962){\makebox(0,0)[r]{\strut{}$t < 0.5\,N$}}%
      \csname LTb\endcsname%
      \put(2178,2742){\makebox(0,0)[r]{\strut{}$t < 8\,N$}}%
      \csname LTb\endcsname%
      \put(2178,2522){\makebox(0,0)[r]{\strut{}$t < 16\,N$}}%
      \csname LTb\endcsname%
      \put(2178,2302){\makebox(0,0)[r]{\strut{}$t < 50\,N$}}%
    }%
    \gplgaddtomacro\gplbacktext{%
      \csname LTb\endcsname%
      \put(4958,440){\makebox(0,0)[r]{\strut{}-0.01}}%
      \put(4958,1114){\makebox(0,0)[r]{\strut{}-0.005}}%
      \put(4958,1788){\makebox(0,0)[r]{\strut{} 0}}%
      \put(4958,2461){\makebox(0,0)[r]{\strut{} 0.005}}%
      \put(4958,3135){\makebox(0,0)[r]{\strut{} 0.01}}%
      \put(5090,220){\makebox(0,0){\strut{}-0.01}}%
      \put(5702,220){\makebox(0,0){\strut{}-0.005}}%
      \put(6315,220){\makebox(0,0){\strut{} 0}}%
      \put(6927,220){\makebox(0,0){\strut{} 0.005}}%
      \put(7539,220){\makebox(0,0){\strut{} 0.01}}%
    }%
    \gplgaddtomacro\gplfronttext{%
    }%
    \gplbacktext
    \put(0,0){\includegraphics{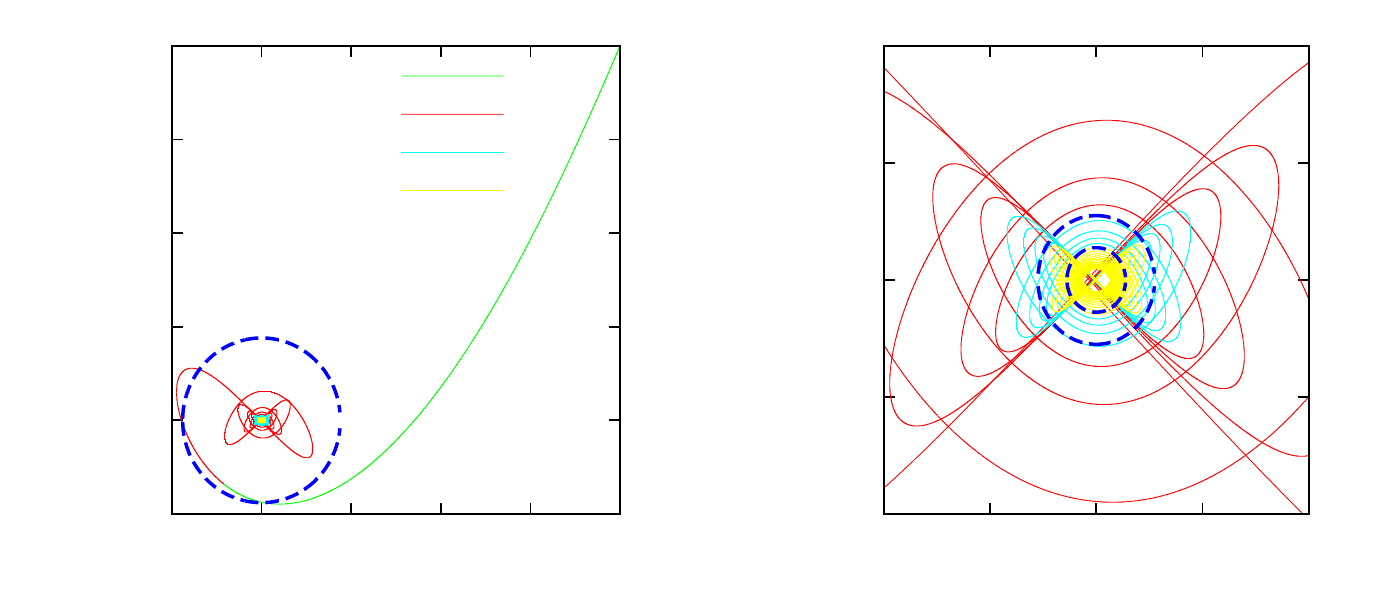}}%
    \gplfronttext
  \end{picture}%
\endgroup

%% file: ph_ablauf.tex
\begingroup
  \makeatletter
  \providecommand\color[2][]{%
    \GenericError{(gnuplot) \space\space\space\@spaces}{%
      Package color not loaded in conjunction with
      terminal option `colourtext'%
    }{See the gnuplot documentation for explanation.%
    }{Either use 'blacktext' in gnuplot or load the package
      color.sty in LaTeX.}%
    \renewcommand\color[2][]{}%
  }%
  \providecommand\includegraphics[2][]{%
    \GenericError{(gnuplot) \space\space\space\@spaces}{%
      Package graphicx or graphics not loaded%
    }{See the gnuplot documentation for explanation.%
    }{The gnuplot epslatex terminal needs graphicx.sty or graphics.sty.}%
    \renewcommand\includegraphics[2][]{}%
  }%
  \providecommand\rotatebox[2]{#2}%
  \@ifundefined{ifGPcolor}{%
    \newif\ifGPcolor
    \GPcolortrue
  }{}%
  \@ifundefined{ifGPblacktext}{%
    \newif\ifGPblacktext
    \GPblacktexttrue
  }{}%
  \let\gplgaddtomacro\g@addto@macro
  \gdef\gplbacktext{}%
  \gdef\gplfronttext{}%
  \makeatother
  \ifGPblacktext
    \def\colorrgb#1{}%
    \def\colorgray#1{}%
  \else
    \ifGPcolor
      \def\colorrgb#1{\color[rgb]{#1}}%
      \def\colorgray#1{\color[gray]{#1}}%
      \expandafter\def\csname LTw\endcsname{\color{white}}%
      \expandafter\def\csname LTb\endcsname{\color{black}}%
      \expandafter\def\csname LTa\endcsname{\color{black}}%
      \expandafter\def\csname LT0\endcsname{\color[rgb]{1,0,0}}%
      \expandafter\def\csname LT1\endcsname{\color[rgb]{0,1,0}}%
      \expandafter\def\csname LT2\endcsname{\color[rgb]{0,0,1}}%
      \expandafter\def\csname LT3\endcsname{\color[rgb]{1,0,1}}%
      \expandafter\def\csname LT4\endcsname{\color[rgb]{0,1,1}}%
      \expandafter\def\csname LT5\endcsname{\color[rgb]{1,1,0}}%
      \expandafter\def\csname LT6\endcsname{\color[rgb]{0,0,0}}%
      \expandafter\def\csname LT7\endcsname{\color[rgb]{1,0.3,0}}%
      \expandafter\def\csname LT8\endcsname{\color[rgb]{0.5,0.5,0.5}}%
    \else
      \def\colorrgb#1{\color{black}}%
      \def\colorgray#1{\color[gray]{#1}}%
      \expandafter\def\csname LTw\endcsname{\color{white}}%
      \expandafter\def\csname LTb\endcsname{\color{black}}%
      \expandafter\def\csname LTa\endcsname{\color{black}}%
      \expandafter\def\csname LT0\endcsname{\color{black}}%
      \expandafter\def\csname LT1\endcsname{\color{black}}%
      \expandafter\def\csname LT2\endcsname{\color{black}}%
      \expandafter\def\csname LT3\endcsname{\color{black}}%
      \expandafter\def\csname LT4\endcsname{\color{black}}%
      \expandafter\def\csname LT5\endcsname{\color{black}}%
      \expandafter\def\csname LT6\endcsname{\color{black}}%
      \expandafter\def\csname LT7\endcsname{\color{black}}%
      \expandafter\def\csname LT8\endcsname{\color{black}}%
    \fi
  \fi
  \setlength{\unitlength}{0.0500bp}%
  \begin{picture}(7936.00,5102.00)%
    \gplgaddtomacro\gplbacktext{%
      \csname LTb\endcsname%
      \put(814,805){\makebox(0,0)[r]{\strut{} 0}}%
      \put(814,1309){\makebox(0,0)[r]{\strut{} 5}}%
      \put(814,1813){\makebox(0,0)[r]{\strut{} 10}}%
      \put(814,2317){\makebox(0,0)[r]{\strut{} 15}}%
      \put(814,2821){\makebox(0,0)[r]{\strut{} 20}}%
      \put(814,3325){\makebox(0,0)[r]{\strut{} 25}}%
      \put(814,3829){\makebox(0,0)[r]{\strut{} 30}}%
      \put(814,4333){\makebox(0,0)[r]{\strut{} 35}}%
      \put(814,4837){\makebox(0,0)[r]{\strut{} 40}}%
      \put(946,484){\makebox(0,0){\strut{} 0}}%
      \put(2045,484){\makebox(0,0){\strut{} 10}}%
      \put(3144,484){\makebox(0,0){\strut{} 20}}%
      \put(4243,484){\makebox(0,0){\strut{} 30}}%
      \put(5341,484){\makebox(0,0){\strut{} 40}}%
      \put(6440,484){\makebox(0,0){\strut{} 50}}%
      \put(7539,484){\makebox(0,0){\strut{} 60}}%
      \put(176,2770){\rotatebox{-270}{\makebox(0,0){\strut{}$\ln{\left(n_k+1\right)}$}}}%
      \put(4242,154){\makebox(0,0){\strut{}$t$ [$N$]}}%
    }%
    \gplgaddtomacro\gplfronttext{%
    }%
    \gplgaddtomacro\gplbacktext{%
      \csname LTb\endcsname%
      \put(4231,1205){\makebox(0,0)[r]{\strut{}-1}}%
      \put(4231,1436){\makebox(0,0)[r]{\strut{} 0}}%
      \put(4231,1667){\makebox(0,0)[r]{\strut{} 1}}%
      \put(4231,1898){\makebox(0,0)[r]{\strut{} 2}}%
      \put(4231,2129){\makebox(0,0)[r]{\strut{} 3}}%
      \put(4231,2359){\makebox(0,0)[r]{\strut{} 4}}%
      \put(4231,2590){\makebox(0,0)[r]{\strut{} 5}}%
      \put(4231,2821){\makebox(0,0)[r]{\strut{} 6}}%
      \put(4231,3052){\makebox(0,0)[r]{\strut{} 7}}%
      \put(4363,985){\makebox(0,0){\strut{} 0}}%
      \put(4958,985){\makebox(0,0){\strut{} 0.2}}%
      \put(5554,985){\makebox(0,0){\strut{} 0.4}}%
      \put(6149,985){\makebox(0,0){\strut{} 0.6}}%
      \put(6745,985){\makebox(0,0){\strut{} 0.8}}%
      \put(7340,985){\makebox(0,0){\strut{} 1}}%
      \put(3945,2128){\rotatebox{-270}{\makebox(0,0){\strut{}}}}%
      \put(5851,699){\makebox(0,0){\strut{}}}%
    }%
    \gplgaddtomacro\gplfronttext{%
    }%
    \gplbacktext
    \put(0,0){\includegraphics{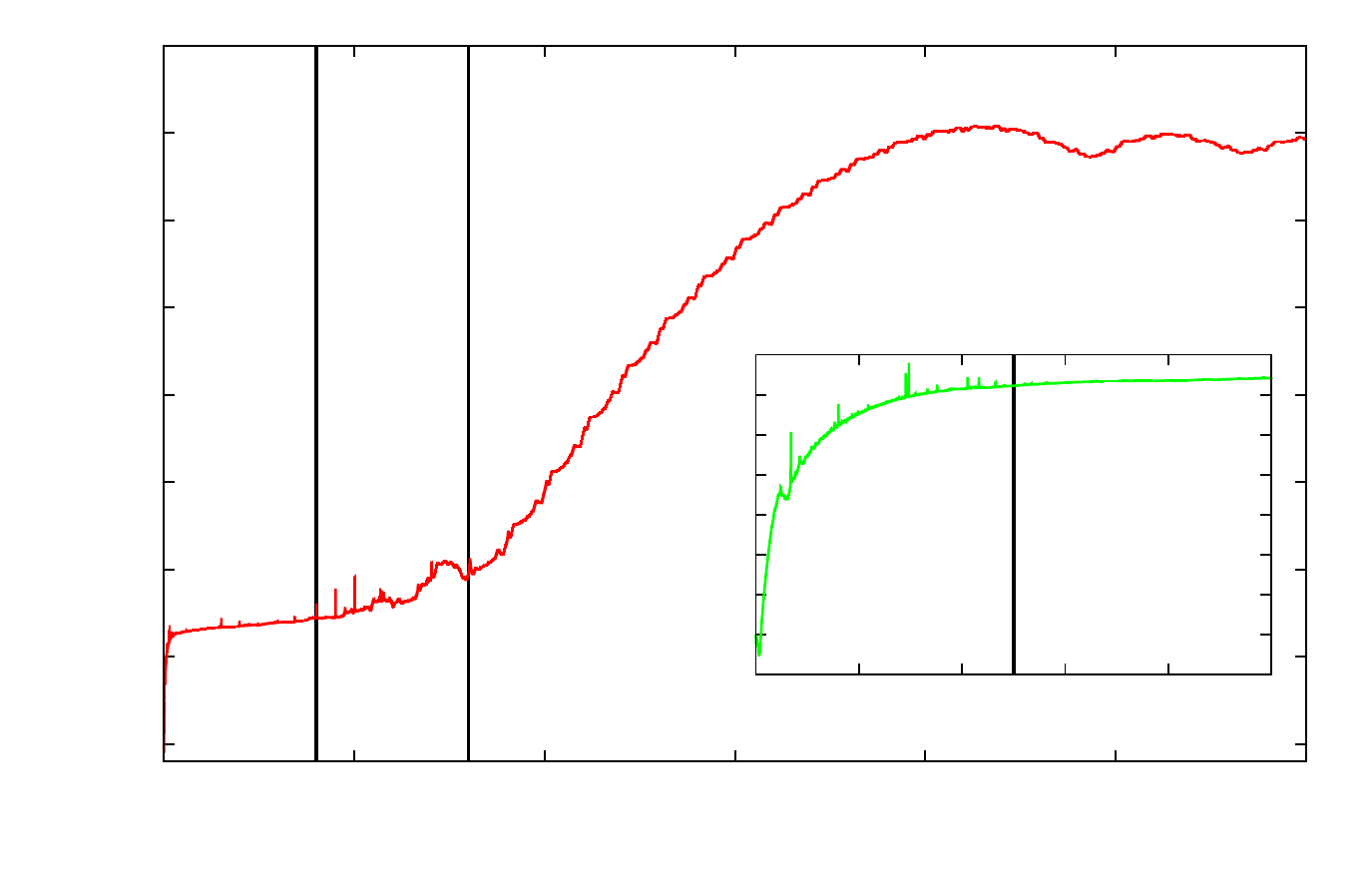}}%
    \gplfronttext
  \end{picture}%
\endgroup

%% file: vel_int.tex
\begingroup
  \makeatletter
  \providecommand\color[2][]{%
    \GenericError{(gnuplot) \space\space\space\@spaces}{%
      Package color not loaded in conjunction with
      terminal option `colourtext'%
    }{See the gnuplot documentation for explanation.%
    }{Either use 'blacktext' in gnuplot or load the package
      color.sty in LaTeX.}%
    \renewcommand\color[2][]{}%
  }%
  \providecommand\includegraphics[2][]{%
    \GenericError{(gnuplot) \space\space\space\@spaces}{%
      Package graphicx or graphics not loaded%
    }{See the gnuplot documentation for explanation.%
    }{The gnuplot epslatex terminal needs graphicx.sty or graphics.sty.}%
    \renewcommand\includegraphics[2][]{}%
  }%
  \providecommand\rotatebox[2]{#2}%
  \@ifundefined{ifGPcolor}{%
    \newif\ifGPcolor
    \GPcolortrue
  }{}%
  \@ifundefined{ifGPblacktext}{%
    \newif\ifGPblacktext
    \GPblacktexttrue
  }{}%
  \let\gplgaddtomacro\g@addto@macro
  \gdef\gplbacktext{}%
  \gdef\gplfronttext{}%
  \makeatother
  \ifGPblacktext
    \def\colorrgb#1{}%
    \def\colorgray#1{}%
  \else
    \ifGPcolor
      \def\colorrgb#1{\color[rgb]{#1}}%
      \def\colorgray#1{\color[gray]{#1}}%
      \expandafter\def\csname LTw\endcsname{\color{white}}%
      \expandafter\def\csname LTb\endcsname{\color{black}}%
      \expandafter\def\csname LTa\endcsname{\color{black}}%
      \expandafter\def\csname LT0\endcsname{\color[rgb]{1,0,0}}%
      \expandafter\def\csname LT1\endcsname{\color[rgb]{0,1,0}}%
      \expandafter\def\csname LT2\endcsname{\color[rgb]{0,0,1}}%
      \expandafter\def\csname LT3\endcsname{\color[rgb]{1,0,1}}%
      \expandafter\def\csname LT4\endcsname{\color[rgb]{0,1,1}}%
      \expandafter\def\csname LT5\endcsname{\color[rgb]{1,1,0}}%
      \expandafter\def\csname LT6\endcsname{\color[rgb]{0,0,0}}%
      \expandafter\def\csname LT7\endcsname{\color[rgb]{1,0.3,0}}%
      \expandafter\def\csname LT8\endcsname{\color[rgb]{0.5,0.5,0.5}}%
    \else
      \def\colorrgb#1{\color{black}}%
      \def\colorgray#1{\color[gray]{#1}}%
      \expandafter\def\csname LTw\endcsname{\color{white}}%
      \expandafter\def\csname LTb\endcsname{\color{black}}%
      \expandafter\def\csname LTa\endcsname{\color{black}}%
      \expandafter\def\csname LT0\endcsname{\color{black}}%
      \expandafter\def\csname LT1\endcsname{\color{black}}%
      \expandafter\def\csname LT2\endcsname{\color{black}}%
      \expandafter\def\csname LT3\endcsname{\color{black}}%
      \expandafter\def\csname LT4\endcsname{\color{black}}%
      \expandafter\def\csname LT5\endcsname{\color{black}}%
      \expandafter\def\csname LT6\endcsname{\color{black}}%
      \expandafter\def\csname LT7\endcsname{\color{black}}%
      \expandafter\def\csname LT8\endcsname{\color{black}}%
    \fi
  \fi
  \setlength{\unitlength}{0.0500bp}%
  \begin{picture}(7936.00,3400.00)%
    \gplgaddtomacro\gplbacktext{%
      \csname LTb\endcsname%
      \put(946,704){\makebox(0,0)[r]{\strut{} 0}}%
      \put(946,1051){\makebox(0,0)[r]{\strut{} 0.1}}%
      \put(946,1399){\makebox(0,0)[r]{\strut{} 0.2}}%
      \put(946,1746){\makebox(0,0)[r]{\strut{} 0.3}}%
      \put(946,2093){\makebox(0,0)[r]{\strut{} 0.4}}%
      \put(946,2440){\makebox(0,0)[r]{\strut{} 0.5}}%
      \put(946,2788){\makebox(0,0)[r]{\strut{} 0.6}}%
      \put(946,3135){\makebox(0,0)[r]{\strut{} 0.7}}%
      \put(1458,484){\makebox(0,0){\strut{} 10}}%
      \put(1881,484){\makebox(0,0){\strut{} 20}}%
      \put(2303,484){\makebox(0,0){\strut{} 30}}%
      \put(2726,484){\makebox(0,0){\strut{} 40}}%
      \put(3148,484){\makebox(0,0){\strut{} 50}}%
      \put(3571,484){\makebox(0,0){\strut{} 60}}%
      \put(176,1919){\rotatebox{-270}{\makebox(0,0){\strut{}$\left<n_0\right>$}}}%
      \put(2324,154){\makebox(0,0){\strut{}$\mathcal{N}$}}%
    }%
    \gplgaddtomacro\gplfronttext{%
      \csname LTb\endcsname%
      \put(2584,2962){\makebox(0,0)[r]{\strut{}$\left<n_0\right>$}}%
      \csname LTb\endcsname%
      \put(2584,2742){\makebox(0,0)[r]{\strut{}$\exp{(-\pi)}$}}%
    }%
    \gplgaddtomacro\gplbacktext{%
      \csname LTb\endcsname%
      \put(4914,704){\makebox(0,0)[r]{\strut{}-160}}%
      \put(4914,974){\makebox(0,0)[r]{\strut{}-140}}%
      \put(4914,1244){\makebox(0,0)[r]{\strut{}-120}}%
      \put(4914,1514){\makebox(0,0)[r]{\strut{}-100}}%
      \put(4914,1784){\makebox(0,0)[r]{\strut{}-80}}%
      \put(4914,2055){\makebox(0,0)[r]{\strut{}-60}}%
      \put(4914,2325){\makebox(0,0)[r]{\strut{}-40}}%
      \put(4914,2595){\makebox(0,0)[r]{\strut{}-20}}%
      \put(4914,2865){\makebox(0,0)[r]{\strut{} 0}}%
      \put(4914,3135){\makebox(0,0)[r]{\strut{} 20}}%
      \put(5046,484){\makebox(0,0){\strut{} 0}}%
      \put(5640,484){\makebox(0,0){\strut{} 5}}%
      \put(6233,484){\makebox(0,0){\strut{} 10}}%
      \put(6827,484){\makebox(0,0){\strut{} 15}}%
      \put(7420,484){\makebox(0,0){\strut{} 20}}%
      \put(4144,1919){\rotatebox{-270}{\makebox(0,0){\strut{}$\ln{\left[\mathcal{F}(\mathcal{N},\alpha)\right]}$}}}%
      \put(6292,154){\makebox(0,0){\strut{}$\mathcal{N}$}}%
    }%
    \gplgaddtomacro\gplfronttext{%
      \csname LTb\endcsname%
      \put(6552,2962){\makebox(0,0)[r]{\strut{}$\alpha = 0.5$}}%
      \csname LTb\endcsname%
      \put(6552,2742){\makebox(0,0)[r]{\strut{}$\alpha = 2$}}%
    }%
    \gplbacktext
    \put(0,0){\includegraphics{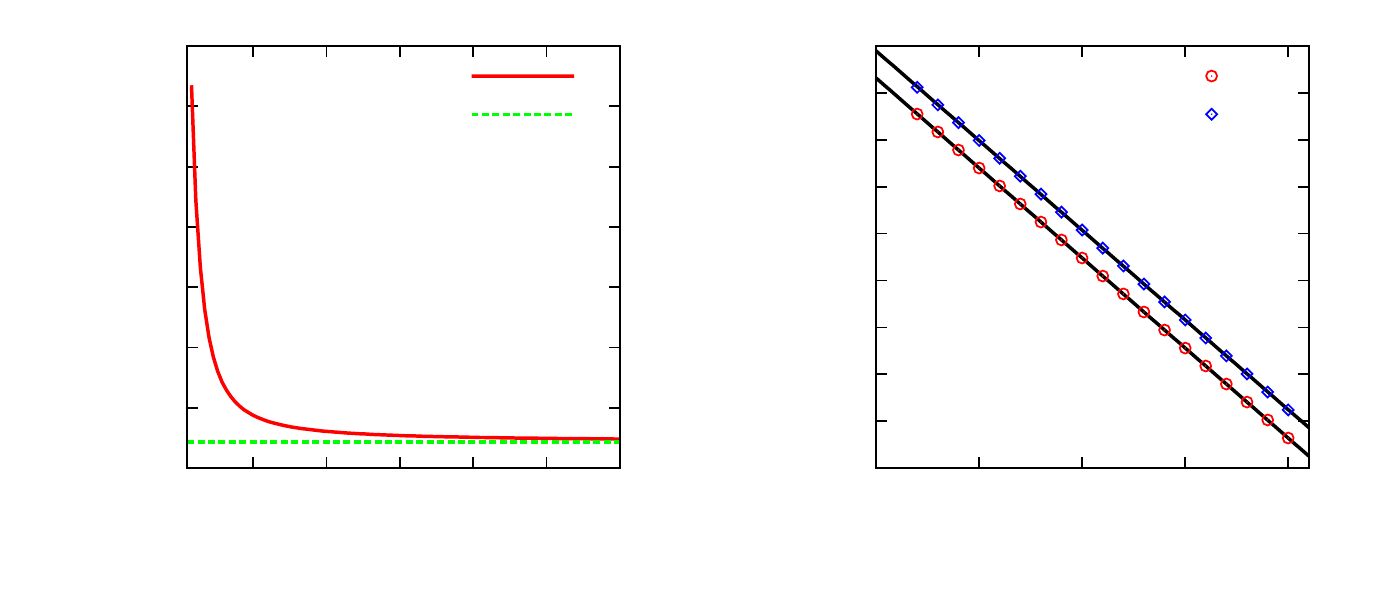}}%
    \gplfronttext
  \end{picture}%
\endgroup

%% file: fi_eff.tex
\begingroup
  \makeatletter
  \providecommand\color[2][]{%
    \GenericError{(gnuplot) \space\space\space\@spaces}{%
      Package color not loaded in conjunction with
      terminal option `colourtext'%
    }{See the gnuplot documentation for explanation.%
    }{Either use 'blacktext' in gnuplot or load the package
      color.sty in LaTeX.}%
    \renewcommand\color[2][]{}%
  }%
  \providecommand\includegraphics[2][]{%
    \GenericError{(gnuplot) \space\space\space\@spaces}{%
      Package graphicx or graphics not loaded%
    }{See the gnuplot documentation for explanation.%
    }{The gnuplot epslatex terminal needs graphicx.sty or graphics.sty.}%
    \renewcommand\includegraphics[2][]{}%
  }%
  \providecommand\rotatebox[2]{#2}%
  \@ifundefined{ifGPcolor}{%
    \newif\ifGPcolor
    \GPcolortrue
  }{}%
  \@ifundefined{ifGPblacktext}{%
    \newif\ifGPblacktext
    \GPblacktexttrue
  }{}%
  \let\gplgaddtomacro\g@addto@macro
  \gdef\gplbacktext{}%
  \gdef\gplfronttext{}%
  \makeatother
  \ifGPblacktext
    \def\colorrgb#1{}%
    \def\colorgray#1{}%
  \else
    \ifGPcolor
      \def\colorrgb#1{\color[rgb]{#1}}%
      \def\colorgray#1{\color[gray]{#1}}%
      \expandafter\def\csname LTw\endcsname{\color{white}}%
      \expandafter\def\csname LTb\endcsname{\color{black}}%
      \expandafter\def\csname LTa\endcsname{\color{black}}%
      \expandafter\def\csname LT0\endcsname{\color[rgb]{1,0,0}}%
      \expandafter\def\csname LT1\endcsname{\color[rgb]{0,1,0}}%
      \expandafter\def\csname LT2\endcsname{\color[rgb]{0,0,1}}%
      \expandafter\def\csname LT3\endcsname{\color[rgb]{1,0,1}}%
      \expandafter\def\csname LT4\endcsname{\color[rgb]{0,1,1}}%
      \expandafter\def\csname LT5\endcsname{\color[rgb]{1,1,0}}%
      \expandafter\def\csname LT6\endcsname{\color[rgb]{0,0,0}}%
      \expandafter\def\csname LT7\endcsname{\color[rgb]{1,0.3,0}}%
      \expandafter\def\csname LT8\endcsname{\color[rgb]{0.5,0.5,0.5}}%
    \else
      \def\colorrgb#1{\color{black}}%
      \def\colorgray#1{\color[gray]{#1}}%
      \expandafter\def\csname LTw\endcsname{\color{white}}%
      \expandafter\def\csname LTb\endcsname{\color{black}}%
      \expandafter\def\csname LTa\endcsname{\color{black}}%
      \expandafter\def\csname LT0\endcsname{\color{black}}%
      \expandafter\def\csname LT1\endcsname{\color{black}}%
      \expandafter\def\csname LT2\endcsname{\color{black}}%
      \expandafter\def\csname LT3\endcsname{\color{black}}%
      \expandafter\def\csname LT4\endcsname{\color{black}}%
      \expandafter\def\csname LT5\endcsname{\color{black}}%
      \expandafter\def\csname LT6\endcsname{\color{black}}%
      \expandafter\def\csname LT7\endcsname{\color{black}}%
      \expandafter\def\csname LT8\endcsname{\color{black}}%
    \fi
  \fi
  \setlength{\unitlength}{0.0500bp}%
  \begin{picture}(7936.00,5102.00)%
    \gplgaddtomacro\gplbacktext{%
      \csname LTb\endcsname%
      \put(814,1584){\makebox(0,0)[r]{\strut{} 0}}%
      \csname LTb\endcsname%
      \put(814,1991){\makebox(0,0)[r]{\strut{} 10}}%
      \csname LTb\endcsname%
      \put(814,2397){\makebox(0,0)[r]{\strut{} 20}}%
      \csname LTb\endcsname%
      \put(814,2804){\makebox(0,0)[r]{\strut{} 30}}%
      \csname LTb\endcsname%
      \put(814,3211){\makebox(0,0)[r]{\strut{} 40}}%
      \csname LTb\endcsname%
      \put(814,3617){\makebox(0,0)[r]{\strut{} 50}}%
      \csname LTb\endcsname%
      \put(814,4024){\makebox(0,0)[r]{\strut{} 60}}%
      \csname LTb\endcsname%
      \put(814,4430){\makebox(0,0)[r]{\strut{} 70}}%
      \csname LTb\endcsname%
      \put(814,4837){\makebox(0,0)[r]{\strut{} 80}}%
      \csname LTb\endcsname%
      \put(1613,1364){\makebox(0,0){\strut{} 10}}%
      \csname LTb\endcsname%
      \put(2353,1364){\makebox(0,0){\strut{} 20}}%
      \csname LTb\endcsname%
      \put(3094,1364){\makebox(0,0){\strut{} 30}}%
      \csname LTb\endcsname%
      \put(3835,1364){\makebox(0,0){\strut{} 40}}%
      \csname LTb\endcsname%
      \put(4576,1364){\makebox(0,0){\strut{} 50}}%
      \csname LTb\endcsname%
      \put(5317,1364){\makebox(0,0){\strut{} 60}}%
      \csname LTb\endcsname%
      \put(6057,1364){\makebox(0,0){\strut{} 70}}%
      \csname LTb\endcsname%
      \put(6798,1364){\makebox(0,0){\strut{} 80}}%
      \csname LTb\endcsname%
      \put(7539,1364){\makebox(0,0){\strut{} 90}}%
      \put(176,3210){\rotatebox{-270}{\makebox(0,0){\strut{}$\ln{\left(n_0^{tot}+1\right)}$}}}%
      \put(4242,1034){\makebox(0,0){\strut{}$\mathcal{N}$}}%
    }%
    \gplgaddtomacro\gplfronttext{%
      \csname LTb\endcsname%
      \put(2300,613){\makebox(0,0)[r]{\strut{}$x = 0.001$}}%
      \csname LTb\endcsname%
      \put(2300,393){\makebox(0,0)[r]{\strut{}$x = 0.003$}}%
      \csname LTb\endcsname%
      \put(2300,173){\makebox(0,0)[r]{\strut{}$x = 0.005$}}%
      \csname LTb\endcsname%
      \put(4475,613){\makebox(0,0)[r]{\strut{}$x = 0.0057$}}%
      \csname LTb\endcsname%
      \put(4475,393){\makebox(0,0)[r]{\strut{}$x = 0.008$}}%
      \csname LTb\endcsname%
      \put(4475,173){\makebox(0,0)[r]{\strut{}$x = 0.01$}}%
      \csname LTb\endcsname%
      \put(6650,613){\makebox(0,0)[r]{\strut{}$x = 0.015$}}%
    }%
    \gplbacktext
    \put(0,0){\includegraphics{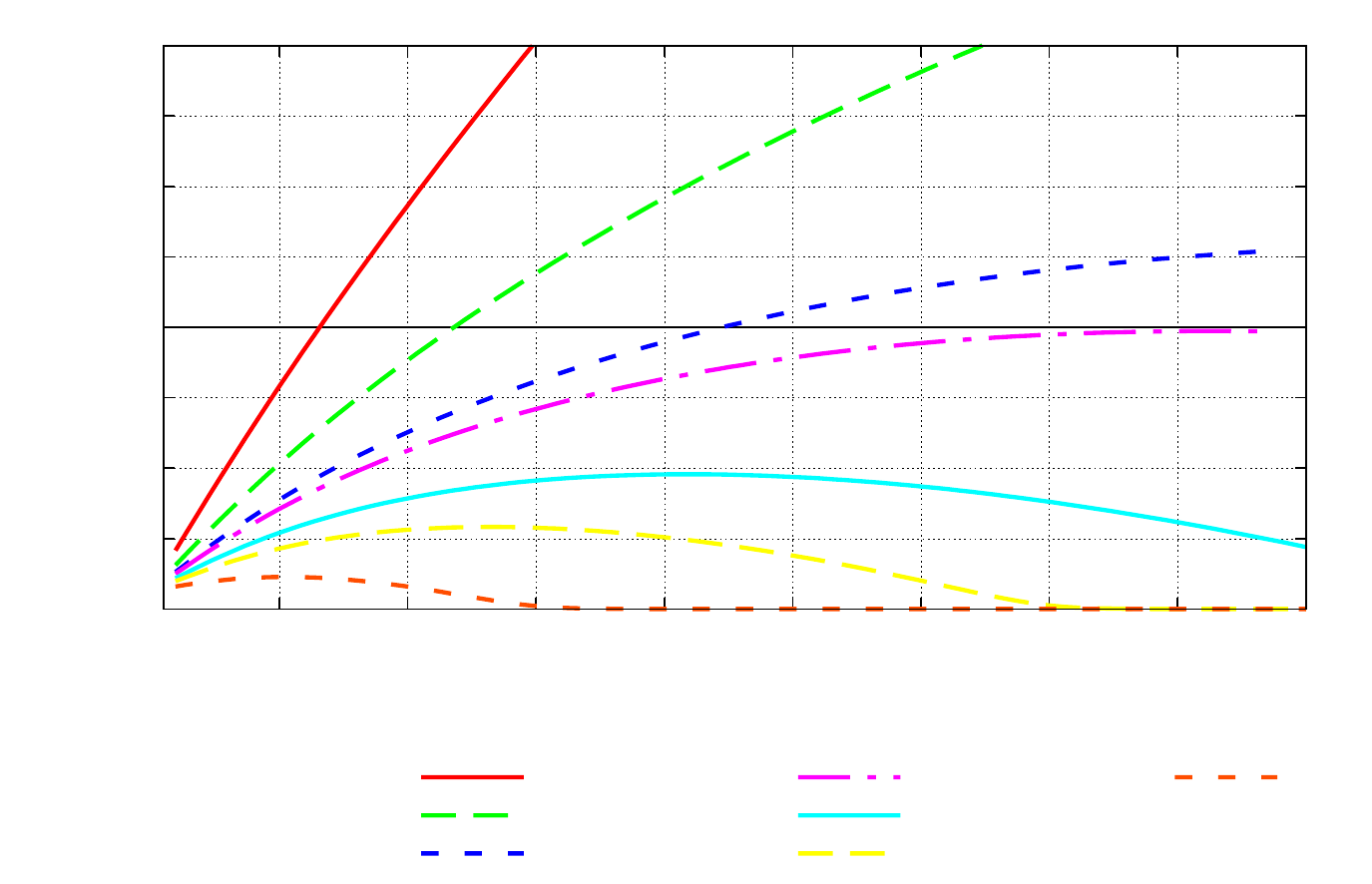}}%
    \gplfronttext
  \end{picture}%
\endgroup

%% file: first_infall2.tex
\begingroup
  \makeatletter
  \providecommand\color[2][]{%
    \GenericError{(gnuplot) \space\space\space\@spaces}{%
      Package color not loaded in conjunction with
      terminal option `colourtext'%
    }{See the gnuplot documentation for explanation.%
    }{Either use 'blacktext' in gnuplot or load the package
      color.sty in LaTeX.}%
    \renewcommand\color[2][]{}%
  }%
  \providecommand\includegraphics[2][]{%
    \GenericError{(gnuplot) \space\space\space\@spaces}{%
      Package graphicx or graphics not loaded%
    }{See the gnuplot documentation for explanation.%
    }{The gnuplot epslatex terminal needs graphicx.sty or graphics.sty.}%
    \renewcommand\includegraphics[2][]{}%
  }%
  \providecommand\rotatebox[2]{#2}%
  \@ifundefined{ifGPcolor}{%
    \newif\ifGPcolor
    \GPcolortrue
  }{}%
  \@ifundefined{ifGPblacktext}{%
    \newif\ifGPblacktext
    \GPblacktexttrue
  }{}%
  \let\gplgaddtomacro\g@addto@macro
  \gdef\gplbacktext{}%
  \gdef\gplfronttext{}%
  \makeatother
  \ifGPblacktext
    \def\colorrgb#1{}%
    \def\colorgray#1{}%
  \else
    \ifGPcolor
      \def\colorrgb#1{\color[rgb]{#1}}%
      \def\colorgray#1{\color[gray]{#1}}%
      \expandafter\def\csname LTw\endcsname{\color{white}}%
      \expandafter\def\csname LTb\endcsname{\color{black}}%
      \expandafter\def\csname LTa\endcsname{\color{black}}%
      \expandafter\def\csname LT0\endcsname{\color[rgb]{1,0,0}}%
      \expandafter\def\csname LT1\endcsname{\color[rgb]{0,1,0}}%
      \expandafter\def\csname LT2\endcsname{\color[rgb]{0,0,1}}%
      \expandafter\def\csname LT3\endcsname{\color[rgb]{1,0,1}}%
      \expandafter\def\csname LT4\endcsname{\color[rgb]{0,1,1}}%
      \expandafter\def\csname LT5\endcsname{\color[rgb]{1,1,0}}%
      \expandafter\def\csname LT6\endcsname{\color[rgb]{0,0,0}}%
      \expandafter\def\csname LT7\endcsname{\color[rgb]{1,0.3,0}}%
      \expandafter\def\csname LT8\endcsname{\color[rgb]{0.5,0.5,0.5}}%
    \else
      \def\colorrgb#1{\color{black}}%
      \def\colorgray#1{\color[gray]{#1}}%
      \expandafter\def\csname LTw\endcsname{\color{white}}%
      \expandafter\def\csname LTb\endcsname{\color{black}}%
      \expandafter\def\csname LTa\endcsname{\color{black}}%
      \expandafter\def\csname LT0\endcsname{\color{black}}%
      \expandafter\def\csname LT1\endcsname{\color{black}}%
      \expandafter\def\csname LT2\endcsname{\color{black}}%
      \expandafter\def\csname LT3\endcsname{\color{black}}%
      \expandafter\def\csname LT4\endcsname{\color{black}}%
      \expandafter\def\csname LT5\endcsname{\color{black}}%
      \expandafter\def\csname LT6\endcsname{\color{black}}%
      \expandafter\def\csname LT7\endcsname{\color{black}}%
      \expandafter\def\csname LT8\endcsname{\color{black}}%
    \fi
  \fi
  \setlength{\unitlength}{0.0500bp}%
  \begin{picture}(7936.00,5102.00)%
    \gplgaddtomacro\gplbacktext{%
      \csname LTb\endcsname%
      \put(1078,704){\makebox(0,0)[r]{\strut{} 0}}%
      \put(1078,1531){\makebox(0,0)[r]{\strut{} 200}}%
      \put(1078,2357){\makebox(0,0)[r]{\strut{} 400}}%
      \put(1078,3184){\makebox(0,0)[r]{\strut{} 600}}%
      \put(1078,4010){\makebox(0,0)[r]{\strut{} 800}}%
      \put(1078,4837){\makebox(0,0)[r]{\strut{} 1000}}%
      \put(1210,484){\makebox(0,0){\strut{} 0.002}}%
      \put(1843,484){\makebox(0,0){\strut{} 0.004}}%
      \put(2476,484){\makebox(0,0){\strut{} 0.006}}%
      \put(3109,484){\makebox(0,0){\strut{} 0.008}}%
      \put(3742,484){\makebox(0,0){\strut{} 0.01}}%
      \put(4375,484){\makebox(0,0){\strut{} 0.012}}%
      \put(5007,484){\makebox(0,0){\strut{} 0.014}}%
      \put(5640,484){\makebox(0,0){\strut{} 0.016}}%
      \put(6273,484){\makebox(0,0){\strut{} 0.018}}%
      \put(6906,484){\makebox(0,0){\strut{} 0.02}}%
      \put(7539,484){\makebox(0,0){\strut{} 0.022}}%
      \put(176,2770){\rotatebox{-270}{\makebox(0,0){\strut{}$n_0^{tot}(t_*)$}}}%
      \put(4374,154){\makebox(0,0){\strut{}$x$ [$M_P$]}}%
    }%
    \gplgaddtomacro\gplfronttext{%
      \csname LTb\endcsname%
      \put(6552,4664){\makebox(0,0)[r]{\strut{}$\mathcal{N}=2$}}%
      \csname LTb\endcsname%
      \put(6552,4444){\makebox(0,0)[r]{\strut{}$\mathcal{N}=3$}}%
      \csname LTb\endcsname%
      \put(6552,4224){\makebox(0,0)[r]{\strut{}$\mathcal{N}=4$}}%
    }%
    \gplbacktext
    \put(0,0){\includegraphics{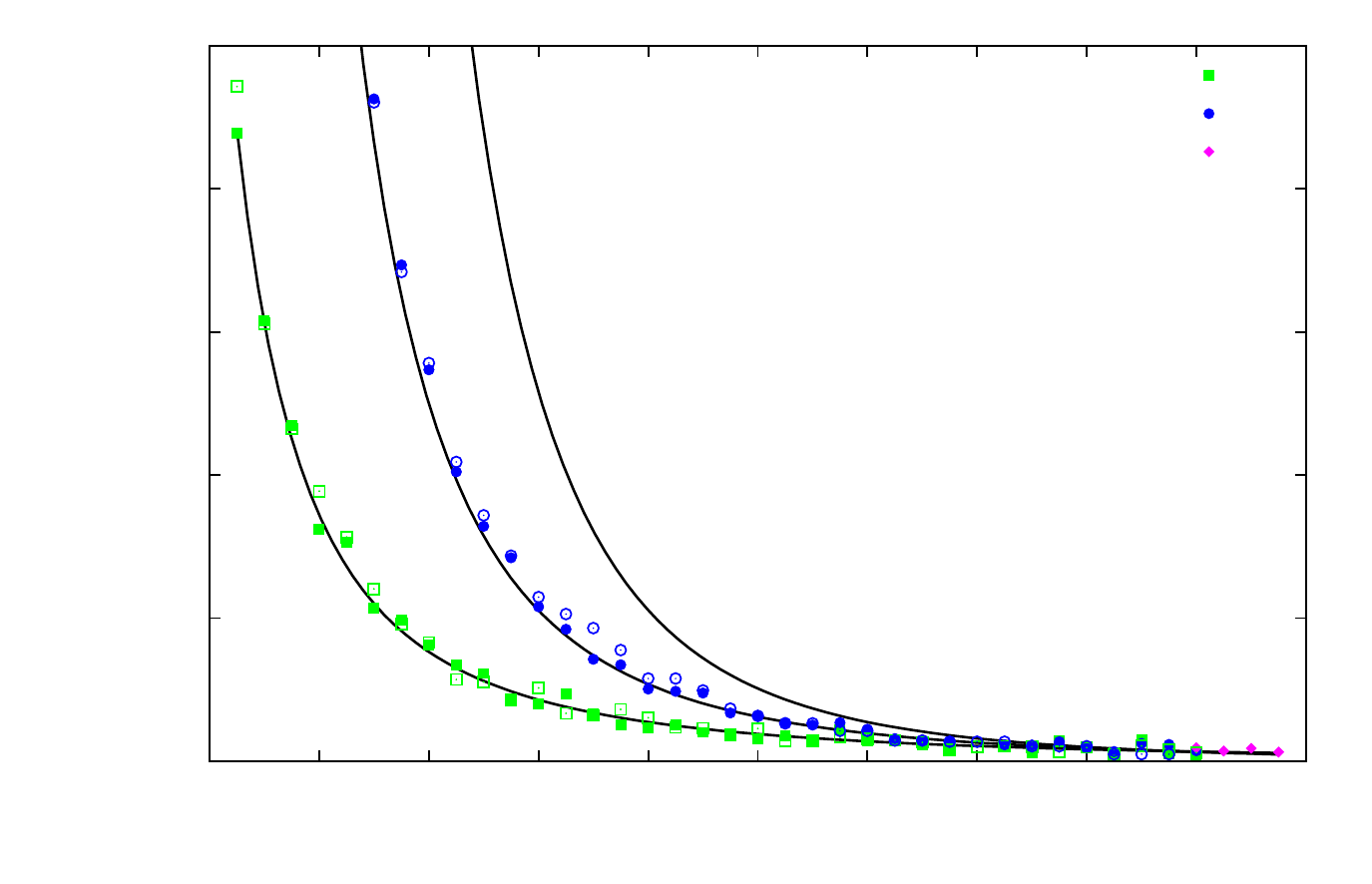}}%
    \gplfronttext
  \end{picture}%
\endgroup

%% file: fi_iv.tex
\begingroup
  \makeatletter
  \providecommand\color[2][]{%
    \GenericError{(gnuplot) \space\space\space\@spaces}{%
      Package color not loaded in conjunction with
      terminal option `colourtext'%
    }{See the gnuplot documentation for explanation.%
    }{Either use 'blacktext' in gnuplot or load the package
      color.sty in LaTeX.}%
    \renewcommand\color[2][]{}%
  }%
  \providecommand\includegraphics[2][]{%
    \GenericError{(gnuplot) \space\space\space\@spaces}{%
      Package graphicx or graphics not loaded%
    }{See the gnuplot documentation for explanation.%
    }{The gnuplot epslatex terminal needs graphicx.sty or graphics.sty.}%
    \renewcommand\includegraphics[2][]{}%
  }%
  \providecommand\rotatebox[2]{#2}%
  \@ifundefined{ifGPcolor}{%
    \newif\ifGPcolor
    \GPcolortrue
  }{}%
  \@ifundefined{ifGPblacktext}{%
    \newif\ifGPblacktext
    \GPblacktexttrue
  }{}%
  \let\gplgaddtomacro\g@addto@macro
  \gdef\gplbacktext{}%
  \gdef\gplfronttext{}%
  \makeatother
  \ifGPblacktext
    \def\colorrgb#1{}%
    \def\colorgray#1{}%
  \else
    \ifGPcolor
      \def\colorrgb#1{\color[rgb]{#1}}%
      \def\colorgray#1{\color[gray]{#1}}%
      \expandafter\def\csname LTw\endcsname{\color{white}}%
      \expandafter\def\csname LTb\endcsname{\color{black}}%
      \expandafter\def\csname LTa\endcsname{\color{black}}%
      \expandafter\def\csname LT0\endcsname{\color[rgb]{1,0,0}}%
      \expandafter\def\csname LT1\endcsname{\color[rgb]{0,1,0}}%
      \expandafter\def\csname LT2\endcsname{\color[rgb]{0,0,1}}%
      \expandafter\def\csname LT3\endcsname{\color[rgb]{1,0,1}}%
      \expandafter\def\csname LT4\endcsname{\color[rgb]{0,1,1}}%
      \expandafter\def\csname LT5\endcsname{\color[rgb]{1,1,0}}%
      \expandafter\def\csname LT6\endcsname{\color[rgb]{0,0,0}}%
      \expandafter\def\csname LT7\endcsname{\color[rgb]{1,0.3,0}}%
      \expandafter\def\csname LT8\endcsname{\color[rgb]{0.5,0.5,0.5}}%
    \else
      \def\colorrgb#1{\color{black}}%
      \def\colorgray#1{\color[gray]{#1}}%
      \expandafter\def\csname LTw\endcsname{\color{white}}%
      \expandafter\def\csname LTb\endcsname{\color{black}}%
      \expandafter\def\csname LTa\endcsname{\color{black}}%
      \expandafter\def\csname LT0\endcsname{\color{black}}%
      \expandafter\def\csname LT1\endcsname{\color{black}}%
      \expandafter\def\csname LT2\endcsname{\color{black}}%
      \expandafter\def\csname LT3\endcsname{\color{black}}%
      \expandafter\def\csname LT4\endcsname{\color{black}}%
      \expandafter\def\csname LT5\endcsname{\color{black}}%
      \expandafter\def\csname LT6\endcsname{\color{black}}%
      \expandafter\def\csname LT7\endcsname{\color{black}}%
      \expandafter\def\csname LT8\endcsname{\color{black}}%
    \fi
  \fi
  \setlength{\unitlength}{0.0500bp}%
  \begin{picture}(7936.00,3400.00)%
    \gplgaddtomacro\gplbacktext{%
      \csname LTb\endcsname%
      \put(682,704){\makebox(0,0)[r]{\strut{}-8}}%
      \put(682,1008){\makebox(0,0)[r]{\strut{}-6}}%
      \put(682,1312){\makebox(0,0)[r]{\strut{}-4}}%
      \put(682,1616){\makebox(0,0)[r]{\strut{}-2}}%
      \put(682,1920){\makebox(0,0)[r]{\strut{} 0}}%
      \put(682,2223){\makebox(0,0)[r]{\strut{} 2}}%
      \put(682,2527){\makebox(0,0)[r]{\strut{} 4}}%
      \put(682,2831){\makebox(0,0)[r]{\strut{} 6}}%
      \put(682,3135){\makebox(0,0)[r]{\strut{} 8}}%
      \put(814,484){\makebox(0,0){\strut{} 0}}%
      \put(1365,484){\makebox(0,0){\strut{} 0.2}}%
      \put(1917,484){\makebox(0,0){\strut{} 0.4}}%
      \put(2468,484){\makebox(0,0){\strut{} 0.6}}%
      \put(3020,484){\makebox(0,0){\strut{} 0.8}}%
      \put(3571,484){\makebox(0,0){\strut{} 1}}%
      \put(176,1919){\rotatebox{-270}{\makebox(0,0){\strut{}$\ln{\left(n^{tot}_0+1\right)}$}}}%
      \put(2192,154){\makebox(0,0){\strut{}$t$ [$N$]}}%
    }%
    \gplgaddtomacro\gplfronttext{%
      \csname LTb\endcsname%
      \put(2584,1757){\makebox(0,0)[r]{\strut{}\small{$(0.2; -0.05)$}}}%
      \csname LTb\endcsname%
      \put(2584,1537){\makebox(0,0)[r]{\strut{}\small{$(0.2; -0.5)$}}}%
      \csname LTb\endcsname%
      \put(2584,1317){\makebox(0,0)[r]{\strut{}\small{$(0; 0.05)$}}}%
      \csname LTb\endcsname%
      \put(2584,1097){\makebox(0,0)[r]{\strut{}\small{$(0; 0.5)$}}}%
      \csname LTb\endcsname%
      \put(2584,877){\makebox(0,0)[r]{\strut{}\small{$(0.2; 0)$}}}%
    }%
    \gplgaddtomacro\gplbacktext{%
      \csname LTb\endcsname%
      \put(4914,704){\makebox(0,0)[r]{\strut{}-160}}%
      \put(4914,1008){\makebox(0,0)[r]{\strut{}-140}}%
      \put(4914,1312){\makebox(0,0)[r]{\strut{}-120}}%
      \put(4914,1616){\makebox(0,0)[r]{\strut{}-100}}%
      \put(4914,1920){\makebox(0,0)[r]{\strut{}-80}}%
      \put(4914,2223){\makebox(0,0)[r]{\strut{}-60}}%
      \put(4914,2527){\makebox(0,0)[r]{\strut{}-40}}%
      \put(4914,2831){\makebox(0,0)[r]{\strut{}-20}}%
      \put(4914,3135){\makebox(0,0)[r]{\strut{} 0}}%
      \put(5545,484){\makebox(0,0){\strut{} 5}}%
      \put(6168,484){\makebox(0,0){\strut{} 10}}%
      \put(6791,484){\makebox(0,0){\strut{} 15}}%
      \put(7414,484){\makebox(0,0){\strut{} 20}}%
      \put(4144,1919){\rotatebox{-270}{\makebox(0,0){\strut{}$\ln{\left[\mathcal{F}(\mathcal{N},0.5)\right]}$}}}%
      \put(6292,154){\makebox(0,0){\strut{}$\mathcal{N}$}}%
    }%
    \gplgaddtomacro\gplfronttext{%
    }%
    \gplbacktext
    \put(0,0){\includegraphics{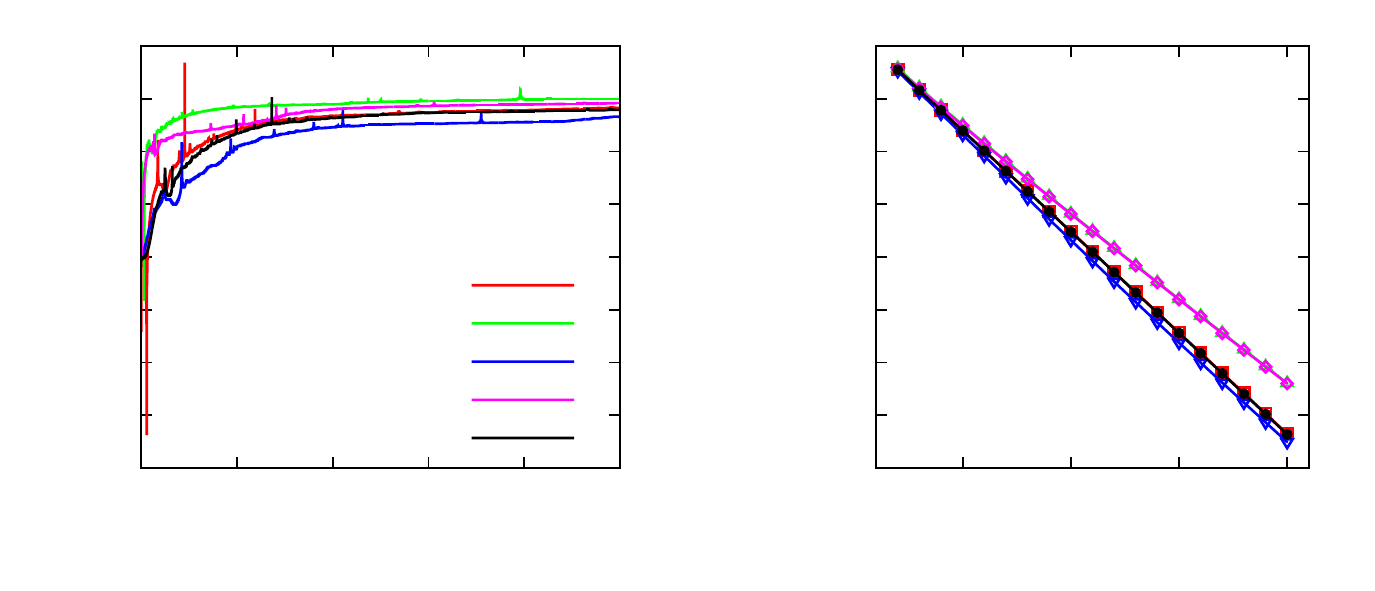}}%
    \gplfronttext
  \end{picture}%
\endgroup

%% file: occ_number2.tex
\begingroup
  \makeatletter
  \providecommand\color[2][]{%
    \GenericError{(gnuplot) \space\space\space\@spaces}{%
      Package color not loaded in conjunction with
      terminal option `colourtext'%
    }{See the gnuplot documentation for explanation.%
    }{Either use 'blacktext' in gnuplot or load the package
      color.sty in LaTeX.}%
    \renewcommand\color[2][]{}%
  }%
  \providecommand\includegraphics[2][]{%
    \GenericError{(gnuplot) \space\space\space\@spaces}{%
      Package graphicx or graphics not loaded%
    }{See the gnuplot documentation for explanation.%
    }{The gnuplot epslatex terminal needs graphicx.sty or graphics.sty.}%
    \renewcommand\includegraphics[2][]{}%
  }%
  \providecommand\rotatebox[2]{#2}%
  \@ifundefined{ifGPcolor}{%
    \newif\ifGPcolor
    \GPcolortrue
  }{}%
  \@ifundefined{ifGPblacktext}{%
    \newif\ifGPblacktext
    \GPblacktexttrue
  }{}%
  \let\gplgaddtomacro\g@addto@macro
  \gdef\gplbacktext{}%
  \gdef\gplfronttext{}%
  \makeatother
  \ifGPblacktext
    \def\colorrgb#1{}%
    \def\colorgray#1{}%
  \else
    \ifGPcolor
      \def\colorrgb#1{\color[rgb]{#1}}%
      \def\colorgray#1{\color[gray]{#1}}%
      \expandafter\def\csname LTw\endcsname{\color{white}}%
      \expandafter\def\csname LTb\endcsname{\color{black}}%
      \expandafter\def\csname LTa\endcsname{\color{black}}%
      \expandafter\def\csname LT0\endcsname{\color[rgb]{1,0,0}}%
      \expandafter\def\csname LT1\endcsname{\color[rgb]{0,1,0}}%
      \expandafter\def\csname LT2\endcsname{\color[rgb]{0,0,1}}%
      \expandafter\def\csname LT3\endcsname{\color[rgb]{1,0,1}}%
      \expandafter\def\csname LT4\endcsname{\color[rgb]{0,1,1}}%
      \expandafter\def\csname LT5\endcsname{\color[rgb]{1,1,0}}%
      \expandafter\def\csname LT6\endcsname{\color[rgb]{0,0,0}}%
      \expandafter\def\csname LT7\endcsname{\color[rgb]{1,0.3,0}}%
      \expandafter\def\csname LT8\endcsname{\color[rgb]{0.5,0.5,0.5}}%
    \else
      \def\colorrgb#1{\color{black}}%
      \def\colorgray#1{\color[gray]{#1}}%
      \expandafter\def\csname LTw\endcsname{\color{white}}%
      \expandafter\def\csname LTb\endcsname{\color{black}}%
      \expandafter\def\csname LTa\endcsname{\color{black}}%
      \expandafter\def\csname LT0\endcsname{\color{black}}%
      \expandafter\def\csname LT1\endcsname{\color{black}}%
      \expandafter\def\csname LT2\endcsname{\color{black}}%
      \expandafter\def\csname LT3\endcsname{\color{black}}%
      \expandafter\def\csname LT4\endcsname{\color{black}}%
      \expandafter\def\csname LT5\endcsname{\color{black}}%
      \expandafter\def\csname LT6\endcsname{\color{black}}%
      \expandafter\def\csname LT7\endcsname{\color{black}}%
      \expandafter\def\csname LT8\endcsname{\color{black}}%
    \fi
  \fi
  \setlength{\unitlength}{0.0500bp}%
  \begin{picture}(7936.00,5102.00)%
    \gplgaddtomacro\gplbacktext{%
      \csname LTb\endcsname%
      \put(946,3299){\makebox(0,0)[r]{\strut{} 0}}%
      \put(946,3519){\makebox(0,0)[r]{\strut{} 50}}%
      \put(946,3738){\makebox(0,0)[r]{\strut{} 100}}%
      \put(946,3958){\makebox(0,0)[r]{\strut{} 150}}%
      \put(946,4178){\makebox(0,0)[r]{\strut{} 200}}%
      \put(946,4398){\makebox(0,0)[r]{\strut{} 250}}%
      \put(946,4617){\makebox(0,0)[r]{\strut{} 300}}%
      \put(946,4837){\makebox(0,0)[r]{\strut{} 350}}%
      \put(1078,3035){\makebox(0,0){\strut{} 0}}%
      \put(1826,3035){\makebox(0,0){\strut{} 0.0015}}%
      \put(2574,3035){\makebox(0,0){\strut{} 0.003}}%
      \put(3322,3035){\makebox(0,0){\strut{} 0.0045}}%
      \put(176,4046){\rotatebox{-270}{\makebox(0,0){\strut{}$\ln{\left(n_0(t_{end})\right)}$}}}%
      \put(2324,2705){\makebox(0,0){\strut{}$x$ [$M_P$]}}%
    }%
    \gplgaddtomacro\gplfronttext{%
      \csname LTb\endcsname%
      \put(2584,4664){\makebox(0,0)[r]{\strut{}$\mathcal{N} = 2$}}%
    }%
    \gplgaddtomacro\gplbacktext{%
      \csname LTb\endcsname%
      \put(4914,3330){\makebox(0,0)[r]{\strut{} 0}}%
      \put(4914,3707){\makebox(0,0)[r]{\strut{} 50}}%
      \put(4914,4084){\makebox(0,0)[r]{\strut{} 100}}%
      \put(4914,4460){\makebox(0,0)[r]{\strut{} 150}}%
      \put(4914,4837){\makebox(0,0)[r]{\strut{} 200}}%
      \put(5046,3035){\makebox(0,0){\strut{} 0}}%
      \put(5794,3035){\makebox(0,0){\strut{} 0.0015}}%
      \put(6542,3035){\makebox(0,0){\strut{} 0.003}}%
      \put(7290,3035){\makebox(0,0){\strut{} 0.0045}}%
      \put(4144,4046){\rotatebox{-270}{\makebox(0,0){\strut{}$\ln{\left(n_0(t_{end})\right)}$}}}%
      \put(6292,2705){\makebox(0,0){\strut{}$x$ [$M_P$]}}%
    }%
    \gplgaddtomacro\gplfronttext{%
      \csname LTb\endcsname%
      \put(6552,4664){\makebox(0,0)[r]{\strut{}$\mathcal{N} = 3$}}%
    }%
    \gplgaddtomacro\gplbacktext{%
      \csname LTb\endcsname%
      \put(946,735){\makebox(0,0)[r]{\strut{} 0}}%
      \put(946,1045){\makebox(0,0)[r]{\strut{} 100}}%
      \put(946,1356){\makebox(0,0)[r]{\strut{} 200}}%
      \put(946,1666){\makebox(0,0)[r]{\strut{} 300}}%
      \put(946,1977){\makebox(0,0)[r]{\strut{} 400}}%
      \put(946,2287){\makebox(0,0)[r]{\strut{} 500}}%
      \put(1078,484){\makebox(0,0){\strut{} 0}}%
      \put(1826,484){\makebox(0,0){\strut{} 0.0015}}%
      \put(2574,484){\makebox(0,0){\strut{} 0.003}}%
      \put(3322,484){\makebox(0,0){\strut{} 0.0045}}%
      \put(176,1495){\rotatebox{-270}{\makebox(0,0){\strut{}$\ln{\left(n_0(t_{end})\right)}$}}}%
      \put(2324,154){\makebox(0,0){\strut{}$x$ [$M_P$]}}%
    }%
    \gplgaddtomacro\gplfronttext{%
      \csname LTb\endcsname%
      \put(2584,2114){\makebox(0,0)[r]{\strut{}$\mathcal{N} = 2$}}%
    }%
    \gplgaddtomacro\gplbacktext{%
      \csname LTb\endcsname%
      \put(4914,755){\makebox(0,0)[r]{\strut{} 0}}%
      \put(4914,1010){\makebox(0,0)[r]{\strut{} 50}}%
      \put(4914,1266){\makebox(0,0)[r]{\strut{} 100}}%
      \put(4914,1521){\makebox(0,0)[r]{\strut{} 150}}%
      \put(4914,1776){\makebox(0,0)[r]{\strut{} 200}}%
      \put(4914,2032){\makebox(0,0)[r]{\strut{} 250}}%
      \put(4914,2287){\makebox(0,0)[r]{\strut{} 300}}%
      \put(5046,484){\makebox(0,0){\strut{} 0}}%
      \put(5794,484){\makebox(0,0){\strut{} 0.0015}}%
      \put(6542,484){\makebox(0,0){\strut{} 0.003}}%
      \put(7290,484){\makebox(0,0){\strut{} 0.0045}}%
      \put(4144,1495){\rotatebox{-270}{\makebox(0,0){\strut{}$\ln{\left(n_0(t_{end})\right)}$}}}%
      \put(6292,154){\makebox(0,0){\strut{}$x$ [$M_P$]}}%
    }%
    \gplgaddtomacro\gplfronttext{%
      \csname LTb\endcsname%
      \put(6552,2114){\makebox(0,0)[r]{\strut{}$\mathcal{N} = 3$}}%
    }%
    \gplbacktext
    \put(0,0){\includegraphics{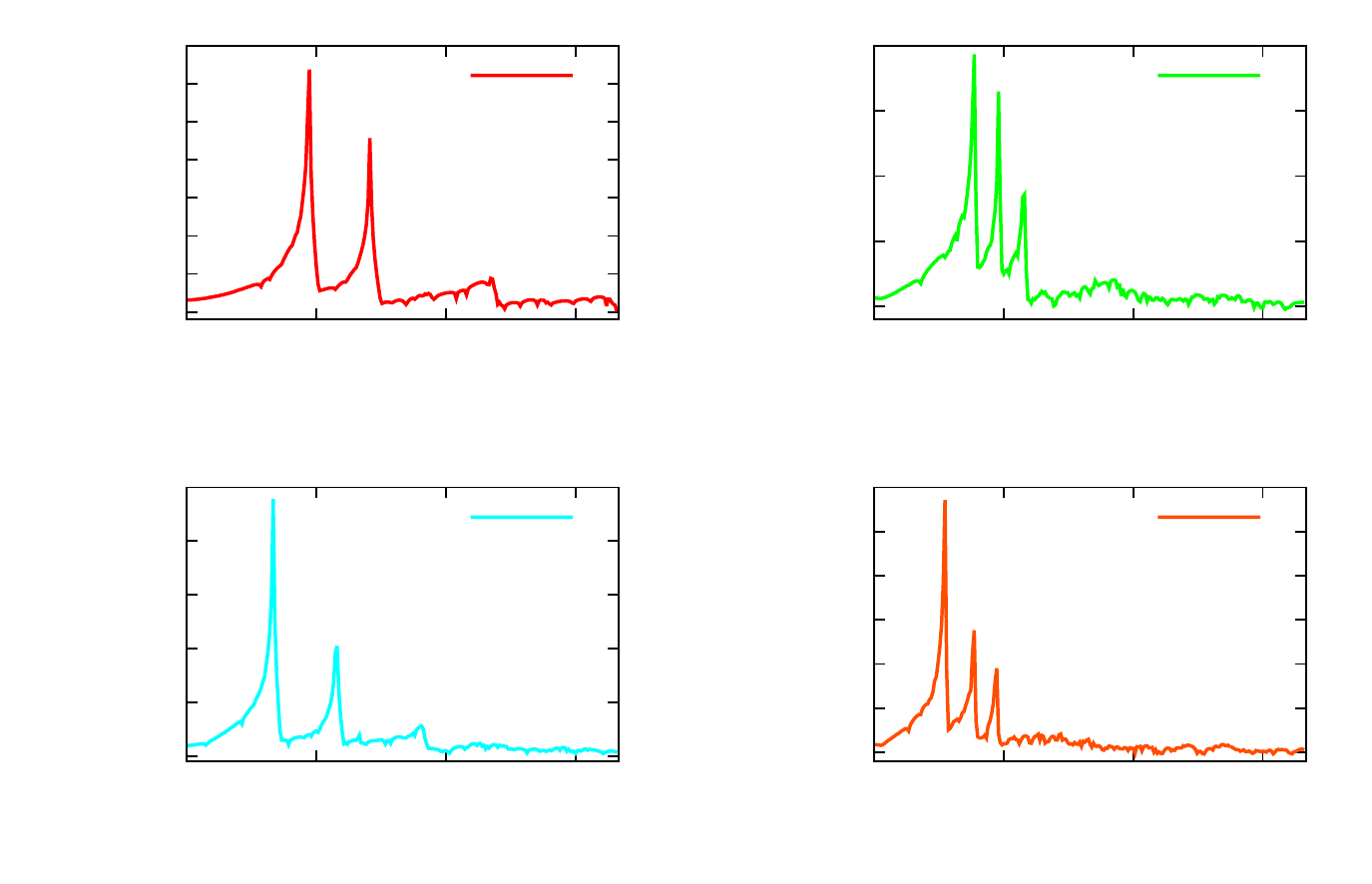}}%
    \gplfronttext
  \end{picture}%
\endgroup

%% file: iband.tex
\begingroup
  \makeatletter
  \providecommand\color[2][]{%
    \GenericError{(gnuplot) \space\space\space\@spaces}{%
      Package color not loaded in conjunction with
      terminal option `colourtext'%
    }{See the gnuplot documentation for explanation.%
    }{Either use 'blacktext' in gnuplot or load the package
      color.sty in LaTeX.}%
    \renewcommand\color[2][]{}%
  }%
  \providecommand\includegraphics[2][]{%
    \GenericError{(gnuplot) \space\space\space\@spaces}{%
      Package graphicx or graphics not loaded%
    }{See the gnuplot documentation for explanation.%
    }{The gnuplot epslatex terminal needs graphicx.sty or graphics.sty.}%
    \renewcommand\includegraphics[2][]{}%
  }%
  \providecommand\rotatebox[2]{#2}%
  \@ifundefined{ifGPcolor}{%
    \newif\ifGPcolor
    \GPcolortrue
  }{}%
  \@ifundefined{ifGPblacktext}{%
    \newif\ifGPblacktext
    \GPblacktexttrue
  }{}%
  \let\gplgaddtomacro\g@addto@macro
  \gdef\gplbacktext{}%
  \gdef\gplfronttext{}%
  \makeatother
  \ifGPblacktext
    \def\colorrgb#1{}%
    \def\colorgray#1{}%
  \else
    \ifGPcolor
      \def\colorrgb#1{\color[rgb]{#1}}%
      \def\colorgray#1{\color[gray]{#1}}%
      \expandafter\def\csname LTw\endcsname{\color{white}}%
      \expandafter\def\csname LTb\endcsname{\color{black}}%
      \expandafter\def\csname LTa\endcsname{\color{black}}%
      \expandafter\def\csname LT0\endcsname{\color[rgb]{1,0,0}}%
      \expandafter\def\csname LT1\endcsname{\color[rgb]{0,1,0}}%
      \expandafter\def\csname LT2\endcsname{\color[rgb]{0,0,1}}%
      \expandafter\def\csname LT3\endcsname{\color[rgb]{1,0,1}}%
      \expandafter\def\csname LT4\endcsname{\color[rgb]{0,1,1}}%
      \expandafter\def\csname LT5\endcsname{\color[rgb]{1,1,0}}%
      \expandafter\def\csname LT6\endcsname{\color[rgb]{0,0,0}}%
      \expandafter\def\csname LT7\endcsname{\color[rgb]{1,0.3,0}}%
      \expandafter\def\csname LT8\endcsname{\color[rgb]{0.5,0.5,0.5}}%
    \else
      \def\colorrgb#1{\color{black}}%
      \def\colorgray#1{\color[gray]{#1}}%
      \expandafter\def\csname LTw\endcsname{\color{white}}%
      \expandafter\def\csname LTb\endcsname{\color{black}}%
      \expandafter\def\csname LTa\endcsname{\color{black}}%
      \expandafter\def\csname LT0\endcsname{\color{black}}%
      \expandafter\def\csname LT1\endcsname{\color{black}}%
      \expandafter\def\csname LT2\endcsname{\color{black}}%
      \expandafter\def\csname LT3\endcsname{\color{black}}%
      \expandafter\def\csname LT4\endcsname{\color{black}}%
      \expandafter\def\csname LT5\endcsname{\color{black}}%
      \expandafter\def\csname LT6\endcsname{\color{black}}%
      \expandafter\def\csname LT7\endcsname{\color{black}}%
      \expandafter\def\csname LT8\endcsname{\color{black}}%
    \fi
  \fi
  \setlength{\unitlength}{0.0500bp}%
  \begin{picture}(7936.00,5668.00)%
    \gplgaddtomacro\gplbacktext{%
      \csname LTb\endcsname%
      \put(1078,3538){\makebox(0,0)[r]{\strut{}-0.05}}%
      \put(1078,3784){\makebox(0,0)[r]{\strut{} 0}}%
      \put(1078,4030){\makebox(0,0)[r]{\strut{} 0.05}}%
      \put(1078,4276){\makebox(0,0)[r]{\strut{} 0.1}}%
      \put(1078,4522){\makebox(0,0)[r]{\strut{} 0.15}}%
      \put(1078,4768){\makebox(0,0)[r]{\strut{} 0.2}}%
      \put(1078,5014){\makebox(0,0)[r]{\strut{} 0.25}}%
      \put(1210,3318){\makebox(0,0){\strut{} 0}}%
      \put(2529,3318){\makebox(0,0){\strut{} 0.0005}}%
      \put(3847,3318){\makebox(0,0){\strut{} 0.001}}%
      \put(5166,3318){\makebox(0,0){\strut{} 0.0015}}%
      \put(6484,3318){\makebox(0,0){\strut{} 0.002}}%
      \put(176,4300){\rotatebox{-270}{\makebox(0,0){\strut{}$\mu$}}}%
      \put(4374,2988){\makebox(0,0){\strut{}$\phi_0$ [$M_P$]}}%
    }%
    \gplgaddtomacro\gplfronttext{%
    }%
    \gplgaddtomacro\gplbacktext{%
      \csname LTb\endcsname%
      \put(1078,1144){\makebox(0,0)[r]{\strut{}-0.05}}%
      \put(1078,1401){\makebox(0,0)[r]{\strut{} 0}}%
      \put(1078,1659){\makebox(0,0)[r]{\strut{} 0.05}}%
      \put(1078,1916){\makebox(0,0)[r]{\strut{} 0.1}}%
      \put(1078,2174){\makebox(0,0)[r]{\strut{} 0.15}}%
      \put(1078,2431){\makebox(0,0)[r]{\strut{} 0.2}}%
      \put(1078,2689){\makebox(0,0)[r]{\strut{} 0.25}}%
      \put(1511,924){\makebox(0,0){\strut{} 10}}%
      \put(3018,924){\makebox(0,0){\strut{} 15}}%
      \put(4525,924){\makebox(0,0){\strut{} 20}}%
      \put(6032,924){\makebox(0,0){\strut{} 25}}%
      \put(7539,924){\makebox(0,0){\strut{} 30}}%
      \put(176,1942){\rotatebox{-270}{\makebox(0,0){\strut{}$\mu$}}}%
      \put(4374,594){\makebox(0,0){\strut{}$t$ [$N$]}}%
    }%
    \gplgaddtomacro\gplfronttext{%
      \csname LTb\endcsname%
      \put(2432,173){\makebox(0,0)[r]{\strut{}$x=0.0$}}%
      \csname LTb\endcsname%
      \put(4607,173){\makebox(0,0)[r]{\strut{}$x=0.000115$}}%
      \csname LTb\endcsname%
      \put(6782,173){\makebox(0,0)[r]{\strut{}$x=0.000515$}}%
    }%
    \gplbacktext
    \put(0,0){\includegraphics{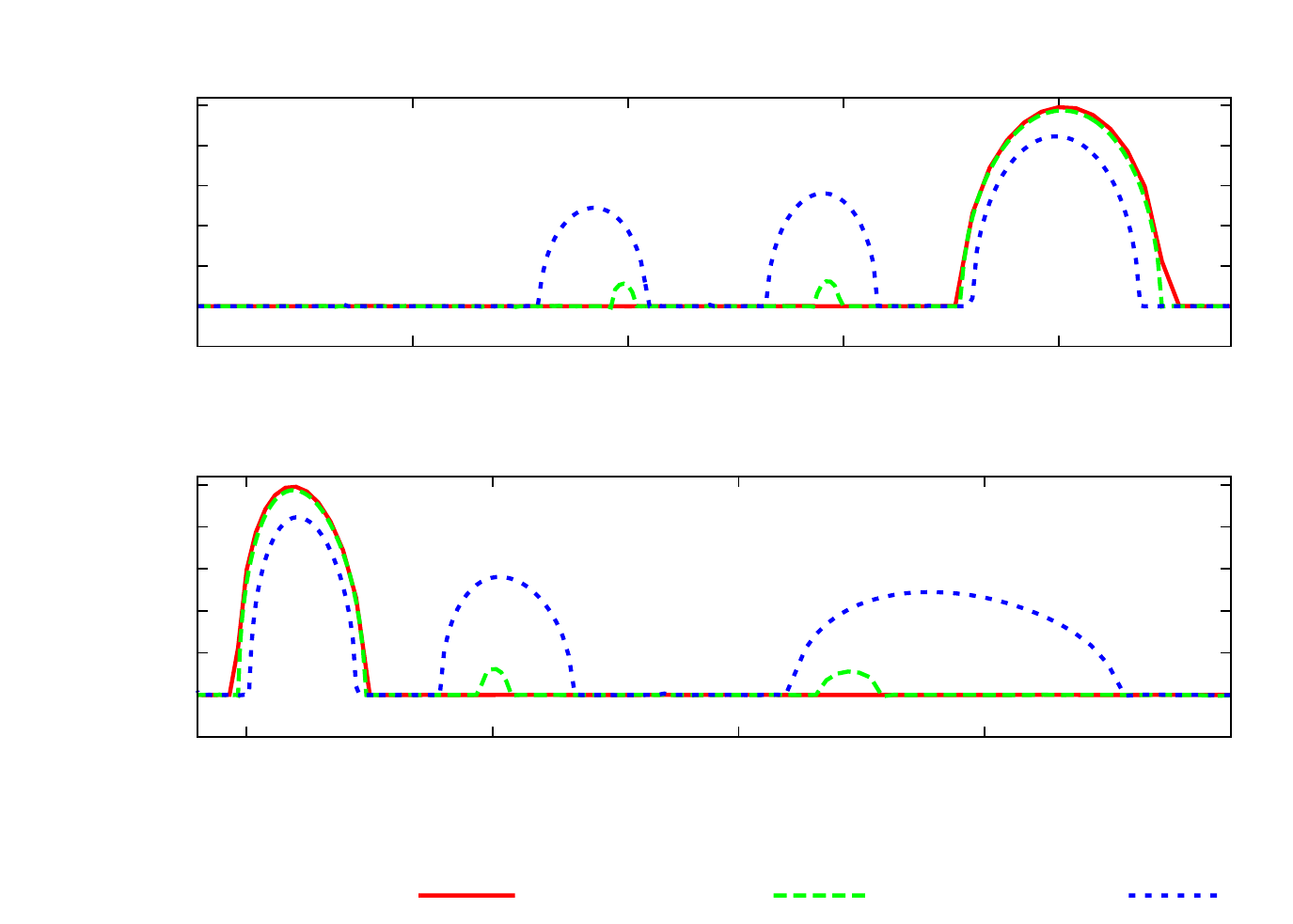}}%
    \gplfronttext
  \end{picture}%
\endgroup

%% file: first_band.tex
\begingroup
  \makeatletter
  \providecommand\color[2][]{%
    \GenericError{(gnuplot) \space\space\space\@spaces}{%
      Package color not loaded in conjunction with
      terminal option `colourtext'%
    }{See the gnuplot documentation for explanation.%
    }{Either use 'blacktext' in gnuplot or load the package
      color.sty in LaTeX.}%
    \renewcommand\color[2][]{}%
  }%
  \providecommand\includegraphics[2][]{%
    \GenericError{(gnuplot) \space\space\space\@spaces}{%
      Package graphicx or graphics not loaded%
    }{See the gnuplot documentation for explanation.%
    }{The gnuplot epslatex terminal needs graphicx.sty or graphics.sty.}%
    \renewcommand\includegraphics[2][]{}%
  }%
  \providecommand\rotatebox[2]{#2}%
  \@ifundefined{ifGPcolor}{%
    \newif\ifGPcolor
    \GPcolortrue
  }{}%
  \@ifundefined{ifGPblacktext}{%
    \newif\ifGPblacktext
    \GPblacktexttrue
  }{}%
  \let\gplgaddtomacro\g@addto@macro
  \gdef\gplbacktext{}%
  \gdef\gplfronttext{}%
  \makeatother
  \ifGPblacktext
    \def\colorrgb#1{}%
    \def\colorgray#1{}%
  \else
    \ifGPcolor
      \def\colorrgb#1{\color[rgb]{#1}}%
      \def\colorgray#1{\color[gray]{#1}}%
      \expandafter\def\csname LTw\endcsname{\color{white}}%
      \expandafter\def\csname LTb\endcsname{\color{black}}%
      \expandafter\def\csname LTa\endcsname{\color{black}}%
      \expandafter\def\csname LT0\endcsname{\color[rgb]{1,0,0}}%
      \expandafter\def\csname LT1\endcsname{\color[rgb]{0,1,0}}%
      \expandafter\def\csname LT2\endcsname{\color[rgb]{0,0,1}}%
      \expandafter\def\csname LT3\endcsname{\color[rgb]{1,0,1}}%
      \expandafter\def\csname LT4\endcsname{\color[rgb]{0,1,1}}%
      \expandafter\def\csname LT5\endcsname{\color[rgb]{1,1,0}}%
      \expandafter\def\csname LT6\endcsname{\color[rgb]{0,0,0}}%
      \expandafter\def\csname LT7\endcsname{\color[rgb]{1,0.3,0}}%
      \expandafter\def\csname LT8\endcsname{\color[rgb]{0.5,0.5,0.5}}%
    \else
      \def\colorrgb#1{\color{black}}%
      \def\colorgray#1{\color[gray]{#1}}%
      \expandafter\def\csname LTw\endcsname{\color{white}}%
      \expandafter\def\csname LTb\endcsname{\color{black}}%
      \expandafter\def\csname LTa\endcsname{\color{black}}%
      \expandafter\def\csname LT0\endcsname{\color{black}}%
      \expandafter\def\csname LT1\endcsname{\color{black}}%
      \expandafter\def\csname LT2\endcsname{\color{black}}%
      \expandafter\def\csname LT3\endcsname{\color{black}}%
      \expandafter\def\csname LT4\endcsname{\color{black}}%
      \expandafter\def\csname LT5\endcsname{\color{black}}%
      \expandafter\def\csname LT6\endcsname{\color{black}}%
      \expandafter\def\csname LT7\endcsname{\color{black}}%
      \expandafter\def\csname LT8\endcsname{\color{black}}%
    \fi
  \fi
  \setlength{\unitlength}{0.0500bp}%
  \begin{picture}(7936.00,11338.00)%
    \gplgaddtomacro\gplbacktext{%
      \csname LTb\endcsname%
      \put(1078,9207){\makebox(0,0)[r]{\strut{}-0.05}}%
      \csname LTb\endcsname%
      \put(1078,9674){\makebox(0,0)[r]{\strut{} 0}}%
      \csname LTb\endcsname%
      \put(1078,10140){\makebox(0,0)[r]{\strut{} 0.05}}%
      \csname LTb\endcsname%
      \put(1078,10607){\makebox(0,0)[r]{\strut{} 0.1}}%
      \csname LTb\endcsname%
      \put(1078,11073){\makebox(0,0)[r]{\strut{} 0.15}}%
      \csname LTb\endcsname%
      \put(1210,8987){\makebox(0,0){\strut{} 0}}%
      \csname LTb\endcsname%
      \put(2282,8987){\makebox(0,0){\strut{} 0.0002}}%
      \csname LTb\endcsname%
      \put(3353,8987){\makebox(0,0){\strut{} 0.0004}}%
      \csname LTb\endcsname%
      \put(4425,8987){\makebox(0,0){\strut{} 0.0006}}%
      \csname LTb\endcsname%
      \put(5496,8987){\makebox(0,0){\strut{} 0.0008}}%
      \put(176,10140){\rotatebox{-270}{\makebox(0,0){\strut{}$\mu$}}}%
      \put(3353,8657){\makebox(0,0){\strut{}$\varphi_0$ [$M_P$]}}%
    }%
    \gplgaddtomacro\gplfronttext{%
      \csname LTb\endcsname%
      \put(6948,10470){\makebox(0,0)[r]{\strut{}$x=0.0013$}}%
      \csname LTb\endcsname%
      \put(6948,10250){\makebox(0,0)[r]{\strut{}$x=0.00135$}}%
      \csname LTb\endcsname%
      \put(6948,10030){\makebox(0,0)[r]{\strut{}$x=0.0014$}}%
      \csname LTb\endcsname%
      \put(6948,9810){\makebox(0,0)[r]{\strut{}$x=0.00145$}}%
    }%
    \gplgaddtomacro\gplbacktext{%
      \csname LTb\endcsname%
      \put(1078,6373){\makebox(0,0)[r]{\strut{}-0.05}}%
      \csname LTb\endcsname%
      \put(1078,6684){\makebox(0,0)[r]{\strut{} 0}}%
      \csname LTb\endcsname%
      \put(1078,6995){\makebox(0,0)[r]{\strut{} 0.05}}%
      \csname LTb\endcsname%
      \put(1078,7306){\makebox(0,0)[r]{\strut{} 0.1}}%
      \csname LTb\endcsname%
      \put(1078,7617){\makebox(0,0)[r]{\strut{} 0.15}}%
      \csname LTb\endcsname%
      \put(1078,7928){\makebox(0,0)[r]{\strut{} 0.2}}%
      \csname LTb\endcsname%
      \put(1078,8239){\makebox(0,0)[r]{\strut{} 0.25}}%
      \csname LTb\endcsname%
      \put(1210,6153){\makebox(0,0){\strut{} 0}}%
      \csname LTb\endcsname%
      \put(2034,6153){\makebox(0,0){\strut{} 0.00025}}%
      \csname LTb\endcsname%
      \put(2858,6153){\makebox(0,0){\strut{} 0.0005}}%
      \csname LTb\endcsname%
      \put(3683,6153){\makebox(0,0){\strut{} 0.00075}}%
      \csname LTb\endcsname%
      \put(4507,6153){\makebox(0,0){\strut{} 0.001}}%
      \csname LTb\endcsname%
      \put(5331,6153){\makebox(0,0){\strut{} 0.00125}}%
      \put(176,7306){\rotatebox{-270}{\makebox(0,0){\strut{}$\mu$}}}%
      \put(3353,5823){\makebox(0,0){\strut{}$\varphi_0$ [$M_P$]}}%
    }%
    \gplgaddtomacro\gplfronttext{%
      \csname LTb\endcsname%
      \put(6948,7636){\makebox(0,0)[r]{\strut{}$x=0.002$}}%
      \csname LTb\endcsname%
      \put(6948,7416){\makebox(0,0)[r]{\strut{}$x=0.00205$}}%
      \csname LTb\endcsname%
      \put(6948,7196){\makebox(0,0)[r]{\strut{}$x=0.00211$}}%
      \csname LTb\endcsname%
      \put(6948,6976){\makebox(0,0)[r]{\strut{}$x=0.00215$}}%
    }%
    \gplgaddtomacro\gplbacktext{%
      \csname LTb\endcsname%
      \put(1078,3538){\makebox(0,0)[r]{\strut{}-0.02}}%
      \csname LTb\endcsname%
      \put(1078,4160){\makebox(0,0)[r]{\strut{} 0}}%
      \csname LTb\endcsname%
      \put(1078,4783){\makebox(0,0)[r]{\strut{} 0.02}}%
      \csname LTb\endcsname%
      \put(1078,5405){\makebox(0,0)[r]{\strut{} 0.04}}%
      \csname LTb\endcsname%
      \put(1210,3318){\makebox(0,0){\strut{} 0}}%
      \csname LTb\endcsname%
      \put(1989,3318){\makebox(0,0){\strut{} 0.0002}}%
      \csname LTb\endcsname%
      \put(2769,3318){\makebox(0,0){\strut{} 0.0004}}%
      \csname LTb\endcsname%
      \put(3548,3318){\makebox(0,0){\strut{} 0.0006}}%
      \csname LTb\endcsname%
      \put(4327,3318){\makebox(0,0){\strut{} 0.0008}}%
      \csname LTb\endcsname%
      \put(5106,3318){\makebox(0,0){\strut{} 0.001}}%
      \put(176,4471){\rotatebox{-270}{\makebox(0,0){\strut{}$\mu$}}}%
      \put(3353,2988){\makebox(0,0){\strut{}$\varphi_0$ [$M_P$]}}%
    }%
    \gplgaddtomacro\gplfronttext{%
      \csname LTb\endcsname%
      \put(6948,4801){\makebox(0,0)[r]{\strut{}$x=0.0027$}}%
      \csname LTb\endcsname%
      \put(6948,4581){\makebox(0,0)[r]{\strut{}$x=0.00275$}}%
      \csname LTb\endcsname%
      \put(6948,4361){\makebox(0,0)[r]{\strut{}$x=0.0028$}}%
      \csname LTb\endcsname%
      \put(6948,4141){\makebox(0,0)[r]{\strut{}$x=0.00285$}}%
    }%
    \gplgaddtomacro\gplbacktext{%
      \csname LTb\endcsname%
      \put(1078,704){\makebox(0,0)[r]{\strut{}-0.05}}%
      \csname LTb\endcsname%
      \put(1078,1015){\makebox(0,0)[r]{\strut{} 0}}%
      \csname LTb\endcsname%
      \put(1078,1326){\makebox(0,0)[r]{\strut{} 0.05}}%
      \csname LTb\endcsname%
      \put(1078,1637){\makebox(0,0)[r]{\strut{} 0.1}}%
      \csname LTb\endcsname%
      \put(1078,1948){\makebox(0,0)[r]{\strut{} 0.15}}%
      \csname LTb\endcsname%
      \put(1078,2259){\makebox(0,0)[r]{\strut{} 0.2}}%
      \csname LTb\endcsname%
      \put(1078,2570){\makebox(0,0)[r]{\strut{} 0.25}}%
      \csname LTb\endcsname%
      \put(1210,484){\makebox(0,0){\strut{} 0}}%
      \csname LTb\endcsname%
      \put(2282,484){\makebox(0,0){\strut{} 0.0005}}%
      \csname LTb\endcsname%
      \put(3353,484){\makebox(0,0){\strut{} 0.001}}%
      \csname LTb\endcsname%
      \put(4425,484){\makebox(0,0){\strut{} 0.0015}}%
      \csname LTb\endcsname%
      \put(5496,484){\makebox(0,0){\strut{} 0.002}}%
      \put(176,1637){\rotatebox{-270}{\makebox(0,0){\strut{}$\mu$}}}%
      \put(3353,154){\makebox(0,0){\strut{}$\varphi_0$ [$M_P$]}}%
    }%
    \gplgaddtomacro\gplfronttext{%
      \csname LTb\endcsname%
      \put(6948,1967){\makebox(0,0)[r]{\strut{}$x=0.00345$}}%
      \csname LTb\endcsname%
      \put(6948,1747){\makebox(0,0)[r]{\strut{}$x=0.0035$}}%
      \csname LTb\endcsname%
      \put(6948,1527){\makebox(0,0)[r]{\strut{}$x=0.00354$}}%
      \csname LTb\endcsname%
      \put(6948,1307){\makebox(0,0)[r]{\strut{}$x=0.0036$}}%
    }%
    \gplbacktext
    \put(0,0){\includegraphics{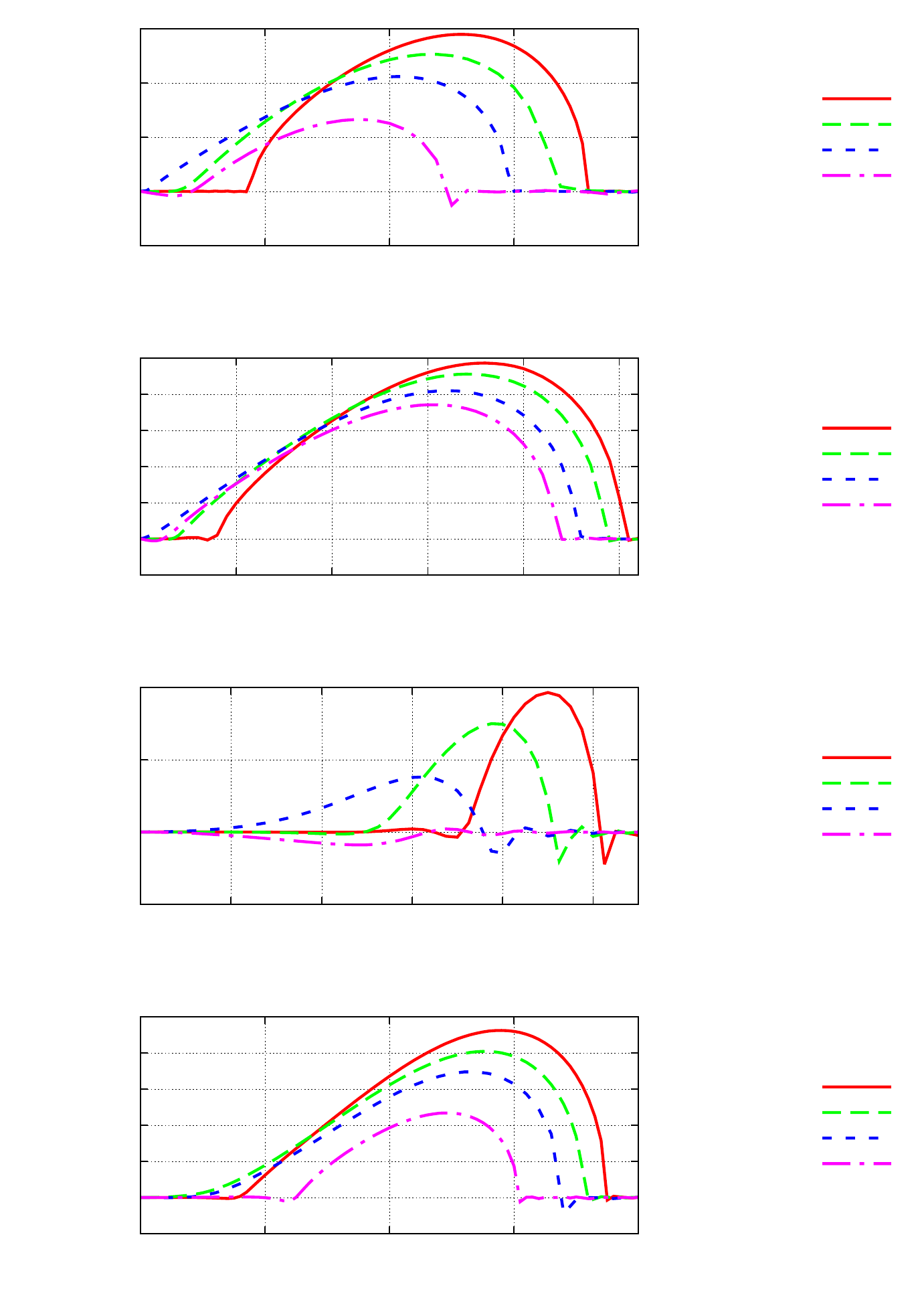}}%
    \gplfronttext
  \end{picture}%
\endgroup

%% file: mass_dep.tex
\begingroup
  \makeatletter
  \providecommand\color[2][]{%
    \GenericError{(gnuplot) \space\space\space\@spaces}{%
      Package color not loaded in conjunction with
      terminal option `colourtext'%
    }{See the gnuplot documentation for explanation.%
    }{Either use 'blacktext' in gnuplot or load the package
      color.sty in LaTeX.}%
    \renewcommand\color[2][]{}%
  }%
  \providecommand\includegraphics[2][]{%
    \GenericError{(gnuplot) \space\space\space\@spaces}{%
      Package graphicx or graphics not loaded%
    }{See the gnuplot documentation for explanation.%
    }{The gnuplot epslatex terminal needs graphicx.sty or graphics.sty.}%
    \renewcommand\includegraphics[2][]{}%
  }%
  \providecommand\rotatebox[2]{#2}%
  \@ifundefined{ifGPcolor}{%
    \newif\ifGPcolor
    \GPcolortrue
  }{}%
  \@ifundefined{ifGPblacktext}{%
    \newif\ifGPblacktext
    \GPblacktexttrue
  }{}%
  \let\gplgaddtomacro\g@addto@macro
  \gdef\gplbacktext{}%
  \gdef\gplfronttext{}%
  \makeatother
  \ifGPblacktext
    \def\colorrgb#1{}%
    \def\colorgray#1{}%
  \else
    \ifGPcolor
      \def\colorrgb#1{\color[rgb]{#1}}%
      \def\colorgray#1{\color[gray]{#1}}%
      \expandafter\def\csname LTw\endcsname{\color{white}}%
      \expandafter\def\csname LTb\endcsname{\color{black}}%
      \expandafter\def\csname LTa\endcsname{\color{black}}%
      \expandafter\def\csname LT0\endcsname{\color[rgb]{1,0,0}}%
      \expandafter\def\csname LT1\endcsname{\color[rgb]{0,1,0}}%
      \expandafter\def\csname LT2\endcsname{\color[rgb]{0,0,1}}%
      \expandafter\def\csname LT3\endcsname{\color[rgb]{1,0,1}}%
      \expandafter\def\csname LT4\endcsname{\color[rgb]{0,1,1}}%
      \expandafter\def\csname LT5\endcsname{\color[rgb]{1,1,0}}%
      \expandafter\def\csname LT6\endcsname{\color[rgb]{0,0,0}}%
      \expandafter\def\csname LT7\endcsname{\color[rgb]{1,0.3,0}}%
      \expandafter\def\csname LT8\endcsname{\color[rgb]{0.5,0.5,0.5}}%
    \else
      \def\colorrgb#1{\color{black}}%
      \def\colorgray#1{\color[gray]{#1}}%
      \expandafter\def\csname LTw\endcsname{\color{white}}%
      \expandafter\def\csname LTb\endcsname{\color{black}}%
      \expandafter\def\csname LTa\endcsname{\color{black}}%
      \expandafter\def\csname LT0\endcsname{\color{black}}%
      \expandafter\def\csname LT1\endcsname{\color{black}}%
      \expandafter\def\csname LT2\endcsname{\color{black}}%
      \expandafter\def\csname LT3\endcsname{\color{black}}%
      \expandafter\def\csname LT4\endcsname{\color{black}}%
      \expandafter\def\csname LT5\endcsname{\color{black}}%
      \expandafter\def\csname LT6\endcsname{\color{black}}%
      \expandafter\def\csname LT7\endcsname{\color{black}}%
      \expandafter\def\csname LT8\endcsname{\color{black}}%
    \fi
  \fi
  \setlength{\unitlength}{0.0500bp}%
  \begin{picture}(7936.00,3400.00)%
    \gplgaddtomacro\gplbacktext{%
      \csname LTb\endcsname%
      \put(1078,704){\makebox(0,0)[r]{\strut{}-0.05}}%
      \put(1078,1051){\makebox(0,0)[r]{\strut{} 0}}%
      \put(1078,1399){\makebox(0,0)[r]{\strut{} 0.05}}%
      \put(1078,1746){\makebox(0,0)[r]{\strut{} 0.1}}%
      \put(1078,2093){\makebox(0,0)[r]{\strut{} 0.15}}%
      \put(1078,2440){\makebox(0,0)[r]{\strut{} 0.2}}%
      \put(1078,2788){\makebox(0,0)[r]{\strut{} 0.25}}%
      \put(1078,3135){\makebox(0,0)[r]{\strut{} 0.3}}%
      \put(1210,484){\makebox(0,0){\strut{} 0}}%
      \put(2053,484){\makebox(0,0){\strut{} 0.0005}}%
      \put(2896,484){\makebox(0,0){\strut{} 0.001}}%
      \put(176,1919){\rotatebox{-270}{\makebox(0,0){\strut{}$\mu$}}}%
      \put(2390,154){\makebox(0,0){\strut{}$\varphi_0$ [$M_P$]}}%
    }%
    \gplgaddtomacro\gplfronttext{%
    }%
    \gplgaddtomacro\gplbacktext{%
      \csname LTb\endcsname%
      \put(5046,704){\makebox(0,0)[r]{\strut{}-0.05}}%
      \put(5046,1109){\makebox(0,0)[r]{\strut{} 0}}%
      \put(5046,1514){\makebox(0,0)[r]{\strut{} 0.05}}%
      \put(5046,1920){\makebox(0,0)[r]{\strut{} 0.1}}%
      \put(5046,2325){\makebox(0,0)[r]{\strut{} 0.15}}%
      \put(5046,2730){\makebox(0,0)[r]{\strut{} 0.2}}%
      \put(5046,3135){\makebox(0,0)[r]{\strut{} 0.25}}%
      \put(5178,484){\makebox(0,0){\strut{} 0}}%
      \put(5965,484){\makebox(0,0){\strut{} 0.0005}}%
      \put(6752,484){\makebox(0,0){\strut{} 0.001}}%
      \put(7539,484){\makebox(0,0){\strut{} 0.0015}}%
      \put(4144,1919){\rotatebox{-270}{\makebox(0,0){\strut{}$\mu$}}}%
      \put(6358,154){\makebox(0,0){\strut{}$\varphi_0$ [$M_P$]}}%
    }%
    \gplgaddtomacro\gplfronttext{%
    }%
    \gplbacktext
    \put(0,0){\includegraphics{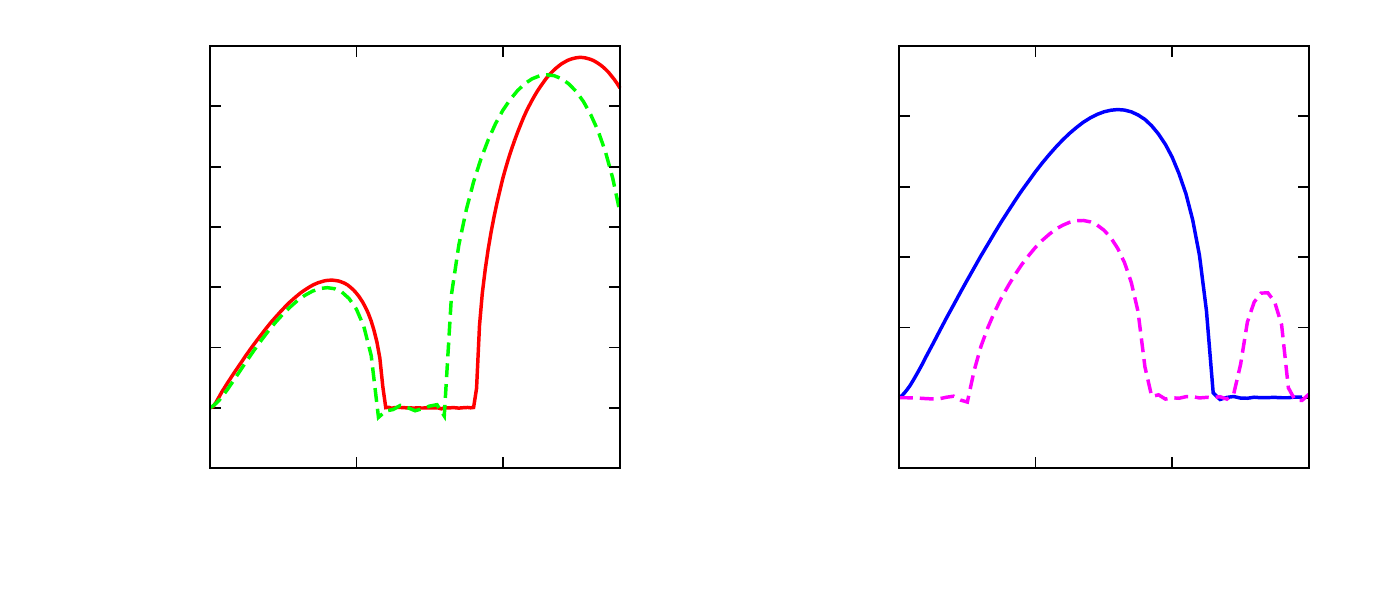}}%
    \gplfronttext
  \end{picture}%
\endgroup